\documentclass[%
 reprint,
 superscriptaddress,
 amsmath,amssymb,
 aps,
prb,
]{revtex4-2}

\usepackage{graphicx} 
\usepackage{dcolumn} 
\usepackage{bm}
\usepackage{hyperref}
\usepackage{tikz}
\usetikzlibrary{calc}

\usepackage[export]{adjustbox}

\usepackage{ellipsis} 

\usepackage{tikz}
\usetikzlibrary{calc}
\newcolumntype{M}[1]{>{\centering\arraybackslash}m{#1}}

\usepackage[final]{changes}

\hypersetup{
    colorlinks=true,
    linkcolor=red,
    filecolor=magenta,      
    urlcolor=cyan,
}
\usepackage{braket}
\usepackage{physics}
\usepackage{chemformula}

\usepackage{bm}

\newcommand{\hscha}{\hat{\mathcal{H}}}

\newcommand{\unlscha}{u}
\newcommand{\phatnlscha}[1]{\hat{p}_{#1}}

\newcommand{\phatbmnlscha}{\hat{\bm{p}}}

\newcommand{\Ynlscha}[2]{\Upsilon_{\text{nl},#1#2}}
\newcommand{\Ytildenlscha}[2]{\widetilde{\Upsilon}_{\text{nl},#1#2}}

\newcommand{\Yinvnlscha}[2]{\overset{-1}{\Upsilon}_{\text{nl},#1#2}}
\newcommand{\Ybmnlscha}{\bm{\Upsilon}_\text{nl}}
\newcommand{\Ytildebmnlscha}{\widetilde{\bm{\Upsilon}}_\text{nl}}

\newcommand{\Anlscha}[2]{A_{\text{nl},#1#2}}
\newcommand{\Atildenlscha}[2]{\widetilde{A}_{\text{nl},#1#2}}

\newcommand{\Abmnlscha}{\bm{A}_\text{nl}}
\newcommand{\Atildebmnlscha}{\widetilde{\bm{A}}_\text{nl}}

\newcommand{\Tnlscha}[2]{\Theta_{\text{nl},#1#2}}
\newcommand{\Ttildenlscha}[2]{\widetilde{\Theta}_{\text{nl},#1#2}}
\newcommand{\Tbmnlscha}{\bm{\Theta}_\text{nl}}
\newcommand{\Ttildebmnlscha}{\widetilde{\bm{\Theta}}_\text{nl}}

\newcommand{\Yovnlscha}[1]{\overline{\Upsilon}_{\text{nl},#1}}

\newcommand{\Aovnlscha}[1]{\overline{A}_{\text{nl},#1}}

\newcommand{\FCnlscha}[2]{\Phi_{\text{nl},#1#2}}
\newcommand{\FCbmnlscha}{\bm{\Phi}_\text{nl}}

\newcommand{\Dnlscha}[2]{D_{\text{nl},#1#2}}
\newcommand{\Dbmnlscha}{\bm{D}_\text{nl}}

\newcommand{\onlscha}[1]{\omega_{\text{nl},#1}}

\newcommand{\nnlscha}[1]{n_{\text{nl},#1}}

\newcommand{\polnlscha}[2]{e_{\text{nl},#1}^{#2}}
\newcommand{\polbmnlscha}[1]{\bm{e}_{\text{nl},#1}}

\newcommand{\dlogJdq}[1]{d_{#1}}
\newcommand{\dtildelogJdq}[1]{\widetilde{d}_{#1}}
\newcommand{\dtildebmlogJdq}{\widetilde{\bm{d}}}
\newcommand{\dlogJdqbm}{\bm{d}}

\newcommand{\Jinvnlscha}[2]{\overset{-1}{J}{}^{#1}_{#2}}
\newcommand{\Jinvbmnlscha}{\overset{-1}{\bm{J}}}

\newcommand{\Jtildeinvnlscha}[2]{\overset{-1}{\widetilde{J}}{}^{#1}_{#2}}
\newcommand{\Jtildeinvbmnlscha}{\overset{-1}{\widetilde{\bm{J}}}}

\newcommand{\gnlscha}[2]{g{}^{#1#2}}
\newcommand{\gtildenlscha}[2]{\widetilde{g}{}^{#1#2}}

\newcommand{\gtildezerobmnlscha}{\widetilde{\bm{g}}_0}
\newcommand{\gbmtildenlscha}{\widetilde{\bm{g}}}
\newcommand{\gbmnlscha}{\bm{g}}

\newcommand{\Jnlscha}[2]{J^{#1}_{#2}}

\newcommand{\Jtildenlscha}[2]{\widetilde{J}{}^{#1}_{#2}}
\newcommand{\Jtildebmnlscha}{\widetilde{\bm{J}}}

\newcommand{\detJnlscha}{\mathcal{J}}

\newcommand{\xinlscha}{\xi}
\newcommand{\xibmnlscha}{\bm{\xi}}

\newcommand{\freeparamnlscha}[1]{\Gamma_{\text{nl},#1}}
\newcommand{\freeparambmnlscha}{\bm{\Gamma}_{\text{nl}}}

\newcommand{\allfreeparameters}{\bm{\mathrm{x}}}

\newcommand{\Fnl}{F_{\text{nl}}}
\newcommand{\Snl}{S_{\text{nl}}}

\newcommand{\rhohatnlscha}{\hat{\rho}_{\text{nl}}}
\newcommand{\rhocartnlscha}{\rho_{\allfreeparameters}}

\newcommand{\gaussnlscha}{\overline{\rho}_{\text{nl}}}
\newcommand{\rhouhatnlscha}{\hat{\overline{\rho}}_{\text{nl}}}
\newcommand{\Hhatnlscha}{\hat{\overline{\mathcal{H}}}_\text{nl}}

\newcommand{\Knlscha}{\mathcal{K}_{\text{nl}}}

\newcommand{\Lonenlscha}[1]{\mathcal{L}^{(1)}_{#1}}
\newcommand{\Ltwonlscha}[2]{\mathcal{L}^{(2)}_{#1#2}}

\newcommand{\Lonebmnlscha}{\bm{\mathcal{L}}^{(1)}}
\newcommand{\Ltwobmnlscha}{\bm{\mathcal{L}}^{(2)}}

\newcommand{\Ktwonlscha}[2]{\mathcal{K}^{(2)}_{#1#2} }
\newcommand{\Konenlscha}[1]{\mathcal{K}^{(1)}_{#1} }

\newcommand{\Ktwobmnlscha}{\bm{\mathcal{K}}^{(2)} }
\newcommand{\Konebmnlscha}{\bm{\mathcal{K}}^{(1)} }

\newcommand{\Kzeronlscha}{\mathcal{K}^{(0)}}

\newcommand{\myint}{\int_{-\infty}^{+\infty} \hspace{-0.2cm}}

\newcommand{\averagegaussnl}[1]{\left\langle #1 \right\rangle_\text{nl}}

\newcommand{\masstns}[1]{\mathcal{M}_{#1}}
\newcommand{\masstnsinv}[1]{\overset{-1}{\mathcal{M}}_{#1}}
\newcommand{\masstnsbm}{\bm{\mathcal{M}}}
\newcommand{\sqrtmasstnsbm}{\sqrt{\bm{\mathcal{M}}}}
\newcommand{\sqrtmasstnsTbm}{\sqrt{\bm{\mathcal{M}}}^T}
\newcommand{\sqrtmasstns}[1]{\sqrt{\mathcal{M}}_{#1}}
\newcommand{\sqrtmasstnsT}[1]{\sqrt{\mathcal{M}}^T_{#1}}
\newcommand{\invsqrtmasstns}[1]{\overset{-1}{\sqrt{\mathcal{M}}}_{#1}}
\newcommand{\invsqrtmasstnsT}[1]{\overset{-T}{\sqrt{\mathcal{M}}}_{#1}}

\newcommand{\Tru}{\overline{\Tr}}
\newcommand{\Kgeom}{\overline{K}}
\newcommand{\Khatgeom}{\hat{\overline{K}}}
\newcommand{\VuBO}{\overline{V}^{\text{(BO)}}}
\newcommand{\VhatuBO}{\hat{\overline{V}}{}^{\text{(BO)}}}
\newcommand{\Vgeom}{\overline{V}_\text{g}}

\newcommand{\Vgeomfunc}{V_\text{g}}
\newcommand{\urhoBO}{\overline{\rho}{}^\text{(BO)}}
\newcommand{\uhatrhoBO}{\hat{\overline{\rho}}{}^\text{(BO)}}
\newcommand{\uhamiltonianBO}{\overline{H}{}^\text{(BO)}}
\newcommand{\uhathamiltonianBO}{\hat{\overline{H}}{}^\text{(BO)}}

\newcommand{\HAhamiltonian}{\overline{\mathcal{H}}}
\newcommand{\HAhathamiltonian}{\hat{\overline{\mathcal{H}}}}
\newcommand{\HAhamiltonianQ}{\overline{\mathcal{H}}{}^\text{(q)}}
\newcommand{\HAFC}[1]{\overline{\Phi}_{#1}}
\newcommand{\HAFCbm}{\overline{\bm{\Phi}}}

\newcommand{\Znlscha}{\overline{\mathcal{Z}}_\text{nl}}

\begin{document}

\preprint{APS/123-QED}

\title{Beyond Gaussian fluctuations of quantum anharmonic nuclei.}
\author{Antonio Siciliano}
\email[]{antonio.siciliano@uniroma1.it}
\affiliation{Dipartimento di Fisica, Università di Roma La Sapienza, Piazzale Aldo Moro 5, 00185 Roma, Italy}
\author{Lorenzo Monacelli}
\affiliation{Dipartimento di Fisica, Università di Roma La Sapienza, Piazzale Aldo Moro 5, 00185 Roma, Italy}
\affiliation{Theory and Simulation of Materials (THEOS), and National Centre for Computational Design
and Discovery of Novel Materials (MARVEL), École Polytechnique Fédérale de Lausanne, 1015
Lausanne, Switzerland}
\author{Francesco Mauri}
\affiliation{Dipartimento di Fisica, Università di Roma La Sapienza, Piazzale Aldo Moro 5, 00185 Roma, Italy}

\date{\today}

\begin{abstract}
The Self-Consistent Harmonic Approximation (SCHA) describes atoms in solids, including quantum fluctuations and anharmonic effects, in a non-perturbative way. It computes ionic free energy variationally, constraining the atomic quantum-thermal fluctuations to be Gaussian. Consequently, the entropy is analytical; there is no need for thermodynamic integration or heavy diagonalization to include finite temperature effects. In addition, as the probability distribution is fixed, SCHA solves all the equations with Monte Carlo integration without employing Metropolis sampling of the quantum phase space. Unfortunately, the Gaussian approximation breaks down for rotational modes and tunneling effects. We show how to describe these non-Gaussian fluctuations using the quantum variational principle at finite temperatures, keeping the main advantage of SCHA: direct access to free energy. Our method, nonlinear SCHA (NLSCHA), employs an invertible nonlinear transformation to map Cartesian coordinates into an auxiliary manifold parametrized by a finite set of variables. So, we adopt a Gaussian \textit{ansatz} for the density matrix in this new coordinate system. The nonlinearity of the mapping ensures that NLSCHA enlarges the SCHA variational subspace, and its invertibility conserves the information encoded in the density matrix. We evaluate the entropy in the auxiliary space, where it has a simple analytical form. As in the SCHA, the variational principle allows for optimizing free parameters to minimize free energy. Finally, we show that, for the first time, NLSCHA gives direct access to the entropy of a crystal with non-Gaussian degrees of freedom. 
\end{abstract}

\maketitle


\section{Introduction}
\label{Introduction}
Simulating materials in a wide temperature and pressure range is required in many fields, e.g.\ from regulation of biological processes and drug development ($T\sim 200-300$ K $P \sim 0.1$ MPa) \cite{CrystalStructuresDrugs}, to diamond anvil cell experiments ($T\sim 0-300$ K $P \sim 0-300$ GPa) \cite{SuperconductivityHighPressureHydrides} and for understanding the inner structures of planets (up to $T\sim 5000$ K $P \sim 600$ GPa) \cite{FeOinnercore}. Reproducing this variety of conditions \textit{insilico} is challenging due to the significant role of anharmonicity when atoms explore extensive regions of the energy landscape. Furthermore, when light atoms are present, as is the case in gas giants' cores or hydrate superconductors, quantum effects completely disrupt classical predictions. Only Path Integral Molecular Dynamics (PIMD) accurately accounts for all these effects at finite temperatures. Despite the increasing computational power available, PIMD is impractical for systematic investigations of phase diagrams. In addition, within PIMD, we can not compute dynamical responses, as probed by Raman and infrared (IR) spectroscopies.

In recent years, the Self-Consistent Harmonic Approximation (SCHA) \cite{HootonSCHA,SCHA_main,TadanoSrTiO3,SCP1,SCPII} has been established as an accurate and efficient free energy method with the possibility of computing \textit{ab-initio} Raman and IR spectra \cite{TDSCHA_monacelli,TDSCHA_mio,LihmTDSCHA}. 
The SCHA constrains the ionic density matrix to have a Gaussian form in Cartesian coordinates parameterized by the average atomic position and the quantum-thermal fluctuations. The SCHA trial density matrix solves an auxiliary harmonic Hamiltonian so it has an analytical expression of the entropy, meaning that free energy calculations do not require thermodynamic integration. In addition, thanks to the quantum variational principle, we optimize the free parameters to get the best free energy estimation \cite{SCHA_errea}. The SCHA describes very well the quantum thermodynamics of anharmonic crystals as shown by many successful applications; from hydrogen-rich compounds \cite{H3S_SSCHA, LH10_SSCHA, black_metal_hydrogen, Monacelli_hydrogen_new,Marco_ICE,cherubini2024quantum}, perovskites \cite{ranalli2022temperature,MonacelliCsSnI3}, charge density wave materials (CDW) \cite{CDW_melt_TiSe2,CDW_NbSe2_new,SSCHA_CDW_VSe2,CDW_NbS2}, and low dimensional materials, nanoclusters \cite{NanoclustersSCHA}, and acetylenic carbon chains \cite{Romanin2021}.

However, SCHA has limitations. Its weaknesses stem from its assumption of Gaussian atomic fluctuations. A clear example is quantum tunneling between two or more minima in the energy landscape. In multi-minima environments, the exact wave function significantly diverges from a Gaussian, manifesting a multi-peak shape \cite{H3S_SSCHA, CherubiniH3S}.
The Gaussian approximation in Cartesian coordinates also fails when roto-librational modes are active, as observed in Refs \cite{ceriotti_anharmonic_free_energies,HydrogenIVMorresi}. The SCHA method leads to an undesirable hybridization of low-energy rotational modes with other high-energy modes; for instance, in the \ch{H2} molecule, it merges roto-librational modes with the vibron \cite{HydrogenIVMorresi}. An accurate description of these modes is particularly crucial for hybrid organic-inorganic perovskites, promising candidates for efficient next-generation solar cells. These compounds feature octahedral structures that accommodate molecular cations, with orientational order/disorder and rotational degrees of freedom impacting optoelectronic properties and structural stability \cite{OrganicCationRotationPerovskites}. 
\begin{figure*}[!htb]
    \centering
    \begin{minipage}[c]{1.0\linewidth}
    \includegraphics[width=1.0\textwidth]{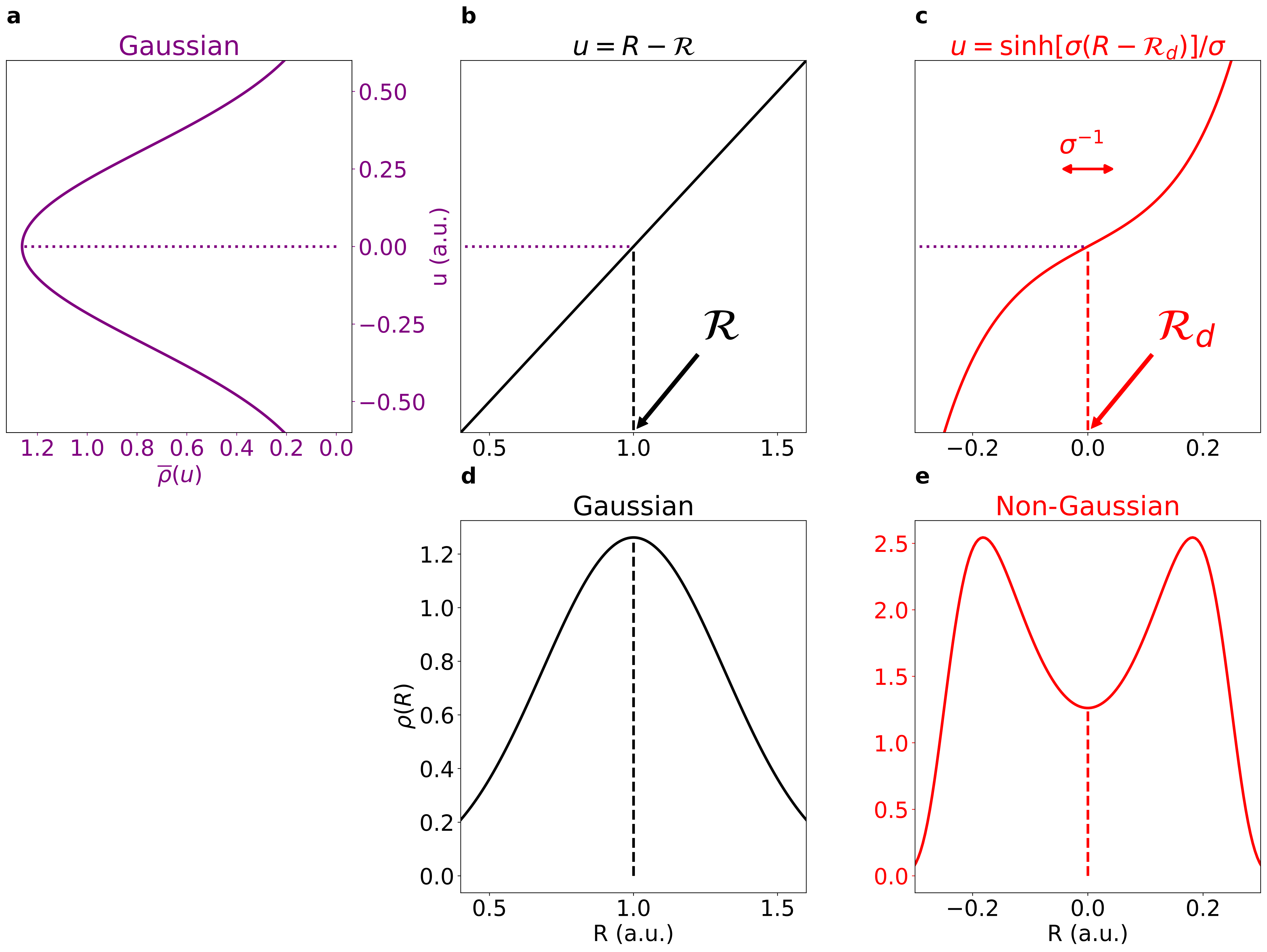}
    \end{minipage}
    \caption{Our starting point to describe the quantum thermodynamics of a crystal is to define a Gaussian trial density matrix in an auxiliary space $\unlscha$. Fig.\ a shows the corresponding probability distribution $\overline{\rho}(u)=\sqrt{\Upsilon/(2\pi)}\exp{-\Upsilon u^2 /2}$. Figs a-b-d illustrate the Self-Consistent Harmonic Approximation (SCHA) method. SCHA transforms $\overline{\rho}(u)$ using a linear transformation of the Cartesian variable $R$, $u =  R -\mathcal{R}$ (Fig.\ b). Fig.\ d reports the SCHA probability distribution in $R$-space. Thanks to the quantum variational principle, SCHA optimizes the free parameters $(\Upsilon,\mathcal{R})$ to best estimate the Born-Oppenheimer free energy within a Gaussian \textit{ansatz} for the density matrix. To describe non-Gaussian fluctuations, we transform $\overline{\rho}(u)$ with a parameterized nonlinear invertible change of variables. This method is called nonlinear SCHA (NLSCHA). Figs a-c-e illustrate an example of NLSCHA combined with a \textit{ad-hoc} parametrization that allows the description of tunneling modes. Fig.\ c shows the transformation employed $u=\xi(R)=\sinh[\sigma(R - \mathcal{R}_\text{d})]/\sigma$. The free parameters are $\mathcal{R}_\text{d}$ and $\sigma>0$, respectively, the center and the width of the hyperbolic distortion (see Fig.\ c). Here, we set $\sigma^{-1}=0.1$ a.u.\ to have non-Gaussian fluctuations (Fig.\ e). As in the SCHA, employing the quantum variational principle, we optimize $(\Upsilon,\mathcal{R}_\text{d},\sigma)$. So, our parametrization activates non-Gaussian fluctuations only if they minimize the ionic free energy. Otherwise, we reproduce the SCHA results from $\sigma \rightarrow 0$.}
    \label{fig: nlscha idea}
\end{figure*} 

Each of these scenarios represents a failure for the SCHA. We propose a solution based on an invertible nonlinear change of variables that maps Cartesian coordinates into an auxiliary manifold parameterized by a finite set of variables (Fig.\ \ref{fig: nlscha idea} a-c). We assume a Gaussian form for the ionic density matrix in this space. The nonlinearity of our mapping ensures that when projected in Cartesian space, the density matrix exhibits non-Gaussian fluctuations (Fig.\ \ref{fig: nlscha idea} e), contrary to SCHA (Fig.\ \ref{fig: nlscha idea} d). 
The new density matrix solves a harmonic Hamiltonian in the nonlinear coordinates, so it has an analytical expression for the entropy. The transformation's invertibility implies information conservation when our density matrix is projected in Cartesian space. Hence, we compute the quantum ionic free energy, including non-Gaussian fluctuations, without extra computational costs related to entropy, maintaining SCHA's main advantage over exact methods: direct access to the free energy. In addition, we employ the quantum variational principle at finite temperatures to optimize the parameters of both the nonlinear transformation and the auxiliary harmonic Hamiltonian. 
Another essential feature of employing an auxiliary Gaussian density matrix is that all the equations can be evaluated by Monte Carlo integration \cite{SCHA_main} and there is no need for Metropolis extractions to converge to equilibrium fluctuations. So, NLSCHA minimizes the ionic free energy with an overall computational cost similar to the standard SCHA, with electronic energies and forces calculations primarily determining this cost.
We also show that our analytical expression for the entropy is minus the temperature derivative of the free energy, so we have direct access to the thermodynamic entropy of crystals with non-Gaussian degrees of freedom.

The structure of our work is organized as follows. In section \ref{SEC: The Self-Consistent Harmonic Approximation}, we define the free energy of nuclei in the Born-Oppenheimer approximation (section \ref{SUBSEC: Ionic free energy of crystals}). Subsequently, we introduce the quantum variational principle as the foundation for approximate methods (section \ref{SUBSEC: The quantum variational principle}), followed by a review of the fundamental concepts of SCHA (section \ref{SUBSEC: The Self-Consistent Harmonic Approximation}). In section \ref{SEC: Quantum mechanics in nonlinear coordinates}, we discuss the class of nonlinear transformations employed (section \ref{SUBSEC: The nonlinear change of variables}). Then, we present an exact mapping of the BO quantum thermodynamics in nonlinear coordinates (section \ref{SUBSEC: Thermodynamics in arbitrary coordinates}). In this framework, we analyze the role of quantum fluctuations in the harmonic approximation, defining the trial Hamiltonian employed by nonlinear SCHA (section \ref{SUBSEC: The harmonic approximation}). 
The underlying Hamiltonian formalism is then crucial as it allows us to evaluate the entropy analytically. In section \ref{SEC: Theory of nonlinear self-consistent harmonic approximation}, we discuss the variational manifold of nonlinear SCHA and its new free parameters (section \ref{SUBSEC: The variational manifold of nonlinear SCHA}). In addition, we show that the entropy is still an analytical quantity in the theory even if there are non-Gaussian fluctuations (section \ref{SUBSEC: The entropy}). The new free energy and its gradients are presented in section \ref{SUBSEC: The free energy}, showing that most of the computational cost is associated with BO energies and forces calculations. Then, in section \ref{SEC: 1D toy model}, a toy model shows the improvements of our generalization with respect to standard SCHA. The anharmonic potential used has a form similar to the one used to model \ch{H3S} transition \cite{CherubiniH3S} where the SCHA misses non-Gaussian fluctuations. In addition, we prove that the NLSCHA entropy coincides with minus the temperature derivative of the free energy, satisfying a fundamental thermodynamic relation (section \ref{SEC: Entropy discussion}). Finally, section \ref{SEC: The stochastic approach} elucidates a stochastic implementation of the method to improve its efficiency.

\section{The Self-Consistent Harmonic Approximation}
\label{SEC: The Self-Consistent Harmonic Approximation}
\replaced{Starting from the BO density matrix (section \ref{SUBSEC: Ionic free energy of crystals}) we introduce the quantum variational principle as the foundational concept for approximate methods (section \ref{SUBSEC: The quantum variational principle}). Then we provide a concise overview of the SCHA emphasizing the origin of its weaknesses (section \ref{SUBSEC: The Self-Consistent Harmonic Approximation}).}{We introduce the focus of our work, which is the crystal free energy within the Born-Oppenheimer (BO) approximation. Recognizing the high computational cost of exact methods, we delve into the quantum free-energy variational principle as the foundational concept for approximate methods. Subsequently, we provide a concise overview of the Self-Consistent Harmonic Approximation (SCHA), emphasizing the origin of its weaknesses.}

\subsection{Ionic free energy of crystals}
\label{SUBSEC: Ionic free energy of crystals}
We consider $N$ interacting distinguishable quantum particles in 3D. Our description of nuclear motion is based on the Born-Oppenheimer (BO) approximation. In this framework, the quantum Hamiltonian for a crystal or molecule is
\begin{equation}
\label{eq: H Born Oppenheimer}
    \hat{H}^{\text{(BO)}}  = \sum_{a=1}^{3N} \frac{\hat{P}^2_a}{2m_a} + V^{\text{(BO)}}(\hat{\bm{R}}) 
\end{equation}
For brevity, we indicate with $a = (A, \alpha)$ a composite index with the atomic index $A$ and Cartesian index $\alpha$, $m_a = m_A$ is the mass of atom $A$, $\hat{P}_{a}=\hat{P}_{A,\alpha}$ is the momentum operator of atom $A$ along the $\alpha$ direction and $\hat{\bm R} = (\hat{\bm R}_1,..,\hat{\bm R}_N)$ is the collection of all position operators with eigenstates $\ket{\bm{R}}=\ket{\bm{R}_1,..,\bm{R}_N}$. According to the BO approximation, $V^{\text{(BO)}}(\hat{\bm R})$, the BO energy surface (BOES) is the ground state energy of the electronic Schrodinger equation at fixed nuclei configuration. This is computed, for example, using Density Functional Theory (DFT), Quantum Monte Carlo (QMC), or appropriately parametrized force fields.
The density matrix describing the system at equilibrium at a temperature $T$ is
\begin{subequations}
\label{eq: exact BO quantities rho Z}
\begin{align}
    & \hat{\rho}^{\text{(BO)}} = \frac{\exp{-\beta  \hat{H}^{\text{(BO)}}}}{Z^{\text{(BO)}}} \label{eq: BO rho}\\
    & Z^{\text{(BO)}} = \Tr\left[\exp{-\beta  \hat{H}^{\text{(BO)}}}\right] \label{eq: BO Z}
\end{align}
\end{subequations}
where $\beta^{-1} = k_\text{B} T$ is the thermal energy, $k_\text{B}$ is the Boltzmann constant and $\Tr[\circ]$ indicates the trace operation over the $N$ atoms Hilbert space. The free energy is then computed either from the partition function or from the density matrix
\begin{equation}
\label{eq: BO free energy}
\begin{aligned}
        F^\text{(BO)} 
        = U^\text{(BO)} - T S^\text{(BO)}
\end{aligned}
\end{equation}
where the internal energy and the entropy are
\begin{subequations}
\begin{align}
    U^\text{(BO)} & = \Tr\left[\hat{\rho}^{\text{(BO)}}\hat{H}^{\text{(BO)}} \right] \label{eq: BO U}\\
    S^\text{(BO)} & = -k_\text{B} \Tr\left[\hat{\rho}^{\text{(BO)}}\log\left(\hat{\rho}^{\text{(BO)}}\right)\right] \label{eq: BO S}
\end{align}
\end{subequations}
The Born-Oppenheimer (BO) free energy, denoted as $ F^\text{(BO)}$, plays a pivotal role in determining the relative stability among different structural phases. Achieving an accuracy of less than $1$ meV per atom in its computation is essential for making quantitative predictions about the stable structure of a crystal or molecule, which, in turn, defines all equilibrium properties. Its significance lies in that the lattice structure is not always experimentally accessible, as is the case for planetary inner cores. This limitation also applies to tabletop experiments on hydrogen-rich compounds with low X-ray cross-sections or compounds subjected to high pressure, where sample sizes can be prohibitively small. Furthermore, from the derivatives of the BO free energy, one derives all thermodynamic properties, including pressure, bulk modulus, specific heat, thermal expansion, etc.

Exact computation of the quantum free energy is not feasible because it requires diagonalizing an N-body Hamiltonian, as represented in Eq.\ \eqref{eq: H Born Oppenheimer}. The computational demands of this process grow exponentially with the number of atoms.
However, there are methods, such as Molecular Dynamics (MD) and Path-Integral MD (PIMD), designed to sample the statistical ensemble generated by the BO Hamiltonian directly. These methods are exact and unbiased in the classical and quantum regimes. Nevertheless, they necessitate extended equilibration times, with time steps depending on the lowest normal modes. Moreover, to incorporate quantum effects, replicating the crystal into several 'beads' is required. In both MD and PIMD, there is no direct access to the entropy. Consequently, there is an urgent need for approximate computationally inexpensive and reliable methods.

\subsection{The quantum variational principle}
\label{SUBSEC: The quantum variational principle}
Full diagonalization of the BO Hamiltonian poses an exceptionally challenging task, and other exact methods, such as MD/PIMD, come with substantial computational costs. To address this bottleneck, more efficient techniques have been developed based on the variational principle for quantum free energy. As a preliminary step, we define the free energy functional $F[\hat{\rho}]$ at temperature $T$, depending on a trial density matrix $\hat{\rho}$, as
\begin{subequations}
\begin{align}
    F[\hat{\rho}] & = U[\hat{\rho}] - T S[\hat{\rho}] \label{eq: free energy functional}\\
    U[\hat{\rho}] & = \Tr\left[\hat{\rho}\hat{H}^{\text{(BO)}} \right] \label{eq: functional U}\\
    S[\hat{\rho}] & = -k_\text{B} \Tr\left[\hat{\rho}\log\left(\hat{\rho}\right)\right] \label{eq: functional S}
\end{align}
\end{subequations}
Here, $U[\hat{\rho}]$ and $S[\hat{\rho}]$ represent the internal energy and entropy functional, respectively. The free energy functional possesses physical meaning only when $\hat{\rho}$ satisfies the conditions of a density matrix: it is normalized, Hermitian, and positive definite 
\begin{subequations}
\label{eq: physical conditions for density matrix}
\begin{align}
    \Tr[\hat{\rho}] & = 1   \\
    \hat{\rho} &= \hat{\rho}^\dagger \\
    \bra{\bm R} \hat{\rho}\ket{\bm R'} &\geq 0 
\end{align}
\end{subequations}
Under these conditions, $F[\hat{\rho}]$ is lower-bounded by the BO free energy (Eq.\ \eqref{eq: BO free energy})
\begin{equation}
    \label{def: variational principle}
    F^\text{(BO)} \leq F[\hat{\rho}]
\end{equation}
This equation represents the quantum free energy variational principle, allowing us to estimate $F^\text{(BO)}$ using a trial density matrix $\hat{\rho}$ with a controlled error. Our choice for $\hat{\rho}$ should be guided by our physical intuition to ensure a reliable approximate method.

\subsection{The Self-Consistent Harmonic Approximation}
\label{SUBSEC: The Self-Consistent Harmonic Approximation}
The Self-Consistent Harmonic Approximation (SCHA) emerges from the intuition that atoms vibrate around fixed equilibrium positions even in the most strongly anharmonic crystals. So, the Gaussian distribution is the least biased quantum distribution with fixed average positions and fluctuations. Given the variational principle, Eq.\ \eqref{def: variational principle}, the SCHA minimizes the free energy functional, Eq.\ \eqref{eq: free energy functional}, plugging a Gaussian density matrix $\hat{\rho} = \hat{\rho}_{\bm{\mathcal{R}},\bm{\Phi}}$. We consider only the subspace of the harmonic Gaussian density matrices
\begin{subequations}
\label{eq: scha rho + Z}
\begin{align}
    \hat{\rho}_{\bm{\mathcal{R}},\bm{\Phi}} &= \frac{\exp{-\beta  \hscha_{\bm{\mathcal{R}},\bm{\Phi}}}}{Z_{\bm{\mathcal{R}},\bm{\Phi}}} \label{eq: rho scha}\\
    Z_{\bm{\mathcal{R}},\bm{\Phi}} &= \Tr\left[\exp{-\beta  \hscha_{\bm{\mathcal{R}},\bm{\Phi}}}\right] \label{eq: Z scha}
\end{align}
\end{subequations}
where $\hscha_{\bm{\mathcal{R}},\bm{\Phi}}$ is the most general stable quadratic Hamiltonian
\begin{equation}
    \hscha_{\bm{\mathcal{R}},\bm{\Phi}} = \sum_{a=1}^{3N} \frac{\hat{P}^2_a}{2m_a} + \frac{1}{2}\sum_{ab=1}^{3N} (\hat{R}_a - \mathcal{R}_a)\Phi_{ab} (\hat{R}_b - \mathcal{R}_b).
\end{equation}
The free parameters in the SCHA are the average atomic positions  $\bm{\mathcal{R}}$, the so-called centroids, and the positive-definite force constant tensor $\bm{\Phi}$ which fixes along with $\beta$ the atomic fluctuations. Note that the eigenvalues of $\bm{\Phi}$ do not yield physical phonons but only auxiliary quantities, just like the Kohn-Scham bands in DFT.
The density matrix, Eq.\ \eqref{eq: scha rho + Z}, is optimized by minimizing the free energy $F[\hat{\rho}_{\bm{\mathcal{R}},\bm{\Phi}}]$ with respect to both $\bm{\mathcal{R}}$ and  $\bm{\Phi}$. The minimum SCHA free energy is reached when the following self-consistent conditions are satisfied 
\begin{subequations}
\label{eq: scha eq conditions}
\begin{align}    \Tr\left[\hat{\rho}^{(0)}_{\bm{\mathcal{R}},\bm{\Phi}}\pdv{V^{\text{(BO)}}}{\hat{R}_a}{ \hat{R}_b} \right] &
    = \Phi^{(0)}_{ab}  
    \label{def: eq condition scha phonons}\\
    \Tr\left[\hat{\rho}^{(0)}_{\bm{\mathcal{R}},\bm{\Phi}}
    \pdv{V^{\text{(BO)}}}{\hat{R}_a} \right] &   = 0
    \label{def: eq condition zero forces}
\end{align}
\end{subequations}
where $\hat{\rho}^{(0)}_{\bm{\mathcal{R}},\bm{\Phi}}$ is the SCHA equilibrium density matrix. Eqs \eqref{eq: scha eq conditions} renormalize the SCHA phonons, defined from $\bm{\Phi}^{(0)}$, and the average atomic positions, $\bm{\mathcal{R}}^{(0)}$, by anharmonicty and quantum fluctuations at finite temperature \cite{Bianco, TDSCHA_mio}.
We have discussed the SCHA method in its original formulation \cite{SCHA_main}, wherein the number of particles, temperature, and volume remain fixed. Indeed, the SCHA can also vary lattice vectors, optimizing the free energy in the isobaric ensemble \cite{monacelli2018pressure,SCP_Pressure}.

The self-consistent description of phonon dynamics (Eqs \eqref{eq: scha eq conditions}) has proven to be very powerful in materials hosting light atoms, such as hydride superconductors \ch{YH6}, \ch{LH10}, \cite{YH6_anomaloussuperconductivity,LH10_SSCHA}, phase XI of water ice \cite{Marco_ICE} and high-pressure hydrogen \cite{Monacelli_hydrogen_new,black_metal_hydrogen}. In these systems, classical approaches are inadequate as the light \ch{H} atoms present huge quantum fluctuations, breaking the harmonic approximation and reshaping the energy landscape.
The SCHA phonons are positive-definite and do not represent the physical excitations of the system, nor do they indicate structural instabilities \cite{TDSCHA_monacelli, TDSCHA_mio}. The second derivative of $F[\hat{\rho}_{\bm{\mathcal{R}},\bm{\Phi}}]$ with respect to the centroids $\bm{\mathcal{R}}$ \cite{Bianco} defines whether the structure is stable or not, and it was used to study charge-density-wave (CDW) materials \cite{CDW_melt_TiSe2,CDW_NbSe2_new,SSCHA_CDW_VSe2,CDW_NbS2}, where the lattice distortion induces a static modulation of the electronic charge.

Is it true that we possess a reliable, computationally inexpensive yet predictive technique? It depends on the material, as the SCHA confines nuclear fluctuations to Gaussians in Cartesian coordinates. As discussed in the section \ref{Introduction}, there are cases where this assumption dramatically breaks, as in rotations and tunneling. In this work, we present for the first time a generalization of SCHA to go beyond Gaussian fluctuations while keeping all the advantages and the same computational cost of the SCHA. 

\section{Quantum mechanics in nonlinear coordinates}
\label{SEC: Quantum mechanics in nonlinear coordinates}
To generalize the Self-Consistent Harmonic Approximation (SCHA), we apply a nonlinear change of variables, discussed in detail in section \ref{SUBSEC: The nonlinear change of variables}. This approach includes non-Gaussian fluctuations in Cartesian space using a Gaussian \textit{ansatz} in the auxiliary space defined by a finite set of free parameters. Notably, the application of nonlinear change of variables has precedent in the solid state community for various purposes. Ref.\ \cite{DFTcurvilinear1} utilized this technique to extend plane-wave-based Density Functional Theory (DFT), while Ref.\ \cite{DFTcurvilinear2} applied it to describe electronic responses to inhomogeneous adiabatic transformations. Also, physical chemistries employ internal coordinates, i.e., \ a specific class of nonlinear transformation, to describe the nuclear dynamics in molecules \cite{VSCF_curvilinear}.

To establish a solid foundation for our new theory, we express the Born-Oppenheimer thermodynamics of a crystal in auxiliary coordinates, as discussed in section \ref{SUBSEC: Thermodynamics in arbitrary coordinates}. Our mapping offers an exact and rigorous approach to redefine the quantum mechanical problem within a nonstandard, non-Cartesian space. Subsequently, in section \ref{SUBSEC: The harmonic approximation}, we discuss the harmonic approximation in this space, emphasizing the distinctions between the quantum and classical cases and setting the stage for nonlinear SCHA.

\subsection{The nonlinear change of variables}
\label{SUBSEC: The nonlinear change of variables}
To generalize the SCHA theory, we introduce a nonlinear change of variables on the atomic Cartesian coordinates $\bm{R}$
\begin{equation}
\label{eq: change of variables}
    \bm{R} = \xibmnlscha(\bm{\unlscha})
\end{equation}
where $\xibmnlscha$ is a general invertible nonlinear transformation that maps the Cartesian coordinates to a set of auxiliary variables $\bm{\unlscha}$ defined in $(-\infty,+\infty)^{3N}$ and with dimension of length.  We assume that $\xibmnlscha$ depends on a set of $S$ free parameters, denoted as $\freeparambmnlscha=\{\freeparambmnlscha^{(i)}\}$, where $i=1,\dots,S$. It is worth noting that each $\freeparambmnlscha^{(i)}$ can be either a $3N$ vector, a $3N\times 3N$ tensor, or higher-order tensors. For example, the SCHA method can be seen as the limit case where $\xibmnlscha$ is a linear translation of the atomic position (see also Fig.\ \ref{fig: nlscha idea})
\begin{equation}
    \bm{R} = \bm{\unlscha} + \bm{\mathcal{R}}
\end{equation}
where the centroid $\bm{\mathcal{R}}$, a $3N$ vector, is the only free parameter, so $S=1$. 
Several examples of nonlinear variables are found in quantum chemistry, where molecular geometry is described by internal variables, i.e.\ bond lengths, bond angles, and dihedral angles.  The choice of internal coordinates simplifies the representation of potential energy surfaces and facilitates the analysis of proteins' molecular structure and reactivity \cite{Mendolicchio2023}.   The key difference with Eq.\ \eqref{eq: change of variables} is that there are no free parameters for this class of transformations ($S=0$) and the internal variables are not defined in $(-\infty,+\infty)^{3N}$.

For what follows, we need some mathematical definitions. The mass-rescaled Jacobian tensors of $\xibmnlscha$ are
\begin{subequations}
\begin{align}
    \Jtildenlscha{a}{b} = \pdv{\widetilde{R}_a}{\widetilde{\unlscha}_b}
    \label{eq: 1-order jacobian} \\
    \Jtildeinvnlscha{a}{b} = \pdv{\widetilde{\unlscha}_a}{\widetilde{R}_b}
    \label{eq: 1-order inverse jacobian}
\end{align}
\end{subequations}
where $\widetilde{\unlscha}_i = \sqrt{m_i} \unlscha_i$ and $\widetilde{R}_i = \sqrt{m_i} R_i$. From the inverse, we introduce the positive definite inverse metric tensor
\begin{equation}
\label{eq: def metric tensor}
    \gtildenlscha{a}{b} = \sum_{i=1}^{3N} \Jtildeinvnlscha{a}{i} \Jtildeinvnlscha{b}{i}
\end{equation}
We denote the determinant of the Jacobian as
\begin{equation}
\label{eq: det of J}
    \detJnlscha = \det\left(\Jtildebmnlscha\right)
\end{equation}
and, as the transformation is invertible, we choose to work always with
\begin{equation}
\label{eq: invertible transformation}
    \detJnlscha > 0
\end{equation}
In the following, we need also the distortion vector
\begin{equation}
\label{eq: d log J d u}
    \dtildelogJdq{a} = \frac{1}{2} \pdv{\log(\detJnlscha)}{\widetilde{\unlscha}_a} 
\end{equation}
which is zero if the transformation is linear.

\subsection{Thermodynamics in nonlinear coordinates}
\label{SUBSEC: Thermodynamics in arbitrary coordinates}
Quantum mechanics is rigorously defined exclusively in Cartesian space, where classical Poisson brackets establish fundamental commutation rules for Cartesian position $\hat{\bm{R}}$ and momentum $\hat{\bm{P}}$. This section demonstrates how to recast BO quantum thermodynamics (see section \ref{SUBSEC: Ionic free energy of crystals}) in nonlinear variables so that the free energy is conserved. We impose that all observables remain invariant under a nonlinear transformation, ensuring that they can be evaluated in $\bm{\unlscha}$-space as
\begin{equation}
\label{eq: Tr O rho R-u} 
    \Tr\left[\hat{O}\hat{\rho}^\text{(BO)}\right]  =
    \Tru\left[\hat{\overline{O}} \hspace{0.05cm}\uhatrhoBO  \right] 
\end{equation}
Here, $\hat{\overline{O}}$ and $\uhatrhoBO$ are the $\bm{\unlscha}$-space counterpart of  $\hat{O}$ and $\hat{\rho}^\text{(BO)}$. $\Tru\left[\hat{\overline{O}}\right]$ is the trace computed assuming $\{\ket{\bm{\unlscha}}\}$ as a complete basis, whereas for $\Tr[\hat{O}]$ we employ the Cartesian basis $\{\ket{\bm{R}}\}$. Eq.\ \eqref{eq: Tr O rho R-u} not only shows how each term in $F^\text{(BO)}$ (Eq.\ \eqref{eq: BO free energy}) transforms in the auxiliary space but also provides a rigorous definition of operators in the new coordinates.

From Eq.\ \eqref{eq: Tr O rho R-u} we derive the explicit expressions of $\overline{O}(\bm{\unlscha}, \bm{\unlscha}')$ and $\urhoBO(\bm{\unlscha},\bm{\unlscha}')$
\begin{equation}
\label{eq: invariance of observables}
\begin{aligned}
    \myint& d\bm{R} \myint d\bm{R}' O(\bm{R}, \bm{R}') \rho^\text{(BO)}(\bm{R}',\bm{R})  \\ 
    = \myint& d\bm{\unlscha} \myint d\bm{\unlscha}' 
    \detJnlscha(\bm{\unlscha}) \detJnlscha(\bm{\unlscha}')
    O(\xibmnlscha(\bm{\unlscha}), \xibmnlscha(\bm{\unlscha}')) \rho^\text{(BO)}(\xibmnlscha(\bm{\unlscha}'),\xibmnlscha(\bm{\unlscha}))  \\
    =
    \myint& d\bm{\unlscha} \myint d\bm{\unlscha}' \overline{O}(\bm{\unlscha}, \bm{\unlscha}') \urhoBO(\bm{\unlscha}',\bm{\unlscha})
\end{aligned}
\end{equation}
At this point, the Jacobians can be repartitioned in many ways among $\overline{O}(\bm{\unlscha}, \bm{\unlscha}')$ and $\urhoBO(\bm{\unlscha}',\bm{\unlscha})$, so there are several equivalent choices to define the new operators in such a way that the observables are invariant. We impose that a Hermitian operator in $\bm{R}$ should be so also in $\bm{\unlscha}$ and that the density matrix transforms as an operator, so the ambiguity disappears and we get
\begin{equation}
\label{eq: O(u,u') general expression}
    \overline{O}(\bm{\unlscha}, \bm{\unlscha}') = \sqrt{\detJnlscha(\bm{\unlscha})}  O(\xibmnlscha(\bm{\unlscha}), \xibmnlscha(\bm{\unlscha}')) \sqrt{\detJnlscha(\bm{\unlscha}')}
\end{equation}
Eq.\ \eqref{eq: O(u,u') general expression} provides a method to get the operators in the auxiliary space using only the Jacobian's determinant (Eq.\ \eqref{eq: det of J}) and the operator's matrix elements in Cartesian space. In addition, the solution of Eq.\ \eqref{eq: invariance of observables} demonstrates \textit{a posteriori} that  $\{\ket{\bm{\unlscha}}\}$  is a complete basis set on which we compute all the observables.

Thanks to Eq.\ \eqref{eq: O(u,u') general expression}, the density matrix becomes
\begin{equation}
	\label{eq: rho BO in u space}
	\urhoBO (\bm{\unlscha},\bm{\unlscha}') = 
	\sqrt{\detJnlscha(\bm{\unlscha}) }
	\rho^\text{(BO)}(\xibmnlscha(\bm{\unlscha}), \xibmnlscha(\bm{\unlscha}')) \sqrt{\detJnlscha(\bm{\unlscha}')}
\end{equation}
meaning that it is a physical density matrix also in $\bm{\unlscha}$-space, i.e.\ $\Tru\left[\uhatrhoBO\right]=1$ $\urhoBO (\bm{\unlscha},\bm{\unlscha}')\geq0$ and $\urhoBO(\bm{\unlscha},\bm{\unlscha}') = \urhoBO(\bm{\unlscha}',\bm{\unlscha})^\dagger$; so Eqs \eqref{eq: physical conditions for density matrix} hold also in the auxiliary manifold. Eq.\ \eqref{eq: O(u,u') general expression} is the rule to transform the nuclear Hamiltonian (see appendix \ref{APP: Quantum mechanics in nonlinear variables})
\begin{equation}
    \label{eq: H BO u u'}   	\uhamiltonianBO(\bm{\unlscha},\bm{\unlscha}') =  \Kgeom(\bm{\unlscha},\bm{\unlscha}')  + \VuBO(\bm{\unlscha},\bm{\unlscha}')
\end{equation}
$\Kgeom(\bm{\unlscha},\bm{\unlscha}')$ is the kinetic operator and contains a differential part and a potential-like term $\Vgeom(\bm{\unlscha})$ \cite{DFTcurvilinear1,DFTcurvilinear2}
\begin{subequations}
	\label{eq: K u u'} 
	\begin{align}
		\hspace{-0.3cm}\Kgeom (\bm{\unlscha},\bm{\unlscha}')  & = 
		\delta(\bm{\unlscha}-\bm{\unlscha}')\left(
		-\frac{\hbar^2}{2}
		\pdv{}{\widetilde{\bm{\unlscha}}} 
		\cdot \gbmtildenlscha  \cdot
		\pdv{}{\widetilde{\bm{\unlscha}}} 
		+ \Vgeom (\bm{\unlscha})\right)
		\label{eq: K u u' diff + V geom} \\
		\Vgeom(\bm{\unlscha})  & = \frac{\hbar^2}{2}\sum_{ab=1}^{3N}\left[\pdv{\left(\gtildenlscha{a}{b} \dtildelogJdq{b}\right)}{\widetilde{\unlscha}_a} 
		+  \dtildelogJdq{a} \gtildenlscha{a}{b} \dtildelogJdq{b}\right] \label{eq: V geom u kinetic}
	\end{align}
\end{subequations}
and the BOES is given by
\begin{equation}
	\label{eq: V BO u u'}
	\VuBO(\bm{\unlscha},\bm{\unlscha}')  = 
	\delta(\bm{\unlscha}-\bm{\unlscha}')
	V^{\text{(BO)}}(\xibmnlscha(\bm{\unlscha})) .
\end{equation}

Interestingly, in appendix \ref{APP: Quantum mechanics in nonlinear variables}, we show that $\uhatrhoBO$ is generated by the Hamiltonian of Eq.\ \eqref{eq: H BO u u'}
\begin{subequations}
 \label{eq: rho hat BO in u space}
\begin{align}
    \uhatrhoBO & = 
    \frac{\exp{-\beta \uhathamiltonianBO}}{\overline{Z}^\text{(BO)}} 
    \label{eq: rho hat BO in u space 1} \\
    \overline{Z}^\text{(BO)} & = \Tru\left[\exp{-\beta \uhathamiltonianBO}\right]
    \label{eq: rho hat BO in u space 2}
\end{align}
\end{subequations}
proving that the invertible nonlinear transformation preserves the structure of the quantum statistical ensemble and its temperature. Note that $\uhathamiltonianBO$
\begin{equation}
    \uhamiltonianBO = \Khatgeom + \VhatuBO
\end{equation}
is derived from its matrix elements (Eq.\ \eqref{eq: H BO u u'}) with the canonical operators in $\bm{\unlscha}$-space
\begin{subequations}
\label{eq: p and u operators}
\begin{align}
    \bra{\bm{\unlscha}'}\phatnlscha{a}
    \ket{\bm{\unlscha}} &=-i\hbar \delta(\bm{\unlscha} - \bm{\unlscha}')\pdv{}{\unlscha_a} \label{eq: p operator}\\
    \bra{\bm{\unlscha}'}\hat{\unlscha}_a 
    \ket{\bm{\unlscha}}  &= \delta(\bm{\unlscha} - \bm{\unlscha}') \unlscha_a
    \label{eq: u operator}
\end{align}
\end{subequations}
which do not correspond to any standard physical observable of quantum mechanics but play an auxiliary role.

The last term in the nuclear free energy is the entropy $S^\text{(BO)}$ (Eq.\ \eqref{eq: BO S}). Plugging $O(\bm{R},\bm{R}')=\bra{\bm{R}}
\log(\hat{\rho}^\text{(BO)})\ket{\bm{R}'}$ in Eq.\ \eqref{eq: invariance of observables} and using Eq.\ \eqref{eq: rho BO in u space} (see appendix \ref{APP: Quantum mechanics in nonlinear variables}) we get that the entropy is invariant in $\bm{\unlscha}$-space
\begin{equation}
\label{eq: entropy in u space}
\begin{aligned}
     S^\text{(BO)} = &-k_\text{B}
    \Tr\left[\hat{\rho}^\text{(BO)}\log(\hat{\rho}^\text{(BO)})\right] \\
    = & - k_\text{B}
    \Tru\left[\uhatrhoBO\log(\uhatrhoBO)\right].
\end{aligned}
\end{equation}
Entropy keeps the same form because it is the measure of the information encoded in $\hat{\rho}^\text{(BO)}$ and the invertibility of $\xibmnlscha$ ensures that information is not destroyed (see section \ref{SUBSEC: The entropy}). 

In conclusion, we build an exact mapping of quantum thermodynamics in $\bm{\unlscha}$-space, conserving the expectation values and the temperature of the BO density matrix. Our approach also shows a rigorous method to quantize nonlinear variables introducing operators corresponding to observables. 

\subsection{The harmonic approximation}
\label{SUBSEC: The harmonic approximation}
In this section, we lay the foundations for nonlinear SCHA by deriving the harmonic approximation (HA) of the BO Hamiltonian in $\bm{\unlscha}$-space Eq.\ \eqref{eq: H BO u u'}. In addition, we highlight the difference with the HA in $\bm{R}$-space. 
In Eq. \eqref{eq: H BO u u'}, we consider just the linear and quadratic terms in $\delta \unlscha_a = \unlscha_a - \unlscha{}_{0,a}$, where $\bm{\unlscha}{}_0$ denotes the minimum of the BO potential. This applies to both the kinetic and BOES contributions leading to
\begin{equation}
\label{eq: H BO u u' harmonic}
\begin{aligned}
    \HAhathamiltonian = & 
    \frac{1}{2}\phatbmnlscha \cdot \gbmtildenlscha{}_0  \cdot \phatbmnlscha + \delta \hat{\bm{\unlscha}} \cdot
    \pdv{\Vgeom} {\bm{\unlscha}} \biggl|_0
    \\
    & 
    +\frac{1}{2} \delta \hat{\bm{\unlscha}} \cdot \left(\overline{\bm{\Phi}} +
    \pdv{\Vgeom}{\bm{\unlscha}}{\bm{\unlscha}} \biggl|_0\right) 
    \cdot \delta \hat{\bm{\unlscha}} 
\end{aligned}
\end{equation}
where $\gbmtildenlscha{}_0$ is the inverse metric tensor evaluated at equilibrium positions and $\HAFCbm$ is the force constant
\begin{equation}
\label{eq: HA FC in u}
    \HAFC{ab} = 
    \pdv{\VuBO}{\unlscha_a}{\unlscha_b} \biggl|_0
\end{equation}
The HA in $\bm{\unlscha}$-space is more complex than its standard $\bm{R}$-space counterpart because of $\Vgeom$'s derivatives that shift $\bm{\unlscha}_0$ and $\overline{\bm{\Phi}}$. A straightforward consequence is that, at the quantum level, the HA's results in $\bm{\unlscha}$-space differs from the ones we get in $\bm{R}$-space \cite{HarmonicDelocalizedInternalCoordinate}. 
This contrasts with the classical case, where HA frequencies remain unaffected by the choice of canonical variables. Classical mechanics has no fluctuations, so it does not account for the presence of an underlying metric or the use of different coordinate systems. 
Conversely, fluctuations due to the uncertainty principle in quantum mechanics enable exploring a broader phase space compared to the classical counterpart. Ultimately, this effect, combined with a nonlinear transformation, captures the distortion caused by the metric encoded by $\Vgeom$, so the results of quantum HA depend on the chosen variables. Indeed, by neglecting the terms with $\Vgeom$, we recover the normal modes of the HA in $\bm{R}$-space (see appendix \ref{APP: Quantum mechanics in nonlinear variables}).

\section{Theory of nonlinear self-consistent harmonic approximation}
\label{SEC: Theory of nonlinear self-consistent harmonic approximation}
This section introduces the Nonlinear Self-Consistent Harmonic Approximation (NLSCHA) theory. Our approach is based on the quantum variational principle and employs a trial density matrix generated by a harmonic Hamiltonian, as the traditional SCHA. Still, with a crucial distinction: it is based on the nonlinear variables $\bm{\unlscha}$.

In section \ref{SUBSEC: The variational manifold of nonlinear SCHA}, we delineate the new variational manifold, defining all the free parameters used to minimize the free energy variationally. Subsequently, in section \ref{SUBSEC: The entropy}, we demonstrate that despite having non-Gaussian fluctuations in the theory, the entropy remains an analytical quantity and preserves its harmonic form. Finally, in section \ref{SUBSEC: The free energy}, we calculate the nonlinear SCHA free energy and its gradients.

\subsection{The variational manifold}
\label{SUBSEC: The variational manifold of nonlinear SCHA}
In nonlinear SCHA, the trial density matrix plugged in the variational principle is generated by the harmonic Hamiltonian in $\bm{\unlscha}$-space (see section \ref{SUBSEC: The harmonic approximation})
\begin{equation}
\label{eq: nlscha Hamiltonian operator}
    \Hhatnlscha = 
    \sum_{ab=1}^{3N} \frac{1}{2}\left( \phatnlscha{a} \masstnsinv{ab} \phatnlscha{b}
    +  \hat{\unlscha}_a \FCnlscha{a}{b} \hat{\unlscha}_b \right)
\end{equation}
Note that the nonlinear SCHA Hamiltonian differs from Eq.\ \eqref{eq: H BO u u' harmonic}. First, we reabsorb $\bm{\unlscha}_0$ in  $\freeparambmnlscha$, i.e., \ the free parameters of the nonlinear transformation. Secondly, we safely neglect the derivatives of $\Vgeom$, which shift both $\bm{\unlscha}_0$ and $\FCbmnlscha$. Indeed, this simplification does not pose an issue, as we compensate by variationally optimizing $\freeparambmnlscha$ and $\FCbmnlscha$ as free parameters.
There is also a more subtle aspect. In standard quantum mechanics, the harmonic approximation does not truncate the kinetic energy operator, which is already quadratic in the momentum. In contrast, in $\bm{\unlscha}$-space, we also approximate this term (Eq.\ \eqref{eq: H BO u u'}) by retaining only the equilibrium metric tensor. To correct this pathology and improve the description of kinetic quantum fluctuations, we introduce in Eq.\ \eqref{eq: nlscha Hamiltonian operator} the additional free parameter $\masstnsbm$, a symmetric positive-definite mass tensor. 
Significantly, having non-exact kinetic energy in the generating Hamiltonian solely impacts the definition of the nonlinear SCHA trial density matrix as we evaluate the exact kinetic energy in the variational principle (see section \ref{SUBSEC: The free energy}). 

So, the density matrix associated with Eq.\ \eqref{eq: nlscha Hamiltonian operator} is
\begin{subequations}
\label{eq: nlscha rho hat}
\begin{align}
    \rhouhatnlscha & = \frac{\exp{-\beta \Hhatnlscha}}{\Znlscha} \label{eq: nlscha rho hat 1} \\
    \Znlscha & = \Tru\left[\exp{-\beta \Hhatnlscha}\right] \label{eq: nlscha rho hat 2}
\end{align}
\end{subequations}
and, using the rule to transform density matrices (Eq.\ \eqref{eq: rho BO in u space}), we get the nonlinear SCHA trial density matrix in Cartesian space (i.e.\ the one used in the variational principle)
\begin{equation}
\label{eq: R rho R' nlscha}
    \bra{\bm{R}} \rhohatnlscha \ket{\bm{R}'} = \frac{\gaussnlscha(\bm{\unlscha}, \bm{\unlscha}')}{\sqrt{\detJnlscha(\bm{\unlscha}) \detJnlscha(\bm{\unlscha}')}}
\end{equation}
the invertibility of $\xibmnlscha$ (Eq.\ \eqref{eq: invertible transformation}) ensures that the trial density matrix is well defined. As our method is based on the harmonic Hamiltonian of Eq.\ \eqref{eq: nlscha Hamiltonian operator}, we have that Eq.\ \eqref{eq: R rho R' nlscha} is fully determined by
\begin{equation}
\label{eq: u rho u' nlscha}
\begin{aligned}
    \gaussnlscha(\bm{\unlscha}, \bm{\unlscha}') &= \sqrt{\det\left(\frac{\Ybmnlscha}{2\pi}\right)} 
    \exp\left\{
    -\frac{1}{4}\sum_{ab=1}^{3N} \unlscha_a \Tnlscha{a}{b} \unlscha_b \right.\\
    &\left.
    -\frac{1}{4}\sum_{ab=1}^{3N} \unlscha'_a \Tnlscha{a}{b} \unlscha'_b 
    + \sum_{ab=1}^{3N} \unlscha'_a \Anlscha{a}{b} \unlscha_b  \right\}
\end{aligned}
\end{equation}
where the $3N \times 3N$ symmetric and real tensors $\Ybmnlscha, \Tbmnlscha, \Abmnlscha$ are connected by
\begin{equation}
\label{eq: Y = T - 2 A}
    \Ybmnlscha =  \Tbmnlscha - 2 \Abmnlscha
\end{equation}
and $\Ybmnlscha$ is positive definite. In addition, we introduce the nonlinear SCHA dynamical matrix $\Dbmnlscha$
\begin{equation}
\label{eq: def nlscha phonons}
    \Dnlscha{a}{b} = \sum_{ij=1}^{3N} \invsqrtmasstnsT{ai}
    \FCnlscha{i}{j}\invsqrtmasstns{jb} = \sum_{\mu=1}^{3N} \onlscha{\mu}^2 \polnlscha{\mu}{a} \polnlscha{\mu}{b}
\end{equation}
The tensors $\Ybmnlscha$, $\Abmnlscha$ satisfy
\begin{subequations}
\label{eq: def Y A nlscha}
\begin{align}
    & \Ynlscha{a}{b} = \sum_{ij=1}^{3N} \sqrtmasstnsT{ai} \Yovnlscha{ij} 
    \sqrtmasstns{jb}
    \label{eq: def Y nlscha}\\
    & \Yovnlscha{ij} =  \sum_{\mu=1}^{3N} \frac{2\onlscha{\mu}}{\hbar(1 +2\nnlscha{\mu})} \polnlscha{\mu}{i} \polnlscha{\mu}{j} \\
    & \Anlscha{a}{b} = \sum_{ij=1}^{3N} \sqrtmasstnsT{ai} \Aovnlscha{ij} 
     \sqrtmasstns{jb}\label{eq: def A nlscha} \\
    & \Aovnlscha{ij}  = \sum_{\mu=1}^{3N} \frac{2\onlscha{\mu} \nnlscha{\mu} (\nnlscha{\mu} + 1)}{\hbar(1 + 2\nnlscha{\mu})} \polnlscha{\mu}{i} \polnlscha{\mu}{j}
\end{align}
\end{subequations}
where all the modes, thanks to Eq.\ \eqref{eq: nlscha rho hat}, are thermally excited by $\beta= (k_{\text{B}}T)^{-1} $
\begin{equation}
\label{eq: BE occupation}
     \nnlscha{\mu} = \frac{1}{e^{\beta\hbar\onlscha{\mu}} - 1}
\end{equation}

We build nonlinear SCHA as a variational method (Eq.\ \eqref{def: variational principle}), employing Eq.\ \eqref{eq: R rho R' nlscha} as the \textit{ansatz} for the density matrix and retaining the property of having an underlying Hamiltonian (Eq.\ \eqref{eq: nlscha Hamiltonian operator}), but defined in $\bm{\unlscha}$-space. The free parameters are $\FCbmnlscha$, the nonlinear force constant, $\masstnsbm$, the mass-tensor, and $\freeparambmnlscha$, the specific parametrization of $\xibmnlscha$ (Eq.\ \eqref{eq: change of variables}). We underly that the dependence on $\freeparambmnlscha$ arises from evaluating the BOES  $V^\text{(BO)}(\bm{R})$ in the new space as $V^\text{(BO)}(\xibmnlscha(\bm{\unlscha}))$ (see section \ref{SUBSEC: The free energy}). On the contrary, $\FCbmnlscha$ and $\masstnsbm$ control the nuclear quantum fluctuations.
The $\freeparambmnlscha$ are the variational parameters that allow to have both Gaussian behavior when $\bm{u}=\bm{R} - \bm{\mathcal{R}}$, and non-Gaussian fluctuations, see Fig.\ \ref{fig: nlscha idea}. Indeed, $\rhohatnlscha$ exhibits a Gaussian form only in the auxiliary space; however, when projected into the physical Cartesian space, it gives rise to new fluctuations that depend on the $\xibmnlscha$ employed. In the case of a linear transformation, we find the SCHA solution, while if $\xibmnlscha$ is nonlinear, we systematically surpass it, improving the variational estimation of $F^{\text{(BO)}}$.

\subsection{The entropy}
\label{SUBSEC: The entropy}
In nonlinear SCHA, $\xibmnlscha$ allows us to explore the quantum phase space characterized by non-Gaussian fluctuations. Importantly, we do not lose direct access to entropy, a key advantage associated with SCHA and missing in MD/PIMD. As our theory is based on an auxiliary harmonic Hamiltonian (Eq.\ \eqref{eq: nlscha Hamiltonian operator}), the entropy of nonlinear SCHA is easily computed in this basis
\begin{equation}
\label{eq: S entropy nlscha expression}
\begin{aligned}
    \Snl & = 
    -k_\text{B} \Tr\left[\rhohatnlscha \log\left(\rhohatnlscha\right)\right] 
    = 
    -k_\text{B} \Tru\left[\rhouhatnlscha \log\left(\rhouhatnlscha\right)\right]\\
    &= k_\text{B} \sum_{\mu=1}^{3N}\left[
    (1 + \nnlscha{\mu}) \log(1 + \nnlscha{\mu}) - \nnlscha{\mu}\log(\nnlscha{\mu})\right]
\end{aligned}
\end{equation}
where the first equality is possible thanks to the exact mapping we build between quantum averages in $\bm{\unlscha}$-space and $\bm{R}$-space (section \ref{SUBSEC: Thermodynamics in arbitrary coordinates}) and the second follows from the definition of $\rhohatnlscha$ in the auxiliary space, where it solves a harmonic Hamiltonian. Further details are in appendix \ref{APP: NEW Nonlinear SCHA entropy}.
Eq.\ \eqref{eq: S entropy nlscha expression} demonstrates that the invertibility of $\xibmnlscha$ ensures that entropy remains accessible and conserves its harmonic form as we modify the density matrix. In addition, 
the entropy of the density matrix depends only on $\FCbmnlscha$ and $\masstnsbm$ and not on the nonlinear transformation $\xibmnlscha$. 
Information theory helps clarify this result: entropy measures the information in the density matrix $\rhohatnlscha$. The matrix elements $\gaussnlscha(\bm{\unlscha},\bm{\unlscha}')$ represent a harmonic oscillator in $\bm{\unlscha}$-space for which the entropy is known. Since $\xibmnlscha$ is invertible, the information stored is conserved even after a nonlinear transformation. Therefore, the change of variables $\bm{\unlscha}\rightarrow\bm{R}$ via $\xibmnlscha$ is a transformation that conserves entropy\deleted{, but not temperature (see section \ref{SUBSEC: The modes' effective temperature})}.

\subsection{The free energy}
\label{SUBSEC: The free energy}
In section \ref{SUBSEC: The variational manifold of nonlinear SCHA}, we identify the variational manifold of nonlinear SCHA, and in section \ref{SUBSEC: The entropy}, we demonstrated that the entropy has an easy form, so we ready to compute the free energy of this method. The quantum free energy variational principle using $ \rhohatnlscha$ (Eq.\ \eqref{eq: R rho R' nlscha}) as trial density matrix reads as
\begin{equation}
\label{eq: nlscha free energy}
    F^{\text{(BO)}} \leq \Fnl 
    = F[\rhohatnlscha] = U\left[ \rhohatnlscha \right] - T \Snl
\end{equation}
where $\Fnl$ is the nonlinear SCHA free energy and $\Snl$ (Eq.\ \eqref{eq: S entropy nlscha expression}) is the entropic contribution.

The internal energy, $U\left[ \rhohatnlscha \right]$, contains the kinetic and BOES contributions, and we compute it in $\bm{\unlscha}$-space thanks to our theoretical developments (section \ref{SUBSEC: Thermodynamics in arbitrary coordinates})
\begin{equation}
\begin{aligned}
     U\left[\rhohatnlscha\right]  
     & = \Tr\left[ \rhohatnlscha \hat{K}\right] +
    \Tr\left[ \rhohatnlscha \hat{V}^{\text{(BO)}}\right] \\
    & =  \Tru\left[ \rhouhatnlscha \Khatgeom\right] + 
    \Tru\left[ \rhouhatnlscha \VhatuBO\right]
\end{aligned}
\end{equation}
where $\Khatgeom$ and $\VhatuBO$ are defined through the exact mapping respectively by Eqs \eqref{eq: K u u'} \eqref{eq: V BO u u'}.
Note that the kinetic energy is evaluated exactly, despite the generating Hamiltonian of nonlinear SCHA has an approximate kinetic term (see section \ref{SUBSEC: The variational manifold of nonlinear SCHA}). \deleted{We remark that the NLSCHA Hamiltonian can be added and subtracted in $U\left[\rhohatnlscha\right]$ to increase numerical accuracy for a stochastic evaluation (see section \ref{SEC: The stochastic approach}).} In appendix \ref{APP: NEW Nonlinear SCHA}, we prove that 
\begin{equation}
\label{eq: nlscha kinetic energy}
\begin{aligned}
    \Tru\left[ \rhouhatnlscha \Khatgeom\right] = 
    &\myint d\bm{\unlscha}\left(
    \Kzeronlscha - \sum_{a=1}^{3N} \Konenlscha{a} \pdv{}{\widetilde{\unlscha}_a}
    \right. \\
    & \left. 
    -\sum_{ab=1}^{3N} \Ktwonlscha{a}{b} \pdv{}{\widetilde{\unlscha}_a}{\widetilde{\unlscha}_b} 
    \right) \gaussnlscha(\bm{\unlscha}) 
\end{aligned}
\end{equation}
The coefficients $\Kzeronlscha$, $\Konebmnlscha$ and  $\Ktwobmnlscha$  depend on the nonlinear change of variables (Eq.\ \eqref{eq: change of variables})
\begin{subequations}
\label{eq: K2 K1 K0 def}
\begin{align}
    \Kzeronlscha  & = 
    \frac{\hbar^2}{2} \Tr\left[\gbmtildenlscha \cdot \left(\frac{\Ytildebmnlscha}{4} + \Atildebmnlscha\right)\right]
    + \frac{\hbar^2}{2}  \dtildebmlogJdq \cdot \gbmtildenlscha \cdot  \dtildebmlogJdq \label{eq: K zero}\\
    \Konebmnlscha & = \frac{\hbar^2}{2} \gbmtildenlscha \cdot \dtildebmlogJdq
    \label{eq: K one}\\
    \Ktwobmnlscha
    & = -\frac{\hbar^2}{8} \gbmtildenlscha \label{eq: K two}
\end{align}
\end{subequations}
Note that the first term of $\Kzeronlscha$, Eq.\ \eqref{eq: K zero}, written in polarization basis (Eq.\ \eqref{eq: def nlscha phonons}) reads as
\begin{equation}
\label{eq: kinetic energy of nonlinear phonons}
\begin{aligned}
    &   \frac{\hbar^2}{2} \Tr\left[\gbmtildenlscha \cdot \left(\frac{\Ytildebmnlscha}{4} + \Atildebmnlscha\right)\right]
    \\
    & = \sum_{\mu=1}^{3N} 
    \left(\sqrtmasstnsbm \cdot \Jinvbmnlscha \cdot \overset{-1}{\bm{m}}\cdot \Jinvbmnlscha {}^T \cdot 
    \sqrtmasstnsTbm\right)_{\mu\mu} \hspace{-0.2cm}
    \frac{\hbar \onlscha{\mu}}{4}\left(1 + 2 \nnlscha{\mu}\right)
\end{aligned}
\end{equation}
This equation represents the kinetic energy of harmonic oscillators with non-diagonal masses $\masstnsbm$, and it is weighted by the metric tensor $\gbmtildenlscha$. The latter indicates that the nonlinear SCHA phonons are defined in an arbitrary curvilinear space. Note that, if $\masstnsbm = \bm{m}$ and the transformation is linear, Eq.\ \eqref{eq: kinetic energy of nonlinear phonons} becomes exactly the result of the SCHA. The additional term in $\Kzeronlscha$ plays the role of a geometrical effective potential; in the case of a spherical transformation, it gives the centrifugal barrier (see also Ref.\ \cite{DFTcurvilinear1}). The other terms, $\Konebmnlscha$, and $\Ktwobmnlscha$ encode the quantum fluctuations compatible with a Gaussian density matrix in a nontrivial metric.
The expectation value of the potential energy in nonlinear SCHA is
\begin{equation}
\label{eq: nlscha potential energy}
    \Tru\left[ \rhouhatnlscha \VhatuBO\right] = \myint d\bm{\unlscha} \gaussnlscha(\bm{\unlscha}) V^{\text{(BO)}}\left(\xibmnlscha(\bm{\unlscha})\right)
\end{equation}
and requires electronic total energy calculations. In this term, $\freeparambmnlscha$ enters explicitly as the BO potential depends on the Cartesian coordinates, but we evaluate it in parameterized $\bm{\unlscha}$-space.

We estimate the BO free energy $F^\text{(BO)}$ with a minimization algorithm of $\Fnl$, Eq.\ \eqref{eq: nlscha free energy}, with respect to the free parameters $\FCbmnlscha$, the auxiliary force constant, $\masstnsbm$, the mass tensor, and $\freeparambmnlscha$, the parameters of the nonlinear transformation. The equilibrium conditions are defined by
\begin{subequations}
\label{eq: nlscha equilibrium conditions}
\begin{align}
    \pdv{\Fnl}{\FCbmnlscha} & = \bm{0} \\
    \pdv{\Fnl}{\masstnsbm} & = \bm{0} \\
    \pdv{\Fnl}{\freeparambmnlscha}  & = \bm{0}
\end{align}
\end{subequations}
where all the gradients require just BO energies and forces, see appendix \ref{APP: NEW Gradient of nonlinear SCHA}. Note that as both $\FCbmnlscha$ and $\masstnsbm$ are positive definite, to enforce this condition the minimization can be done using $\sqrt{\FCbmnlscha}$ \cite{monacelli2018pressure} and $\sqrtmasstnsbm$.
Once Eqs \eqref{eq: nlscha equilibrium conditions} are reached, the system's real interacting normal modes are mapped, in a self-consistent framework, on harmonic oscillators defined in nonlinear variables. This allows to have non-Gaussian fluctuations.
To have an initial reasonable guess for the minimization $\xibmnlscha$, we must reproduce the linear transformation on which SCHA is based. Hence, we initialize the minimization in this limit so that we employ the outcome of the SCHA at temperature $T$ $(\bm{\mathcal{R}}^{(0)},\bm{\Phi}^{(0)})$
\begin{subequations} 
\label{eq: initial condition for minimization}
\begin{align}
    \Tr\left[\rhohatnlscha \hat{\bm{R}}\right] & = \bm{\mathcal{R}} ^{(0)}
    \label{eq: xi initialization} \\
    \FCbmnlscha & = \bm{\Phi}^{(0)} 
    \label{eq: FC initialization} \\
    \masstnsbm & = \bm{m} \label{eq: masstns initilization}
\end{align}
\end{subequations}
where $\bm{m} = \delta_{ab} m_a$.
In this way, nonlinear SCHA acts as a post-processing tool for SCHA minimization.
We emphasize that while it is possible to initialize the nonlinear parameters randomly, starting from Eqs \eqref{eq: initial condition for minimization} is preferable. This choice is based on the expectation that the majority of degrees of freedom within the crystal are already accurately described by the SCHA. Only a few modes, such as tunneling in high-pressure ice \cite{RamanHighPressureIce,cherubini2024quantum} and rotating modes of metal-organic perovskites \cite{OrganicCationRotationPerovskites}, require NLSCHA. In our self-consistent theory, we treat the interaction of these modes with the Gaussian-like modes in a non-perturbative manner.

\section{1D toy model}
\label{SEC: 1D toy model}
To show the performance of NLSCHA, we consider the quantum mechanical problem of a particle with mass $m$ in a 1D double-well potential
\begin{equation}
\label{eq: toy model potential}
    V^\text{(BO)}(R) = -a R^2 + b R^4 + \frac{a^2}{4b}
\end{equation}
A potential like Eq.\ \eqref{eq: toy model potential} has been used to model the proton shuttling mode between two sulfur (\ch{S}) atoms leading to the \ch{H3S} R3m/Im$\overline{3}$m transition \cite{CherubiniH3S}. SCHA misses the transition pressure due to non-Gaussian fluctuations. Here, we show that NLSCHA systematically improves the description of nuclear equilibrium properties.

We combine NLSCHA with the following invertible change of variables (see also Fig.\ \ref{fig: nlscha idea})
\begin{equation}
\label{eq: toy model transformation}
    \hspace{-0.3cm}\unlscha=  \sigma^{-1}\sinh\left\{\frac{\sigma (R - \mathcal{R}_\text{d})}{2}
    \left[1 + \frac{1}{1 + \eta^2 (R - \mathcal{R}_\text{d})^2}\right]\right\}
\end{equation}
The free parameters are $\freeparambmnlscha = (\mathcal{R}_\text{d},\eta, \sigma)$, where $\mathcal{R}_\text{d}$ is the center of the nonlinear deformation, $\eta$ and $\sigma$ respectively give the strength of the polynomial and hyperbolic transformation.  In the $\eta,\sigma\rightarrow 0$ limit, we recover $\unlscha = R - \mathcal{R}_\text{d}$, which is the linear SCHA transformation. For the potential of Eq.\ \eqref{eq: toy model potential} $\mathcal{R}_\text{d}=0$ Bohr by symmetry so we have just two free parameters ($S=2$).
\begin{figure}[!htb]
    \centering
    \begin{minipage}[c]{1.0\linewidth}
    \includegraphics[width=1.0\textwidth]{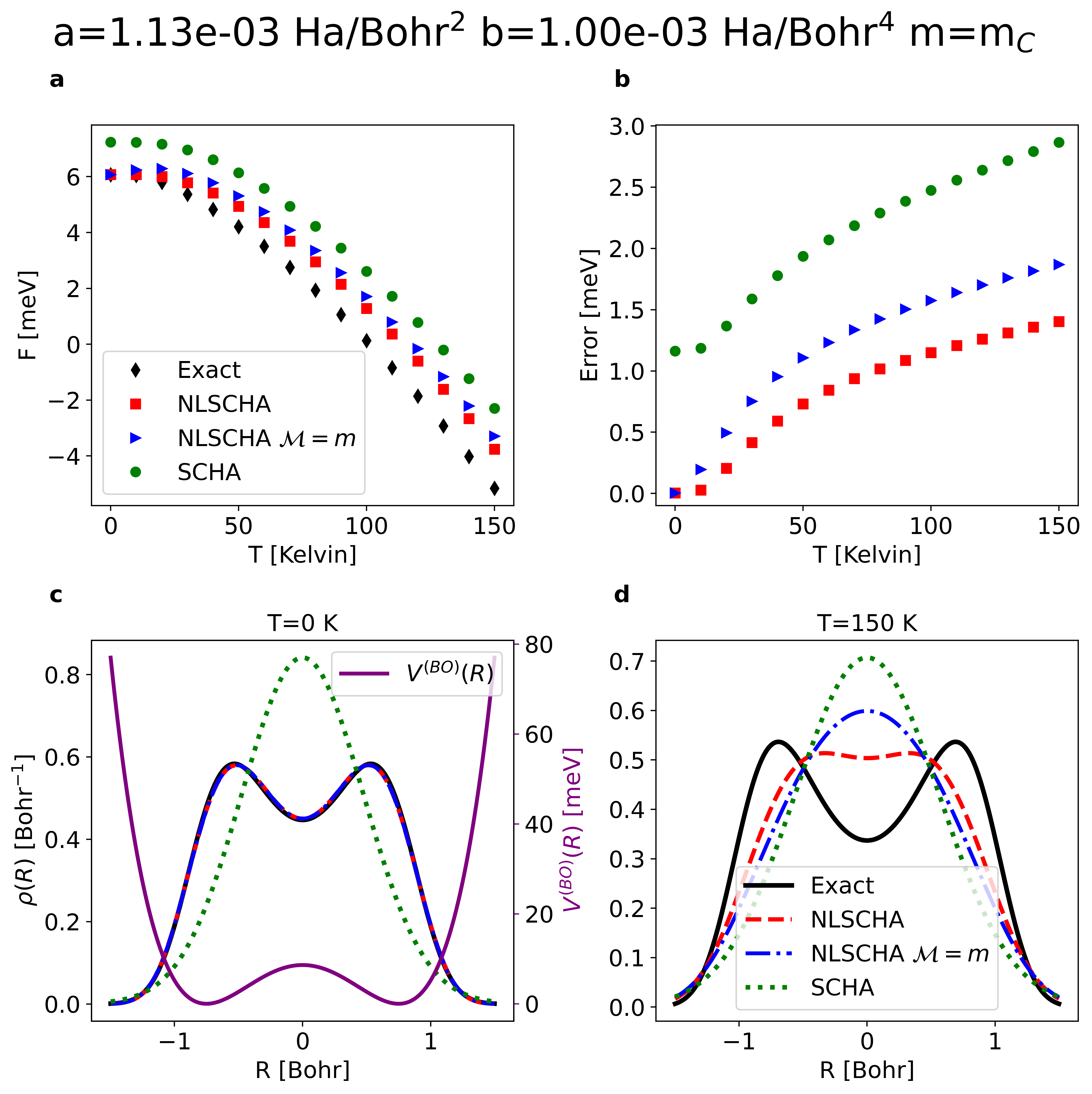}
    \end{minipage}
    \caption{The parameters used are $a = 1.13 \cdot 10^{-3}$ Ha/Bohr$^2$, $b = 1.00 \cdot 10^{-3}$ Ha/Bohr$^4$ (see Eq.\ \eqref{eq: toy model potential}), and $m=m_\text{\ch{C}}$ (Carbon mass). The energy levels' spacings are $E_1 - E_\text{GS} =35$ K, $E_2 - E_\text{GS} =155$ K, and $E_3 - E_\text{GS} =280$ K.
    Fig.\ a reports the exact, SCHA and NLSCHA free energies in the temperature range $0-150$ K.  Fig.\ b compares the SCHA and NLSCHA errors. Figs c-d compare the exact, SCHA and NLSCHA probability distributions at $T=0,150$ K. We present two NLSCHA solutions with the nonlinear transformation of Eq.\ \eqref{eq: toy model transformation}: one where we optimize $\masstns{}$ and the other one where $\masstns{}=m_\text{\ch{C}}$.}
    \label{fig:results 1}
\end{figure}

To benchmark NLSCHA we consider a Carbon (\ch{C}) atom ($m=m_\text{\ch{C}}$) in the potential of Eq.\ \eqref{eq: toy model potential} with $a = 1.13 \cdot 10^{-3}$ Ha/Bohr$^2$ and $b = 1.00 \cdot 10^{-3}$ Ha/Bohr$^4$.  Fig.\ \ref{fig:results 1} a shows the exact, SCHA and NLSCHA free energies from $0$ to $150$ K, and Fig.\ \ref{fig:results 1} b compares the error of SCHA and NLSCHA. In addition, Figs \ref{fig:results 1} c-d show the exact, SCHA and NLSCHA probability distributions at $0$ K and $150$ K. We report all the simulations' details in appendix \ref{APP: Computational details}. 

We present two possible NLSCHA solutions, one where we optimize $\masstns{}$ and the other one where $\masstns{}=m_\text{\ch{C}}$. At $0$ K, the optimization of $\masstns{}$ does not affect the result; it is just a rescaling of $\omega_{\text{nl}}$. Note that at $0$ K, NLSCHA is almost exact as a Gaussian shape combined with the nonlinear transformation of Eq.\ \eqref{eq: toy model transformation} fits perfectly the exact distribution value at $R=0$ Bohr and $R=\pm \sqrt{a/(2b)}$ (the minima of Eq.\ \eqref{eq: toy model potential}). Indeed, for large $R$, where the potential is steeper, the hyperbolic transformation ($u\simeq \sinh(\sigma R)/\sigma$) reproduces the tails of the exact solution (see Fig.\ \ref{fig: NLSCHA 1} c). For small $R$ the polynomial transformation ($u\simeq R + (\sigma^2/3 - \eta^2) R^3/2 + O(R^5)$) describes the deformation due to the tunneling between the minima (see Fig.\ \ref{fig: NLSCHA 1} c). Consequently, at $0$ K, the NLSCHA error of $4.1$ $\mu$eV is three orders of magnitude smaller than the SCHA resolution of $1.2$ meV. 

At finite temperatures, $\masstns{}$ enters non-linearly in $\Upsilon_\text{nl}$ and $A_\text{nl}$ (Eqs \eqref{eq: def Y nlscha} \eqref{eq: def A nlscha}) through the Bose-einstein occupations (Eq.\ \eqref{eq: BE occupation}). Consequently, optimizing $\masstns{}$ expands the variational subspace and improves the NLSCHA results (Fig.\ \ref{fig:results 1} b). If $\masstns{}=m_\text{\ch{C}}$, the error is below $1$ meV up to $40$ K, whereas if $\masstns{}\neq m_\text{\ch{C}}$, up to $70$ K. Then above, $70$ K we improve the SCHA resolution by $50\%$.

The NLSCHA error follows a trend similar to SCHA at high temperatures (Fig.\ \ref{fig:results 1} b). This shared behavior arises, in both cases, from the underlying harmonic Hamiltonian constraining the variational manifold to a subspace where both $\Ybmnlscha$ and $\Abmnlscha$ depend on $\FCbmnlscha$ and $\masstnsbm$ (or $\masstnsbm=\bm{m}$). Consequently, we do not treat $\Ybmnlscha$ and $\Abmnlscha$ as independent free parameters. Whether our choice is the best for a variational theory remains an open question for future works. Nevertheless, there is an enormous advantage in adopting the harmonic framework as the entropy is analytical.

\section{Is the entropy a thermodynamic quantity?}
\label{SEC: Entropy discussion}
As discussed in section \ref{SUBSEC: The entropy}, the NLSCHA entropy $\Snl$ (Eq.\ \eqref{eq: S entropy nlscha expression}) is an analytical function. So, as we variationally optimize $\Fnl$ there is direct access to the entropy $\Snl$ and its derivatives (Eqs \eqref{eq: nlscha equilibrium conditions}). The following question arises: does $\Snl$ coincide with $-\frac{d \Fnl}{d T}$ at the NLSCHA minimum? If true, we can compute analytically the many-body entropy for crystals and molecules with non-Gaussian degrees of freedom. In appendix \ref{APP: dF dT}, we prove that
\begin{equation}
\label{eq: dF dT = - T S}
    \frac{d \Fnl}{d T} \biggl|_{(0)} = \sum_{\mu=1}^{3N} \left(\pdv{\Fnl}{\nnlscha{\mu}}  \pdv{\nnlscha{\mu}}{T}  \right)\biggl|_{(0)} -  \Snl |_{(0)}
\end{equation}
where the subscript $(0)$ indicates that in all the expressions we plug the parameters that self-consistently minimize $\Fnl$ (Eqs \eqref{eq: nlscha equilibrium conditions}). Notably, Ref.\ \cite{TDSCHA_monacelli} already proved that the minimum SCHA free energy temperature derivative coincides with the SCHA entropy changed by the sign.
In NLSCHA, we satisfy $\frac{d \Fnl}{d T}|_{(0)} = - \Snl |_{(0)}$ in two ways.

The first solution is to optimize $\Fnl$ with respect to $\nnlscha{\mu}$, i.e.\ setting $\pdv{\Fnl}{\nnlscha{\mu}}|_{(0)}=0$ in Eq.\ \eqref{eq: dF dT = - T S}. In NLSCHA, $\nnlscha{\mu}$ is not an independent variable but depends on $\Dbmnlscha$. Thus, $\pdv{\Fnl}{\nnlscha{\mu}}|_{(0)}=0$ means relaxing our hypothesis that the NLSCHA density matrix is generated by a harmonic Hamiltonian in $\bm{\unlscha}$-space, thus enlarging the variational manifold. However, this approach would be disadvantageous, as the entropy would no longer be analytical.

The second solution is to introduce different thermal energies $\beta^{-1}_{\mu} > 0$ for each mode in the Bose-Einstein factors 
\begin{equation}
\label{eq: BE occupation tmode}
    \nnlscha{\mu} = \frac{1}{e^{\beta_\mu \hbar \onlscha{\mu}}- 1}
\end{equation}
so that $\pdv{\nnlscha{\mu}}{T}=0$ in Eq.\ \eqref{eq: dF dT = - T S}. Optimizing mode-dependent effective temperatures is not the best choice from a computational point of view, as the corresponding derivative is not well-defined when two auxiliary modes cross\deleted{, i.e.\ when $|\onlscha{\mu} - \onlscha{\nu}|\rightarrow 0$}. In appendix \ref{APP: dF dT}, we prove that optimizing $\beta_\mu$ is equivalent to minimizing the effective mass tensor $\masstnsbm$, keeping the physical thermal energy $\beta^{-1}$ in the Bose-Einstein occupations (Eq.\ \eqref{eq: BE occupation}). Note that $\masstnsbm$ is an atomic tensor so its optimization is similar to the one of $\FCbmnlscha$ and it is always well defined. Hence, in NLSCHA, only when we optimize all the free parameters $\FCbmnlscha$, $\masstnsbm$, and $\freeparambmnlscha$ (Eqs \eqref{eq: nlscha equilibrium conditions})
\begin{equation}
\label{eq: dF dT = - T S NLSCHA}
    \frac{d \Fnl}{d T} \biggl|_{(0)} = - \Snl |_{(0)}
\end{equation}
and there is direct access to the physical many-body entropy. Still, we remark that the violation of Eq.\ \eqref{eq: dF dT = - T S NLSCHA} does not spoil the variational nature of NLSCHA.

\begin{figure}[!htb]
    \centering
    \begin{minipage}[c]{1.0\linewidth}
    \includegraphics[width=1.0\textwidth]{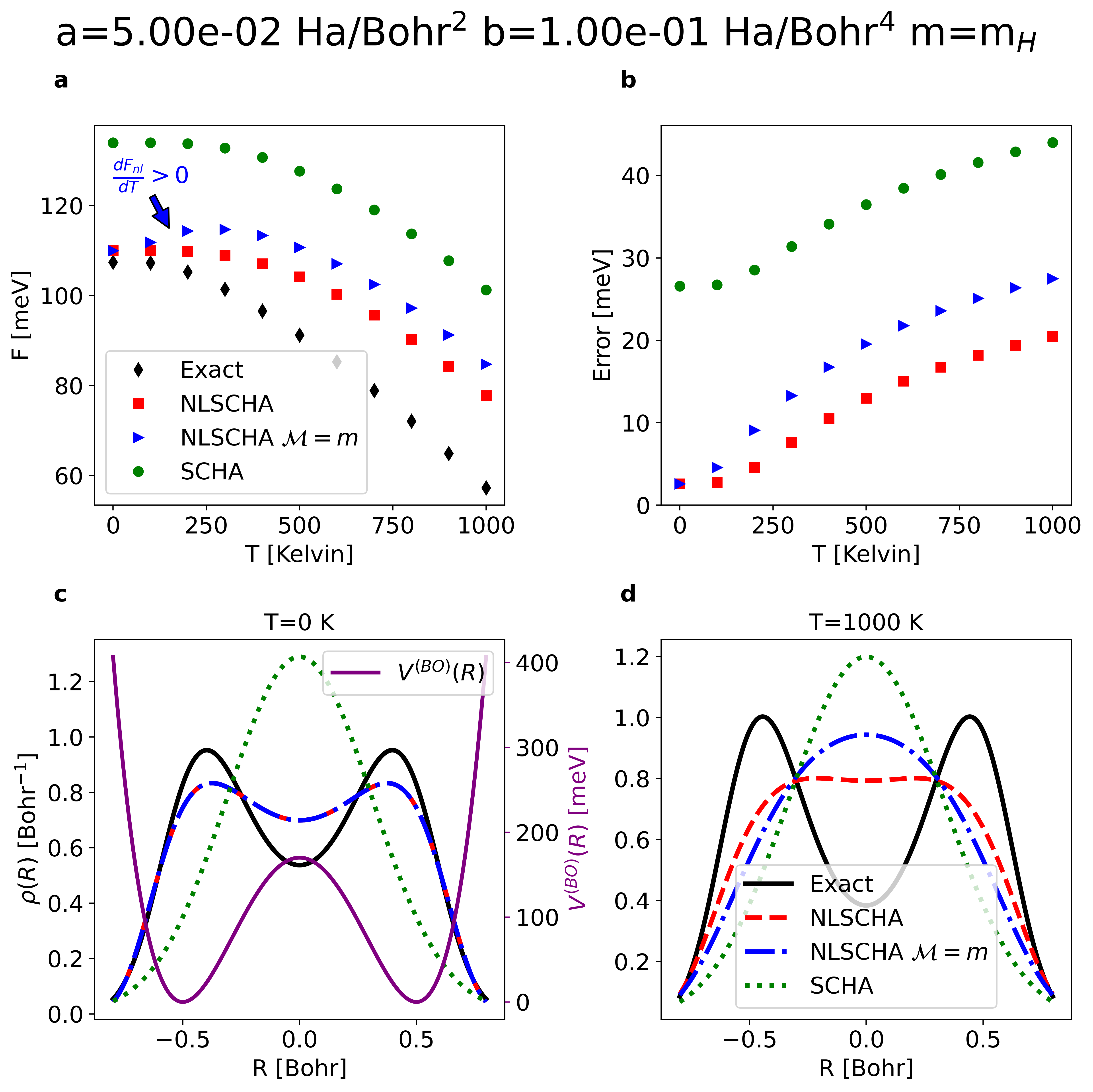}
    \end{minipage}
    \caption{Fig.\ a reports the exact, SCHA and NLSCHA free energies of a Hydrogen (\ch{H}) atom ($m=m_\text{\ch{H}}$) in the potential of Eq.\ \eqref{eq: toy model potential} with $a =0.05$ Ha/Bohr$^2$ and $b=0.1$ Ha/Bohr$^4$. For these parameters, the barrier height $a^2/(4 b)$ is $2000$ K and the energy spacing between the levels is high ($E_1 - E_\text{GS} = 403$ K as $E_2 - E_\text{GS} = 2277$ K), so $1000$ K only populates the first excited state. We combine NLSCHA with Eq.\ \eqref{eq: toy model transformation} where $\eta= 0$ Bohr$^{-1}$. Fig.\ b compares the SCHA and NLSCHA errors and Figs c-d show the probability distributions at $0$,$1000$ K. Note that when $\masstns{}=m_\text{\ch{H}}$ we get a positive temperature derivative of the free energy (see arrow in Fig.\ a), i.e.\ negative entropy.}
    \label{fig:results 2}
\end{figure}

To clarify the above discussion, we consider a hydrogen (\ch{H}) atom, $m=m_\text{\ch{H}}$, in the potential of Eq.\ \eqref{eq: toy model potential} with $a=0.05$ Ha/Bohr$^2$ $b=0.1$ Ha/Bohr$^4$. This time, we combine NLSCHA with the transformation Eq.\ \eqref{eq: toy model transformation} setting $\eta = 0$ Bohr$^{-1}$. In Fig.\ \ref{fig:results 2} we summarize the exact, SCHA and NLSCHA results in the range $0-1000$ K. Note that with $\eta=0$ Bohr$^{-1}$ in Eq.\ \eqref{eq: toy model transformation} we can not fit properly the double well shape of the exact distribution (Fig.\ \ref{fig:results 2} c). Still, in this specific case, the violation of Eq.\ \eqref{eq: dF dT = - T S NLSCHA} when $\masstnsbm=\bm{m}$ emerges more evidently.

\begin{figure}[!htb]
    \centering
    \begin{minipage}[c]{0.72\linewidth}
    \includegraphics[width=1.0\textwidth]{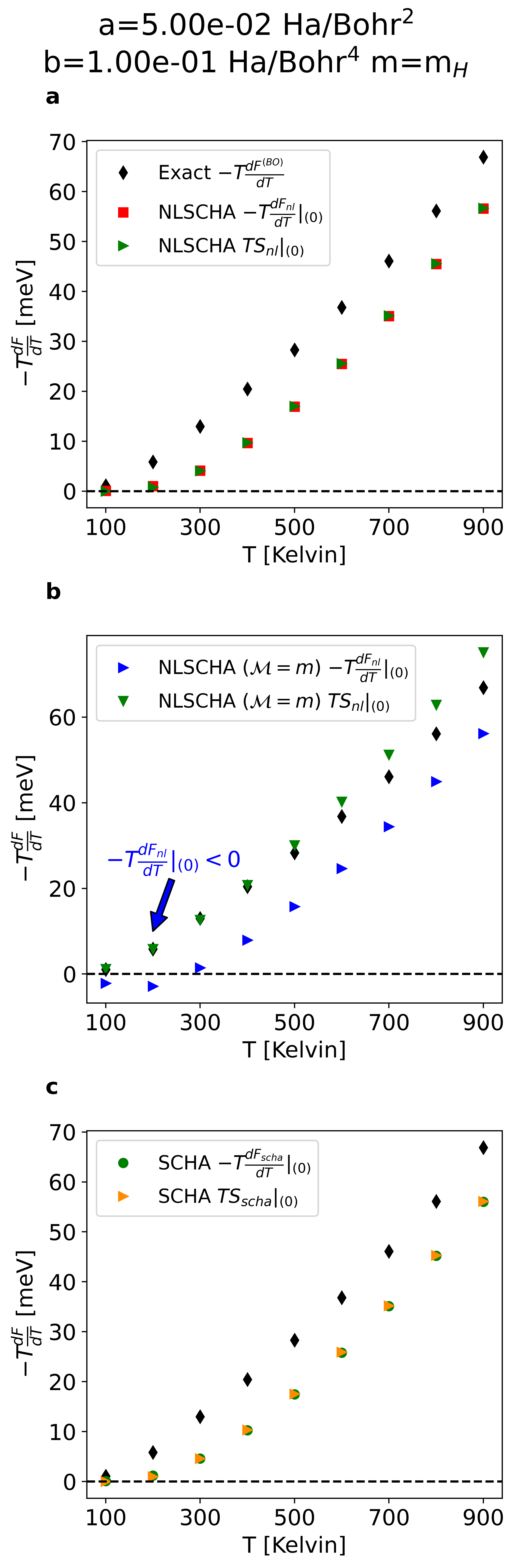}
    \end{minipage}
    \caption{The parameters used in the potential (Eq.\ \eqref{eq: toy model potential}) are $a =0.05$ Ha/Bohr$^2$ and $b=0.1$ Ha/Bohr$^4$ $m=m_\text{\ch{H}}$. Fig.\ a-b compares $-T\frac{d \Fnl}{d T} |_{(0)}$ with $-T \Snl |_{(0)}$ when $\masstns{}$ is optimized (Fig.\ a) or not (in Fig.\ b $\masstns{}=m_\text{\ch{H}}$). In addition, we report $-T\frac{d F^\text{(BO)}}{d T}$. Only when $\masstns{}$ is optimized we recover $-T\frac{d \Fnl}{d T} |_{(0)}=-T \Snl |_{(0)}$. Fig.\ c shows that for SCHA $-T\frac{d F_\text{scha}}{d T} |_{(0)} = T S_\text{scha} |_{(0)}$ holds always.}
    \label{fig: negative entropy}
\end{figure}
In Fig.\ \ref{fig: negative entropy} we compare $-T\frac{d F^\text{(BO)}}{d T}$ with the NLSCHA and SCHA values. In addition, for NLSCHA and SCHA, we report the analytical expression of the entropy multiplied by the temperature. For NLSCHA (Figs \ref{fig: negative entropy} a-b), only if we optimize the effective mass $\masstns{}$, we find Eq. \eqref{eq: dF dT = - T S} (see Fig.\ \ref{fig: negative entropy} a). Otherwise, if $\masstns{}=m_\text{\ch{H}}$, we end up with $-T\frac{d \Fnl}{d T} |_{(0)} < 0$ (see arrow in Fig.\ \ref{fig: negative entropy} b). So if we do not relax all the free parameters and use $\frac{d \Fnl}{d T} |_{(0)}$ to estimate the entropy we get an unphysical result, i.e. negative entropy (see Fig.\ \ref{fig:results 2} a and Fig.\ \ref{fig: negative entropy} b). Still, NLSCHA remains a variational theory and performs remarkably well compared to SCHA as its error is 50 $\%$ smaller.  Finally, for SCHA, we verify that the entropy is minus the temperature derivative of the free energy (see Fig.\ \ref{fig: negative entropy} c). 

\section{The stochastic approach}
\label{SEC: The stochastic approach}
In this section, we show that all the stochastic techniques presented in the SCHA implementation presented in Ref.\ \cite{SCHA_main} make our generalization more efficient. All necessary quantities to express $\Fnl$ are either analytical (as the entropy $\Snl$) or are expressed as averages in real space (Eqs \eqref{eq: nlscha kinetic energy} \eqref{eq: nlscha potential energy})
\begin{equation}
\label{eq: general form of nlscha equation}
    \myint d\bm{R} \rhocartnlscha(\bm{R}) 
    O_{\allfreeparameters}(\bm{R})
\end{equation}
where $O_{\allfreeparameters}(\bm{R})$ might depend on BO energies/forces and $\rhocartnlscha(\bm{R})$ indicates the diagonal elements of the density matrix (Eq.\ \eqref{eq: R rho R' nlscha})
\begin{equation}
    \rhocartnlscha(\bm{R}) = \bra{\bm{R}} \rhohatnlscha \ket{\bm{R}}
\end{equation}
and $\allfreeparameters$ contains all the free parameters $\allfreeparameters = (\freeparambmnlscha, \masstnsbm, \FCbmnlscha)$. Contrary to the SCHA, the kinetic energy integral (Eq.\ \eqref{eq: nlscha kinetic energy}) is not analytical; still, it can be computed using many configurations as it does not require any electronic calculations. Note that to improve the sampling of both the potential and kinetic energy, we can add and subtract in the nonlinear SCHA internal energy (Eq.\ \eqref{eq: nlscha free energy}) the reference Hamiltonian (Eq.\ \eqref{eq: nlscha Hamiltonian operator}).

We evaluate Eq.\ \eqref{eq: general form of nlscha equation} with a Monte Carlo sampling of the probability distribution $\rhocartnlscha(\bm{R})$ using $N_\text{C}$ supercell configurations
\begin{equation}
\label{eq: MC sampling}
    \frac{1}{N_\text{C}} \sum_{\mathcal{I}=1}^{N_\text{C}}O_{\allfreeparameters}
    (\bm{R}_{(\mathcal{I})})
\end{equation}
where the Cartesian configurations $\bm{R}_{(\mathcal{I})}$ are obtained by extracting the nonlinear variables $\bm{\unlscha}_{(\mathcal{I})}$  according to the Gaussian distribution $\gaussnlscha(\bm{\unlscha})$ then transforming them \replaced{according to $\bm{R}_{(\mathcal{I})}=\xibmnlscha(\bm{\unlscha}_{(\mathcal{I})})$ (Eq.\ \eqref{eq: change of variables})}{$R_{a,(\mathcal{I})} = \sum_{b=1}^{3N}
\Jinvnlscha{a}{b}{}_{0,\freeparambmnlscha}
\xinlscha_{b,\freeparambmnlscha}(\bm{\unlscha}_{(\mathcal{I})})$, where we explicitly indicate the dependence on the free parameters $\freeparambmnlscha$}. We compute BO energies and forces on these displaced configurations in the supercell. 

To reduce the computational cost further, we employ histogram reweighting in Cartesian coordinates. Indeed, each minimization step modifies the nonlinear SCHA free parameters $\allfreeparameters^{(0)} \rightarrow \allfreeparameters^{(1)} \ldots \rightarrow \allfreeparameters^{(n)}$. So at the $n$-th iteration, according to Eq.\ \eqref{eq: MC sampling}, we should extract a new ensemble of configurations $\bm{R}_{(\mathcal{I})}^{(n)}$ according to $\rhocartnlscha^{(n)}(\bm{R})$ and recompute the BO energies and forces. 
This would enormously increase the computational cost, so we extract the Cartesian configurations $\bm{R}_{(\mathcal{I})}^{(0)}$ only at the first step using $\rhocartnlscha^{(0)}(\bm{R})$ with free parameters $\allfreeparameters^{(0)}$. Then, at the $n$-step, all equations are computed as
\begin{equation}
    \frac{\sum_{\mathcal{I}=1}^{N_\text{C}} \rho_{(\mathcal{I})} 
    O_{\allfreeparameters^{(n)}}(\bm{R}_{(\mathcal{I})}^{(0)})}{\sum_{\mathcal{I}=1}^{N_\text{C}}  \rho_{(\mathcal{I})}}
\end{equation}
The weights $\rho_{(\mathcal{I})} $ are defined as the ratio between the current and the initial probability distribution
\begin{equation}
    \rho_{(\mathcal{I})} = 
    \frac{\rhocartnlscha^{(n)}(\boldsymbol{R}_\mathcal{I}^{(0)})}
    {\rhocartnlscha^{(0)}(\boldsymbol{R}_\mathcal{I} ^{(0)})}
\end{equation}
where the new probability distribution is
\begin{subequations}
\begin{align}
    \rhocartnlscha^{(n)}(\boldsymbol{R}_\mathcal{I}^{(0)})
    & = \frac{\sqrt{\det\left(\frac{\Ybmnlscha^{(n)}}{2\pi}\right)}
    \mathrm{e}^{-\frac{1}{2} \bm{\unlscha}^{(n)}_{\mathcal{I}}\cdot \Ybmnlscha^{(n)} \cdot \bm{\unlscha}^{(n)}_{\mathcal{I}}}}{\detJnlscha^{(n)}
    (\bm{\unlscha}^{(n)}_{\mathcal{I}})}
    \\
    \bm{\unlscha}^{(n)}_{\mathcal{I}} & = \xibmnlscha^{-1}_{(n)}
    \left(\bm{R}_{\mathcal{I}}^{(0)}\right)
\end{align}
\end{subequations}
and the $(n)$ indicates the dependence on the free parameters $\allfreeparameters^{(n)}$. Note that the inversion of the nonlinear transformation might be non-analytical, requiring numerical methods.
\deleted{We remark that to compute  $\rhocartnlscha^{(n)}(\bm{R}_{(\mathcal{I})}^{(0)})$ we need the auxiliary variables, hence the inverse of the nonlinear transformation, $\xibmnlscha^{-1}_{\freeparambmnlscha^{(n)}}(\bm{R}_{(\mathcal{I})}^{(0)})$.} 
As the minimization proceeds $\rhocartnlscha^{(n)}(\bm{R})$ will start to deviate from $\rhocartnlscha^{(0)}(\bm{R})$. To check the statistical accuracy of the reweighting technique the Kong-Liu ratio is very helpful, as discussed in Ref.\ \cite{monacelli2018pressure}.

\section{Conclusions}
\label{SEC: Conclusions}
The SCHA method encounters limitations in various scenarios, such as materials with rotational degrees of freedom and tunneling effects. Thus, we have introduced a generalization of the method. This extension involves a nonlinear change of variables, incorporating non-Gaussian fluctuations so that the real interacting normal modes of a crystal are mapped onto quantum harmonic oscillators defined in arbitrary coordinates. The enhanced flexibility of our theory is achieved by introducing new free parameters. First, there is the specific parameterization of the nonlinear transformation $\xibmnlscha$ that allows non-Gaussian degrees of freedom, for example, invertible neural network transformations. As in SCHA, there is the force constant matrix $\FCbmnlscha$, which describes the auxiliary phonons. In addition, we have the non-diagonal mass tensor $\masstnsbm$, which improves the description of quantum fluctuations (section \ref{SEC: 1D toy model}) and ensures that the entropy coincides with the temperature derivative of the free energy.

The variational estimation of the BO free energy involves minimizing $\Fnl$ using only BO energies and forces computed on supercell configurations. No other additional computational cost is required as the entropy is analytical, contrary to MD and PIMD, and it does not necessitate the diagonalization of large matrices or thermodynamic integrations. We showed that all equations can be evaluated with stochastic sampling along with histogram reweighting, as discussed in Ref.\ \cite{SCHA_main}. The latter significantly reduces the overall computational cost, especially in the case of \textit{ab-inito} calculation of energies and forces.
Remarkably, if the number of $\freeparambmnlscha$ scales with the system size $N$, the overall computational cost of the minimization is primarily governed, as in SCHA, by the optimization of $\FCbmnlscha$ and $\masstnsbm$, i.e.\ two $3N\times 3N$ tensors. Consequently, the nonlinear generalization remains as computationally efficient as the standard SCHA but surpasses it in describing non-Gaussian fluctuations, opening new applications of the SCHA theory \cite{INPREPARAZIONE}. In particular, our work paves the way for integrating SCHA with internal coordinates' formalism, leveraging both approaches' strengths. The SCHA offers the advantage of analytical entropy and kinetic energy \cite{AutomaticKEO} and incorporating internal coordinates enhances conformational sampling.

\section*{Acknowledgements}
A.S.\ and F.M.\ acknowledge support from the European Union under project ERC-SYN MORE-TEM (grant agreement No 951215).

\appendix

\section{Quantum mechanics in nonlinear variables}
\label{APP: Quantum mechanics in nonlinear variables}
In this appendix, we prove the exact mapping of quantum thermodynamics in nonlinear coordinates presented in section \ref{SEC: Quantum mechanics in nonlinear coordinates}. In appendix \ref{APP SUBSEC: NEW The nonlinear transformation: useful formulas}, we give some preliminary definitions regarding the nonlinear transformation (Jacobians, metric tensors etc.) and present useful formulas for the subsequent derivations. Then, in appendix \ref{APP SUBSEC: NEW The mappsing in nonlinear variables}, we discuss the exact mapping of quantum thermodynamics in nonstandard coordinates. In appendix \ref{APP SUBSEC: NEW The harmonic approximation}, we derive the quantum harmonic approximation in this new space. In appendix \ref{APP SUBSEC: NEW The classical harmonic approximation}, we review the classical HA to highlight the differences with the quantum case.

\subsection{The nonlinear transformation: useful formulas}
\label{APP SUBSEC: NEW The nonlinear transformation: useful formulas}
We employ mass-rescaled variables
\begin{equation}
    \widetilde{R}_a = \sqrt{m_a} R_a \qquad \widetilde{\unlscha}_a = \sqrt{m_a} \unlscha_a
\end{equation}
First, we define the Jacobian tensor of the nonlinear transformation
\begin{equation}
\label{eq app: 1-order jacobian}
    \Jtildenlscha{a}{b} = \pdv{\widetilde{R}_a}{\widetilde{\unlscha}_b}
\end{equation}
and the higher-order tensor
\begin{equation}
\label{eq app: n-order jacobian}
    \Jtildenlscha{a}{b_1..b_n} = \frac{\partial^n \widetilde{R}_a}{\partial \widetilde{\unlscha}_{b_1}..\widetilde{\unlscha}_{b_n}}
\end{equation}
where the lower indices denote the symmetric ones. 
The inverse of the Jacobian tensor, Eq.\ \eqref{eq app: 1-order jacobian}, is
\begin{equation}
\label{eq app: 1-order inv jacobian}
    \Jtildeinvnlscha{a}{b} = \pdv{\widetilde{\unlscha}_a}{\widetilde{R}_b}
\end{equation}
so that
\begin{equation}
    \Jtildebmnlscha\cdot \Jtildeinvbmnlscha = 
    \sum_{i=1}^{3N} \Jtildenlscha{a}{i} \Jtildeinvnlscha{i}{b} = \delta_{ab}
\end{equation}
From Eq.\ \eqref{eq app: 1-order inv jacobian}, we define the inverse (symmetric) metric tensor as
\begin{equation}
\label{eq app: def metric tensor}
    \gtildenlscha{a}{b} = \sum_{i=1}^{3N} \Jtildeinvnlscha{a}{i} \Jtildeinvnlscha{b}{i}
\end{equation}
Similarly to Eq.\ \eqref{eq app: n-order jacobian}, we define the high-order inverse Jacobian
\begin{equation}
\label{eq app: n-order inv jacobian}
    \Jtildeinvnlscha{a}{b_1..b_n}  = \frac{\partial^n \widetilde{\unlscha}_a}{\partial \widetilde{R}_{b_1}..\widetilde{R}_{b_n}}
\end{equation}
\deleted{In the following, we need also the rule for the $\bm{\unlscha}$-derivatives of Eq.\ \eqref{eq app: n-order inv jacobian}
\begin{equation}
\label{eq app: rule for derivative of Jinv wrt u}
    \pdv{\Jtildeinvnlscha{a}{b_1..b_n}}{\widetilde{\unlscha}_i}  = \sum_{j=1}^{3N}
    \Jtildenlscha{j}{i} \Jtildeinvnlscha{a}{jb_1..b_n} 
\end{equation}}

The determinant of the Jacobian tensor, Eq.\ \eqref{eq app: 1-order jacobian}, is
\begin{equation}
    \detJnlscha = \det\left(\Jtildebmnlscha\right)
\end{equation}
and we define the distortion vector $\dtildebmlogJdq$
\begin{equation}
\label{eq app: diff log J}
    \dtildelogJdq{a} = \frac{1}{2} \pdv{}{\widetilde{\unlscha}_a} \log(\detJnlscha)
\end{equation}
\deleted{so that also, we express the derivative $1/\sqrt{\detJnlscha}$ as
\begin{equation}
\label{eq app: derivative of inv sqrt det J}
    \pdv{}{\unlscha_a} \frac{1}{\sqrt{\detJnlscha}} = - \frac{\dlogJdq{a}}{\sqrt{\detJnlscha}}
\end{equation}}
Note that $\dtildebmlogJdq$ can be computed using
\begin{equation}
\label{eq app: equivalent expression for d}
    \dtildelogJdq{a} = \frac{1}{2\detJnlscha} \sum_{i=1}^{3N} \Jtildenlscha{i}{a} \pdv{\detJnlscha}{\widetilde{R}_i}
\end{equation}
where we use the Jacobi formula for the determinant's derivative
\begin{equation}
\label{eq app: jacobia formula for diff J wrt R}
    \pdv{\detJnlscha}{\widetilde{R}_k}  = - \detJnlscha \sum_{ij=1}^{3N} \Jtildenlscha{j}{i}  \Jtildeinvnlscha{i}{jk}
\end{equation}

For what follows, we need the first derivative of $\dtildebmlogJdq$ 
\begin{equation}
\label{eq app: first derivative of d}
    \pdv{\dtildelogJdq{a}}{\widetilde{\unlscha}_b}  = 
    -\frac{1}{2\detJnlscha^2} \pdv{\detJnlscha}{\widetilde{\unlscha}_a}   
    \pdv{\detJnlscha}{\widetilde{\unlscha}_b}
    +\frac{1}{2\detJnlscha} \pdv{\detJnlscha}{\widetilde{\unlscha}_a}{\widetilde{\unlscha}_b}
\end{equation}
and again we use the Jacobi formula to get
\begin{subequations}
\begin{align}
    \pdv{\detJnlscha}{\widetilde{\unlscha}_a}  & =  \detJnlscha \sum_{ij=1}^{3N} \Jtildenlscha{j}{ia}  \Jtildeinvnlscha{i}{j}
    \label{eq app: jacobian formula for diff J wrt u 1}\\
    \pdv{\detJnlscha}{\widetilde{\unlscha}_a}{\widetilde{\unlscha}_b}
    & = \pdv{\detJnlscha}{\widetilde{\unlscha}_b} \sum_{ij=1}^{3N} \Jtildenlscha{j}{ia}  \Jtildeinvnlscha{i}{j}
    +  \detJnlscha \pdv{}{\widetilde{\unlscha}_b}
    \left(\sum_{ij=1}^{3N} \Jtildenlscha{j}{ia}  \Jtildeinvnlscha{i}{j}\right)
    \label{eq app: jacobian formula for diff J wrt u 2}
\end{align}
\end{subequations}
where in Eq.\ \eqref{eq app: jacobian formula for diff J wrt u 2} the following formula comes in help
\begin{equation}
    \pdv{}{\widetilde{\unlscha}_a} \Jtildeinvnlscha{i}{j} = \sum_{k=1}^{3N}\Jtildenlscha{k}{a} \Jtildeinvnlscha{i}{jk}
\end{equation}
We need also the first derivative of the metric tensor $\gbmtildenlscha$ 
\begin{equation}
\label{eq app: diff inv metric tensor wrt u}
    \pdv{\gtildenlscha{a}{b} }{\widetilde{\unlscha}_c} = 
    \sum_{ij=1}^{3N} \Jtildenlscha{j}{c}\left(
    \Jtildeinvnlscha{b}{i}\Jtildeinvnlscha{a}{ij}+
    \Jtildeinvnlscha{a}{i}\Jtildeinvnlscha{b}{ij}
    \right) 
\end{equation}

\deleted{
Note that $\dlogJdqbm$ can be computed using two equivalent expressions
\begin{subequations}
\begin{align}
    &  \dlogJdq{a} = \frac{1}{2}
    \Tr\left[\Jtildeinvbmnlscha 
    \pdv{}{\widetilde{\unlscha}_a}
    \Jtildebmnlscha\right]
    =  \frac{1}{2}
    \sum_{ij=1}^{3N} \Jtildeinvnlscha{i}{j} \Jtildenlscha{j}{ia} 
    \label{eq app: diff log J explicit 1}\\
    & \dlogJdq{a} = - \frac{1}{2}
    \Tr\left[\Jtildebmnlscha \pdv{}{\widetilde{\unlscha}_a} \Jtildeinvbmnlscha\right] 
    = - \frac{1}{2} \sum_{ijk=1}^{3N} 
    \Jtildenlscha{i}{j}  \Jtildeinvnlscha{j}{ik} \Jtildenlscha{k}{a}
    \label{eq app: diff log J explicit 2}
\end{align}
\end{subequations}
where in the last line we use Eq.\ \eqref{eq app: rule for derivative of Jinv wrt u}.

From Eq.\ \eqref{eq app: diff log J explicit 2} we get the following relation
for the $\bm{\unlscha}$-derivative of Eq.\ \eqref{eq app: 1-order inv jacobian}
\begin{equation}
\label{eq app: J inv A}
    \frac{1}{2} \sum_{j=1}^{3N}
    \pdv{\Jtildeinvnlscha{j}{b}}{\widetilde{\unlscha}_j}
    = \frac{1}{2} \sum_{ij=1}^{3N} \Jtildenlscha{i}{j} 
    \Jtildeinvnlscha{j}{ib} 
    = -\sum_{a=1}^{3N} \Jtildeinvnlscha{a}{b} \dlogJdq{a} 
\end{equation}
To simplify the nonlinear SCHA equations we need also
\begin{equation}
\label{eq app: J inv high order from g and A}
\begin{aligned}
     \sum_{a=1}^{3N} \pdv{\Jtildeinvnlscha{i}{a}}{\widetilde{R}_a} 
    & = \sum_{a=1}^{3N} \pdv{}{\widetilde{R}_a}\left[\sum_{j=1}^{3N} \gnlscha{i}{j} \Jtildenlscha{a}{j}\right]  = \sum_{acj=1}^{3N} \Jtildeinvnlscha{c}{a} \pdv{}{\widetilde{\unlscha}_c}\left[ \gnlscha{i}{j} \Jtildenlscha{a}{j}\right] \\
    & = \sum_{acj=1}^{3N} \Jtildeinvnlscha{c}{a} 
    \pdv{\gnlscha{i}{j}}{\widetilde{\unlscha}_c}  \Jtildenlscha{a}{j} 
    + \sum_{acj=1}^{3N} \gnlscha{i}{j}  \Jtildeinvnlscha{c}{a} 
    \pdv{\Jtildenlscha{a}{j}}{\widetilde{\unlscha}_c} \\
    & = \sum_{j=1}^{3N} 
    \pdv{\gnlscha{i}{j}}{\widetilde{\unlscha}_j} 
    + \sum_{j=1}^{3N} \gnlscha{i}{j} \sum_{ac=1}^{3N}
    \Jtildeinvnlscha{c}{a}  \Jtildenlscha{a}{cj} \\
    & = \sum_{j=1}^{3N} \left(
    \pdv{\gnlscha{i}{j}}{\unlscha_j} 
    + 2 \gnlscha{i}{j} \dlogJdq{j}\right)
\end{aligned}
\end{equation}
where in the last line we used Eq.\ \eqref{eq app: diff log J explicit 1}.

The nonlinear SCHA equations require the first derivative of $\dlogJdq{a}$ (Eq.\ \eqref{eq app: diff log J explicit 1}) with respect to $\bm{\unlscha}$
\begin{equation}
\label{eq app: first derivative of A}
    \pdv{\dlogJdq{a}}{\widetilde{\unlscha}_b}  = 
    \frac{1}{2}
    \sum_{ij=1}^{3N} 
    \left(\Jtildeinvnlscha{i}{j} \Jtildenlscha{j}{iab}
    + \sum_{k=1}^{3N} 
    \Jtildenlscha{k}{b} \Jtildeinvnlscha{i}{jk} \Jtildenlscha{j}{ia} 
    \right) 
\end{equation}
and the first derivative of the metric tensor $\gnlscha{a}{b}$ with respect to $\bm{\unlscha}$
\begin{equation}
\label{eq app: diff g wrt u}
    \pdv{\gnlscha{a}{b} }{\widetilde{\unlscha}_c} = 
    \sum_{ij=1}^{3N} \Jtildenlscha{j}{c}\left(
    \Jtildeinvnlscha{b}{i}\Jtildeinvnlscha{a}{ij}+
    \Jtildeinvnlscha{a}{i}\Jtildeinvnlscha{b}{ij}
    \right) 
\end{equation}
Contracting the indices in Eq.\ \eqref{eq app: diff g wrt u} and using Eq.\ \eqref{eq app: J inv high order from g and A}, we get
\begin{equation}
\label{eq app: diff g wrt q contracted}
    \sum_{a=1}^{3N}\pdv{\gnlscha{a}{b} }{\widetilde{\unlscha}_a}  
    = \sum_{i=1}^{3N} \left( \Jtildeinvnlscha{b}{ii}
    - 2\gnlscha{b}{i}\dlogJdq{i} \right)
\end{equation}}

\subsection{The mapping in nonlinear variables}
\label{APP SUBSEC: NEW The mappsing in nonlinear variables}
First, we show that the kinetic energy computed in the 
auxiliary basis
\begin{equation}
\label{eq app: kinetic average in u}
    \hspace{-0.3cm}
    \Tru \left[\hat{\Kgeom}\hspace{0.1cm} \uhatrhoBO \right] = 
    \myint d\bm{\unlscha} \myint d\bm{\unlscha}' \Kgeom(\bm{\unlscha},\bm{\unlscha}') 
    \urhoBO (\bm{\unlscha}',\bm{\unlscha})
\end{equation}
gives the correct result
\begin{equation}
    - \frac{\hbar^2}{2} \sum_{i=1}^{3N} \frac{1}{m_i} \myint d\bm{R} \frac{\partial^2 \rho^\text{(BO)}(\bm{R},\bm{R}')}{\partial R'_i {}^2} \biggl|_{\bm{R}'=\bm{R}}
\end{equation}
Here, we employ an equivalent expression for the kinetic operator in $\bm{\unlscha}$-space (Eq.\ \eqref{eq: K u u'}) 
\begin{equation}
\label{eq app: K u u'}
    \Kgeom(\bm{\unlscha},\bm{\unlscha}')  = -\frac{\hbar^2}{2}
    \delta(\bm{\unlscha} - \bm{\unlscha}') \hspace{-0.1cm}
    \sum_{ab=1}^{3N} \hspace{-0.1cm}
    \left(\pdv{}{\widetilde{\unlscha}_a} + \dtildelogJdq{a} \right)
    \gtildenlscha{a}{b} \hspace{-0.1cm}
    \left(\pdv{}{\widetilde{\unlscha}_b} - \dtildelogJdq{b}\right)
\end{equation}

We use the expressions of $\Kgeom(\bm{\unlscha},\bm{\unlscha}')$ (Eq.\ \eqref{eq app: K u u'}) and $\urhoBO (\bm{\unlscha}',\bm{\unlscha})$ (Eq.\ \eqref{eq: rho BO in u space}) along with the relation for the Dira-$\delta$
\begin{equation}
    \delta(\bm{\unlscha} - \bm{\unlscha}') = \detJnlscha(\bm{R}) \delta(\bm{R} - \bm{R}')  
\end{equation}
to rewrite Eq.\ \eqref{eq app: kinetic average in u} as
\begin{equation}
\label{eq app: kinetic average}
\begin{aligned}
    -\frac{\hbar^2}{2}\myint d\bm{R} \myint d\bm{R}' &
    \frac{\delta(\bm{R} - \bm{R}')}{\sqrt{\detJnlscha(\bm{R})}}   \\
    &  \overline{\mathcal{K}} \left(\rho^\text{(BO)}(\bm{R},\bm{R}') \sqrt{\detJnlscha(\bm{R})} \right)
\end{aligned}
\end{equation}
where we introduced the operator $\overline{\mathcal{K}}$ that depends only on the $\bm{R}$-derivatives
\begin{equation}
\label{eq app: k operator}
    \hspace{-0.2cm} \overline{\mathcal{K}} = \hspace{-0.2cm} \sum_{abi=1}^{3N}\left(\sum_{k=1}^{3N}\Jnlscha{k}{a}\nabla_{R_k} + \dlogJdq{a}\right) 
    \frac{\gnlscha{a}{b}_i}{m_i} 
    \left(\sum_{l=1}^{3N}\Jnlscha{l}{b}\nabla_{R_l} - \dlogJdq{b}\right)
\end{equation}
with
\begin{equation}
\label{eq app: def g abi} 
\gnlscha{a}{b}_i = \Jnlscha{a}{i} \Jnlscha{b}{i}
\end{equation}
and $\nabla_{R_i} = \pdv{}{R_i}$. Expanding $\overline{\mathcal{K}}$ (Eq.\ \eqref{eq app: k operator}) we get
\begin{equation}
\label{eq app: kinetic exact 1}
\begin{aligned}
    & \overline{\mathcal{K}} = 
    \sum_{abkli=1}^{3N}\Jnlscha{k}{a} \frac{\gnlscha{a}{b}_i}{m_i} 
    \Jnlscha{l}{b} \nabla_{R_k}\nabla_{R_l}\\
    &+  \sum_{abkli=1}^{3N}\Jnlscha{k}{a}\left[\nabla_{R_k} \left(\frac{\gnlscha{a}{b}_i}{m_i} 
    \Jnlscha{l}{b}\right)\right]\nabla_{R_l} \\
    &
    - \sum_{abki=1}^{3N}\Jnlscha{k}{a}\left[\nabla_{R_k} \left(\frac{\gnlscha{a}{b}_i}{m_i} \dlogJdq{b} \right) \right] 
    -\sum_{abi=1}^{3N}\dlogJdq{a} \frac{\gnlscha{a}{b}_i}{m_i} \dlogJdq{b}
\end{aligned}
\end{equation}
where we use the square brackets to identify the terms that will not act on the density matrix in Eq.\ \eqref{eq app: kinetic average}.
Then we use the definition of the inverse metric tensor $\gbmnlscha$ Eq.\ \eqref{eq app: def metric tensor} and the expression of $\dlogJdqbm$ given in Eq.\ \eqref{eq app: equivalent expression for d} to have only $\bm{R}$-derivatives in $\overline{\mathcal{K}}$
\begin{equation}
\label{eq app: kinetic exact 2}
\begin{aligned}
    &  \overline{\mathcal{K}} 
    = \sum_{i=1}^{3N} \frac{\nabla^2_{R_i}}{m_i} 
    + \sum_{aki=1}^{3N}\frac{\Jnlscha{k}{a}\Jinvnlscha{a}{ki}}{m_i}  
    \nabla_{R_i} \\
    & - \sum_{aki=1}^{3N}\frac{\Jnlscha{k}{a}}{m_i} \left[\nabla_{R_k} \left(  \frac{\Jinvnlscha{a}{i}}{2\detJnlscha} \nabla_{R_i} \detJnlscha \right)\right]
    - \sum_{i=1}^{3N} \frac{\left(\nabla_{R_i} \detJnlscha \right)^2}{4m_i\detJnlscha^2}  \\
    & = \sum_{i=1}^{3N}\left[\frac{\nabla^2_{R_i}}{m_i} 
    - \frac{\nabla_{R_i} \detJnlscha}{m_i \detJnlscha} \nabla_{R_i}
    - \frac{\nabla^2_{R_i} \detJnlscha}{2 m_i\detJnlscha}  
    + \frac{3\left(\nabla_{R_i} \detJnlscha \right)^2}{4m_i\detJnlscha^2}   \right]
\end{aligned}
\end{equation}
where in the last line we employ Eq.\ \eqref{eq app: jacobia formula for diff J wrt R}.
Now we plug the new expression for $\overline{\mathcal{K}}$, Eq.\ \eqref{eq app: kinetic exact 2}, in Eq.\ \eqref{eq app: kinetic average} and we get  the correct kinetic energy
\begin{equation}
\label{eq app: kinetic u = kinetic R}
\begin{aligned}
    & \Tru \left[\hat{\Kgeom}\hspace{0.1cm} \uhatrhoBO \right] = \\
    &-\frac{\hbar^2}{2}\myint d\bm{R} \myint d\bm{R}' 
    \frac{\delta(\bm{R} - \bm{R}')}{\sqrt{\detJnlscha}} \sum_{i=1}^{3N}\left[\frac{\nabla^2_{R_i}}{m_i} 
    - \frac{(\nabla_{R_i} \detJnlscha)}
    {m_i \detJnlscha} \nabla_{R_i} \right.\\
    &\left.
    - \frac{\left(\nabla^2_{R_i} \detJnlscha\right)}{2 m_i\detJnlscha}  
    + \frac{3\left(\nabla_{R_i} \detJnlscha \right)^2}{4m_i\detJnlscha^2}   \right] 
    \rho^\text{(BO)}(\bm{R},\bm{R}') \sqrt{\detJnlscha} \\
    & = -\frac{\hbar^2}{2}\myint d\bm{R} \myint d\bm{R}' \delta(\bm{R} - \bm{R}')
    \sum_{i=1}^{3N}\left[
    \frac{\nabla^2_{R_i}}{m_i} 
    +  \frac{\nabla^2_{R_i} \sqrt{\detJnlscha}}{m_i \sqrt{\detJnlscha}} 
    \right. \\
    & \left.
    - \frac{(\nabla^2_{R_i} \detJnlscha)}{2 m_i\detJnlscha}  
    + \frac{\left(\nabla_{R_i} \detJnlscha \right)^2}{4m_i\detJnlscha^2}\right] 
    \rho^\text{(BO)}(\bm{R},\bm{R}') \\
    & = -\frac{\hbar^2}{2} \myint d\bm{R} 
    \sum_{i=1}^{3N}\frac{\nabla^2_{R_i}}{m_i}  \rho^\text{(BO)}(\bm{R},\bm{R}') \biggl|_{\bm{R}' = \bm{R}}
\end{aligned}
\end{equation}

The potential energy is simpler
\begin{equation}
\label{eq app: BO u = BO R}
\begin{aligned}
    & \Tru\left[ \hat{\overline{V}}^\text{(BO)}\uhatrhoBO\right]  = 
    \myint d\bm{\unlscha} \myint d\bm{\unlscha}' \VuBO(\bm{\unlscha},\bm{\unlscha}') \urhoBO (\bm{\unlscha}',\bm{\unlscha}) \\
    & = \myint d\bm{R} \myint d\bm{R}' \delta(\bm{R} -\bm{R}')  V^\text{(BO)}(\bm{R}) \rho^\text{(BO)}(\bm{R}',\bm{R})  
\end{aligned}
\end{equation}
So, with Eqs \eqref{eq app: BO u = BO R} \eqref{eq app: kinetic u = kinetic R} we prove that the Hamiltonian in the auxiliary space is the one of Eq.\ \eqref{eq: H BO u u'} as it conserves the expectation values (Eq.\ \eqref{eq: invariance of observables}).

In addition, we show that the operatorial expression for the BO density matrix in $\bm{\unlscha}$-space is Eq.\  \eqref{eq: rho hat BO in u space}. 
\deleted{We expand the Cartesian matrix elements of $\hat{\rho}^\text{(BO)} $
\begin{equation}
    \bra{\bm{R}} \hat{\rho}^\text{(BO)} \ket{\bm{R}'} = \frac{1}{ Z^\text{(BO)}}
    \sum_{n=0}^{+\infty} \frac{(-\beta)^n}{n!} \bra{\bm{R}} \left(\hat{H}^\text{(BO)}\right)^n \ket{\bm{R}'}
\end{equation}
where the matrix element $\bra{\bm{R}} \left(\hat{H}^\text{(BO)}\right)^n \ket{\bm{R}'}$ is computed in Cartesian basis
\begin{widetext}
\begin{equation}
\begin{aligned}
    & 
    \myint d\bm{R}_1 ..  \myint d\bm{R}_{n-1}  H^\text{(BO)}(\bm{R},\bm{R}_1) H^\text{(BO)}(\bm{R}_1,\bm{R}_2)...
     H^\text{(BO)}(\bm{R}_{n-1},\bm{R}')\\
    & = \myint d\bm{\unlscha}_1 ..  \myint d\bm{\unlscha}_{n-1}  \detJnlscha(\bm{\unlscha}_1)...
    \detJnlscha(\bm{\unlscha}_{n-1}) 
    H^\text{(BO)}(\xibmnlscha(\bm{\unlscha}),\xibmnlscha(\bm{\unlscha}_1))...
    H^\text{(BO)}(\xibmnlscha(\bm{\unlscha}_{n-1}),\xibmnlscha(\bm{\unlscha}'))
\end{aligned}
\end{equation}
The density matrix in the auxiliary space (Eq.\ \eqref{eq: rho BO in u space})
\begin{equation}
\begin{aligned}
    & \urhoBO(\bm{\unlscha},\bm{\unlscha}')  = \sqrt{\detJnlscha(\bm{\unlscha}) \detJnlscha(\bm{\unlscha}')}\frac{1}{Z^\text{(BO)}}
    \sum_{n=0}^{+\infty} \frac{(-\beta)^n}{n!} \bra{\bm{R}} \left(\hat{H}^\text{(BO)}\right)^n \ket{\bm{R}'} \\
    & = \frac{1}{Z^\text{(BO)}}
    \sum_{n=0}^{+\infty} \frac{(-\beta)^n}{n!}   \myint d\bm{\unlscha}_1 ...  \myint d\bm{\unlscha}_{n-1}  
     \sqrt{\detJnlscha(\bm{\unlscha})}
     H^\text{(BO)}(\xibmnlscha(\bm{\unlscha}),\xibmnlscha(\bm{\unlscha}_1))
      \sqrt{\detJnlscha(\bm{\unlscha}_1)}
      ... \sqrt{\detJnlscha(\bm{\unlscha}_{n-1}) } 
     H^\text{(BO)}(\xibmnlscha(\bm{\unlscha}_{n-1}),\xibmnlscha(\bm{\unlscha}')) 
     \sqrt{\detJnlscha(\bm{\unlscha}')} \\
    & = \frac{1}{Z^\text{(BO)}}
    \sum_{n=0}^{+\infty} \frac{(-\beta)^n}{n!}   \myint d\bm{\unlscha}_1 ...  \myint d\bm{\unlscha}_{n-1}  
     \uhamiltonianBO(\bm{\unlscha},\bm{\unlscha}_1) \uhamiltonianBO(\bm{\unlscha}_1,\bm{\unlscha}_2) ...
     \uhamiltonianBO(\bm{\unlscha}_{n-1},\bm{\unlscha}')
\end{aligned}
\end{equation}
\end{widetext}
}
Deriving $\hat{\rho}^\text{(BO)} = \exp{-\beta \hat{H}^\text{(BO)}} /Z^\text{(BO)}$ with respect to $\beta$ and project in Cartesian basis we get
\begin{equation}
\label{eq app: diff beta rho 1}
\begin{aligned}
    &- \pdv{\rho^\text{(BO)}(\bm{R},\bm{R}')}{\beta} 
    - \rho^\text{(BO)}(\bm{R},\bm{R}') \pdv{\log(Z^\text{(BO)})}{\beta} \\
    & = \myint d\bm{R}'' 
    H^\text{(BO)}(\bm{R},\bm{R}'') \rho^\text{(BO)}(\bm{R}'',\bm{R}') 
\end{aligned}
\end{equation}
Eq.\ \eqref{eq app: diff beta rho 1} written in the $\bm{\unlscha}$-space reads
\begin{equation}
\label{eq app: diff beta rho 2}
\begin{aligned}
    &  - \pdv{}{\beta} \frac{\urhoBO(\bm{\unlscha},\bm{\unlscha}')}{\sqrt{\detJnlscha(\bm{\unlscha}) \detJnlscha(\bm{\unlscha}')}}
    -  \frac{\urhoBO(\bm{\unlscha},\bm{\unlscha}')}{\sqrt{\detJnlscha(\bm{\unlscha}) \detJnlscha(\bm{\unlscha}')}} \pdv{\log(Z^\text{(BO)})}{\beta}\\
    & =  -\frac{1}{\sqrt{\detJnlscha(\bm{\unlscha}')\detJnlscha(\bm{\unlscha})}} \frac{\hbar^2}{2}\sum_{a=1}^{3N} \myint d\bm{\unlscha}'' \delta(\bm{\unlscha} - \bm{\unlscha}'')
    \\
    & \left[\sqrt{\detJnlscha(\bm{\unlscha}'')} \frac{\partial^2 }{\partial \widetilde{R}''_a{}^2}\biggl|_{\bm{R}''=\xibmnlscha(\bm{\unlscha}'')}
    \frac{1}{\sqrt{\detJnlscha(\bm{\unlscha}'') }} \right] \urhoBO(\bm{\unlscha}'',\bm{\unlscha}') \\
    & + 
    \delta(\bm{\unlscha}-\bm{\unlscha}')V^\text{(BO)}(\xibmnlscha(\bm{\unlscha}))\frac{\urhoBO(\bm{\unlscha},\bm{\unlscha}')}{\sqrt{\detJnlscha(\bm{\unlscha}) \detJnlscha(\bm{\unlscha}')}} \\
    & =  \frac{1}{\sqrt{\detJnlscha(\bm{\unlscha}')\detJnlscha(\bm{\unlscha})}}  \myint d\bm{\unlscha}'' \Kgeom(\bm{\unlscha},\bm{\unlscha}'') \urhoBO(\bm{\unlscha}'',\bm{\unlscha}') \\
    & + 
    \delta(\bm{\unlscha}-\bm{\unlscha}')V^\text{(BO)}(\xibmnlscha(\bm{\unlscha}))\frac{\urhoBO(\bm{\unlscha},\bm{\unlscha}')}{\sqrt{\detJnlscha(\bm{\unlscha}) \detJnlscha(\bm{\unlscha}')}}
\end{aligned}
\end{equation}
Then, we multiply Eq.\ \eqref{eq app: diff beta rho 2} by $\sqrt{\detJnlscha(\bm{\unlscha}) \detJnlscha(\bm{\unlscha}')}$ to get
\begin{equation}
\label{eq app: diff beta rho 3}
\begin{aligned}
    &  - \pdv{}{\beta} \urhoBO(\bm{\unlscha},\bm{\unlscha}')
    -  \urhoBO(\bm{\unlscha},\bm{\unlscha}') \pdv{\log(Z^\text{(BO)})}{\beta}\\
    & =  \myint d\bm{\unlscha}'' \left(\Kgeom(\bm{\unlscha},\bm{\unlscha}'') + \VuBO(\bm{\unlscha},\bm{\unlscha}'')\right) \urhoBO(\bm{\unlscha}'',\bm{\unlscha}') 
\end{aligned}
\end{equation}
Eq.\ \eqref{eq app: diff beta rho 3} shows that the density matrix in the auxiliary space is
\begin{equation}
    \uhatrhoBO = \frac{e^{-\beta \uhathamiltonianBO}}{Z^\text{(BO)}}
\end{equation}
where $Z^\text{(BO)}=\overline{Z}^\text{(BO)}$ is invariant by construction, so we find Eq.\ \eqref{eq: rho hat BO in u space}. 

Next, we discuss how to compute the entropy in the $\bm{\unlscha}$-space. We start with its standard definition
\begin{equation}
\label{eq app: entropy BO 1}
\begin{aligned}
    S^\text{(BO)} = & -k_\text{B} \Tr\left[\hat{\rho}^\text{(BO)} \log\left(\hat{\rho}^\text{(BO)}\right)\right] \\
    = & -k_\text{B} \myint d\bm{R} \myint d\bm{R}' \rho^\text{(BO)}(\bm{R},\bm{R}') \\
    & \bra{\bm{R}'}\log\left[\hat{1} + \left(\hat{\rho}^\text{(BO)} - \hat{1}\right)\right] \ket{\bm{R}} 
\end{aligned}
\end{equation}
then we expand  $\log(1+\hat{O})$
\begin{equation}
\label{eq app: entropy BO 2}
\begin{aligned}
    S^\text{(BO)} = & -k_\text{B} \myint d\bm{R} \myint d\bm{R}' \rho^\text{(BO)}(\bm{R},\bm{R}') \\
    & \sum_{k=0}^{+\infty} \frac{(-1)^k}{k+1}
    \bra{\bm{R}'}\left(\hat{\rho}^\text{(BO)} - \hat{1}\right)^{k+1}\ket{\bm{R}} 
\end{aligned}
\end{equation}
 and use the binomial coefficients
\begin{equation}
\label{eq app: entropy BO 3}
\begin{aligned}
    & S^\text{(BO)} =  -k_\text{B} \myint d\bm{R} \myint d\bm{R}' \rho^\text{(BO)}(\bm{R},\bm{R}') \\
    & \sum_{k=0}^{+\infty} \sum_{n=0}^{k}  \frac{(-1)^k}{k+1}
    \binom{k}{n} (-1)^{k-n}
    \bra{\bm{R}'}\hat{\rho}^\text{(BO)} {}^n \ket{\bm{R}} \\
    & = -k_\text{B} \sum_{k=0}^{+\infty} \sum_{n=0}^{k}  \frac{(-1)^{2k-n}}{k+1}
    \binom{k}{n} \\
    & \myint d\bm{R}  \myint d\bm{R}'   \myint d\bm{R}_1 ..  \myint d\bm{R}_{n-1} \rho^\text{(BO)}(\bm{R},\bm{R}')  \\
    & 
    \rho^\text{(BO)}(\bm{R}',\bm{R}_1) \rho^\text{(BO)}(\bm{R}_1,\bm{R}_2) 
    .. \rho^\text{(BO)}(\bm{R}_{n-1},\bm{R}) 
\end{aligned}
\end{equation}
Changing variables in Eq.\ \eqref{eq app: entropy BO 3} we get
\begin{equation}
\label{eq app: entropy BO 4}
\begin{aligned}
    & S^\text{(BO)} =
    -k_\text{B} \sum_{k=0}^{+\infty} \sum_{n=0}^{k}  \frac{(-1)^{2k-n}}{k+1}
    \binom{k}{n} \myint d\bm{\unlscha}   \myint d\bm{\unlscha}'   \\
    &  \myint d\bm{\unlscha}_1 ..  \myint d\bm{\unlscha}_{n-1} \detJnlscha(\bm{\unlscha}) \detJnlscha(\bm{\unlscha}') \detJnlscha(\bm{\unlscha}_1) .. \detJnlscha(\bm{\unlscha}_{n-1}) \\
    & \rho^\text{(BO)}(\xibmnlscha(\bm{\unlscha}),\xibmnlscha(\bm{\unlscha}')) 
    \rho^\text{(BO)}(\xibmnlscha(\bm{\unlscha}'),\xibmnlscha(\bm{\unlscha}_1)) ..  \rho^\text{(BO)}(\xibmnlscha(\bm{\unlscha}_{n-1}),\xibmnlscha(\bm{\unlscha})) \\
\end{aligned}
\end{equation}
then using the definition of Eq.\ \eqref{eq: rho BO in u space} in Eq.\ \eqref{eq app: entropy BO 4} we get
\begin{equation}
\label{eq app: entropy BO final}
\begin{aligned}
     &S^\text{(BO)} = 
    -k_\text{B} \sum_{k=0}^{+\infty} \sum_{n=0}^{k}  \frac{(-1)^{2k-n}}{k+1}
    \binom{k}{n} \\
    & \myint d\bm{\unlscha}  d\bm{\unlscha}'   d\bm{\unlscha}_1 .. d\bm{\unlscha}_{n-1} \urhoBO(\bm{\unlscha},\bm{\unlscha}') \\
    &  
    \urhoBO(\bm{\unlscha}',\bm{\unlscha}_1) \urhoBO(\bm{\unlscha}_1,\bm{\unlscha}_2) ..  \urhoBO(\bm{\unlscha}_{n-1},\bm{\unlscha}) \\
    & = -k_\text{B} \Tru\left[\uhatrhoBO \log\left(\uhatrhoBO\right)\right]
\end{aligned}
\end{equation}
and we prove Eq.\ \eqref{eq: entropy in u space}.

\subsection{The harmonic approximation}
\label{APP SUBSEC: NEW The harmonic approximation}
The BO Hamitlonian in nonlinear variables reads as
\begin{equation}
    \uhamiltonianBO(\bm{\unlscha},\bm{\unlscha}') 
    = \Kgeom(\bm{\unlscha},\bm{\unlscha}')
    +\VuBO(\bm{\unlscha},\bm{\unlscha}')
\end{equation}
The HA is obtained considering linear and quadratic fluctuations around the equilibrium positions $\bm{\unlscha}_0$
\begin{equation}
\label{eq app: HA Hamiltonian u u'}
\begin{aligned}
     \HAhamiltonian (\bm{\unlscha},\bm{\unlscha}') = & 
    \delta(\bm{\unlscha}-\bm{\unlscha}')\left(
    -\frac{\hbar^2}{2}\pdv{}{\widetilde{\bm{\unlscha}}} \cdot \gtildezerobmnlscha \cdot
    \pdv{}{\widetilde{\bm{\unlscha}}}  
    + \frac{1}{2} \delta\bm{\unlscha}\cdot
    \HAFCbm \cdot \delta\bm{\unlscha}
    \right)\\
    &+ \HAhamiltonianQ(\bm{\unlscha},\bm{\unlscha}')
\end{aligned}
\end{equation}
where the subscript $0$ indicates that the quantities are evaluated at $\bm{\unlscha}_0$, $\delta \bm{\unlscha} = \bm{\unlscha} - \bm{\unlscha}_0$, and $\HAhamiltonianQ(\bm{\unlscha},\bm{\unlscha}')$ is
\begin{subequations}
\begin{align}
    &\HAhamiltonianQ(\bm{\unlscha},\bm{\unlscha}') = 
    \delta(\bm{\unlscha} -\bm{\unlscha}') \left[ 
    \pdv{\Vgeom}{\widetilde{\bm{\unlscha}}}\biggl|_0 \cdot \delta\widetilde{\bm{\unlscha}} + 
    \frac{1}{2} \delta\widetilde{\bm{\unlscha}}\cdot
    \pdv{\Vgeom}{\widetilde{\bm{\unlscha}}}{\widetilde{\bm{\unlscha}}}\biggl|_0 \cdot \delta\widetilde{\bm{\unlscha}}\right] 
    \label{eq app: H Q u u'}
\end{align}
\end{subequations}
In $\HAhamiltonianQ(\bm{\unlscha},\bm{\unlscha}')$ we include the HA of the potential-like term ($\Vgeom(\bm{\unlscha})$) of Eq.\ \eqref{eq: K u u'}. 
If $\HAhamiltonianQ(\bm{\unlscha},\bm{\unlscha}')=0$ we retrieve the harmonic frequencies and eigenvectors obtained in quantum/classical mechanics in $\bm{R}$-space. Indeed, the normal modes of
\begin{equation}
    \HAhathamiltonian = \frac{1}{2} \phatbmnlscha \cdot \gtildezerobmnlscha \cdot \phatbmnlscha + \frac{1}{2} \hat{\bm{\unlscha}} \cdot \HAFCbm \cdot \hat{\bm{\unlscha}}
\end{equation}
are given by
\begin{equation}
\begin{aligned}
    & \overline{D}_{ij} = \frac{1}{\sqrt{m_i m_j}} \sum_{kl=1}^{3N} \Jinvnlscha{k}{i}{}_0 \HAFC{kl}
    \Jinvnlscha{l}{j}{}_0 
\end{aligned}
\end{equation}
Then we use the following relation for the force constant matrix in $\bm{\unlscha}$ and $\bm{R}$ space
\begin{equation}
    \HAFC{kl} = \sum_{mn=1}^{3N} \Jnlscha{m}{k}{}_0 \pdv{V^\text{(BO)}}{R_m}{R_n} \biggl|_0 
    \Jnlscha{n}{l}{}_0
\end{equation}
to show that if $\HAhamiltonianQ(\bm{\unlscha},\bm{\unlscha}')=0$ the normal modes are invariant in the HA
\begin{equation}
    \overline{D}_{ij} = \frac{1}{\sqrt{m_i m_j}}
    \pdv{V^\text{(BO)}}{R_i}{R_j} \biggl|_0
\end{equation}

\subsection{The classical harmonic approximation}
\label{APP SUBSEC: NEW The classical harmonic approximation}
Here, we discuss the classical HA in nonlinear coordinates starting from the standard Lagrangian 
\begin{equation}
\label{eq app: Lagrangian classical}
    \mathcal{L}^\text{(BO)} = \frac{1}{2} \sum_{i=1}^{3N} m_i \dot{R}^2_i - V^\text{(BO)}(\bm{R})
\end{equation}
We apply the nonlinear change of variables of Eq.\ \eqref{eq: change of variables} to transform Eq.\ \eqref{eq app: Lagrangian classical} into
\begin{equation}
\label{eq app: Lagrangian classical u}
    \overline{\mathcal{L}}^\text{(BO)} = \frac{1}{2}  
    \dot{\bm{\unlscha}} \cdot \bm{M} \cdot
    \dot{\bm{\unlscha}}- V^\text{(BO)}(\xibmnlscha(\bm{\unlscha}))
\end{equation}
In Eq.\ \eqref{eq app: Lagrangian classical u}, we define the mass tensor $\bm{M}$
\begin{equation}
    M_{ab} = \sum_{i=1}^{3N} \Jnlscha{i}{a} m_i \Jnlscha{i}{b}
\end{equation}
and its square root
\begin{equation}
    \sqrt{M}_{ai} = 
    \Jnlscha{i}{a} \sqrt{m}_i 
\end{equation}
so that
\begin{equation}
    \bm{M} =  
    \sqrt{\bm{M}} \cdot \sqrt{\bm{M}}{}^T
\end{equation}
The HA of the Lagrangian in $\bm{\unlscha}$-space (Eq.\ \eqref{eq app: Lagrangian classical u}) gives
\begin{equation}
\label{eq app: Lagrangian harmonic u}
    \overline{\mathcal{L}} = \frac{1}{2}  
    \dot{\bm{\unlscha}} \cdot \bm{M}_0 \cdot
    \dot{\bm{\unlscha}}
    - \frac{1}{2}\delta \bm{\unlscha} \cdot 
    \HAFCbm  \cdot\delta \bm{\unlscha}
\end{equation}
where $\HAFCbm$ is defined in Eq.\ \eqref{eq: HA FC in u} and the subscript $0$ indicates that the quantities are evaluated at $\bm{\unlscha}_0$. The equations of motion derived from Eq.\ \eqref{eq app: Lagrangian harmonic u} are
\begin{equation}
\label{eq app: HA classical eqm}
    \bm{M}_0 \cdot
    \delta \ddot{\bm{\unlscha}}  + \HAFCbm  \cdot\delta \bm{\unlscha} = \bm{0}
\end{equation}
In Fourier space, Eq.\ \eqref{eq app: HA classical eqm} transforms into
\begin{equation}
\label{eq app: classical secular equation}
    \det\left(\omega^2  - \sqrt{\bm{M}}_0{}^{-1} 
    \cdot \HAFCbm\cdot
    \sqrt{\bm{M}}_0{}^{-T} \right) = 0
\end{equation}
where $-T$ indicates the inverse of the transpose. 

The HA's results do not depend on the particular choice of the coordinates. Indeed, thanks to an exact cancellation of Jacobians in $\sqrt{\bm{M}}_0$ and in $\HAFCbm$, the dynamical matrix of Eq.\ \eqref{eq app: classical secular equation} coincides with the standard one
\begin{equation}
    \left(\sqrt{\bm{M}}_0{}^{-1} 
    \cdot \pdv{V^\text{(BO)}}{\bm{\unlscha}}{\bm{\unlscha}} \biggl|_0 \cdot
    \sqrt{\bm{M}}_0{}^{-T}\right)_{ab} = \frac{\Phi_{ab}}{\sqrt{m_a m_b}}
\end{equation}
where $\bm{\Phi}$ is defined by
\begin{equation}
\label{eq app: FC standard}
    \Phi_{ab} = \pdv{V^\text{(BO)}}{R_a}{R_b} \biggl|_0
\end{equation}and we used
\begin{equation}
    \pdv{V^\text{(BO)}}{\unlscha_a}{\unlscha_b} \biggl|_0
    = \sum_{ij=1}^{3N}\Jnlscha{i}{a}{}_0\Jnlscha{j}{b}{}_0 \Phi_{ij}
\end{equation}
The fact that classically the HA's frequencies and polarization vectors do not depend on the coordinates used means that without fluctuations we do not explore the effects of a nontrivial metric tensor.

\section{The entropy in nonlinear SCHA}
\label{APP: NEW Nonlinear SCHA entropy}
In this appendix,  we prove that the entropy in nonlinear SCHA is harmonic and does not depend on the particular change of variables $\xibmnlscha$ adopted. 
The proof presented is based on the form of the trial density matrix (Eqs \eqref{eq: R rho R' nlscha} \eqref{eq: u rho u' nlscha} \eqref{eq: Y = T - 2 A} \eqref{eq: def Y A nlscha}), on the invertibility of $\xibmnlscha$ (Eq.\ \eqref{eq: invertible transformation}) and on the fact that $\bm{\unlscha}$ are defined in the range $(-\infty,+\infty)^{3N}$.

The nonlinear SCHA trial density matrix $\rhohatnlscha$ (Eq.\ \eqref{eq: R rho R' nlscha}) is a physical density matrix , i.e.\ it satisfies Eqs \eqref{eq: physical conditions for density matrix}, so it has a spectral decomposition
\begin{equation}
    \rhohatnlscha = \sum_{i=1}^{+\infty} p_i \ket{\psi_{\text{nl},i}} \bra{\psi_{\text{nl},i}}
\end{equation}
with $p_i$ real and positive definite $p_i \geq 0$, $\sum_{i=1}^{+\infty} p_i = 1$. Thus, the corresponding entropy is
\begin{equation}
\label{eq app: sum pi log pi}
    -k_\text{B}\Tr\left[\rhohatnlscha
    \log(\rhohatnlscha)\right] = 
    -k_\text{B} \sum_{i=1}^{+\infty} p_i \log(p_i)
\end{equation}
The eigenvalues of $\rhohatnlscha$  are extracted from
\begin{equation}
\label{eq app: eigenvalues eq}
    \rhohatnlscha \ket{\psi_{\text{nl},i}} = p_i \ket{\psi_{\text{nl},i}}
\end{equation}
To show that Eq.\ \eqref{eq app: sum pi log pi} has the same form of the harmonic entropy and does not depend on $\xibmnlscha$, we project Eq.\ \eqref{eq app: eigenvalues eq} on a complete set, i.e.\ the Cartesian representation,
\begin{equation}
\begin{aligned}
    & \myint \hspace{-0.08cm} d\bm{R}'  \bra{\bm{R}} \rhohatnlscha\ket{\bm{R}'} \bra{\bm{R}'} \ket{\psi_{\text{nl},i}} = 
    p_i  \bra{\bm{R}}\ket{\psi_{\text{nl},i}} \\
    & \myint \hspace{-0.08cm} d\bm{R}'  \bra{\bm{R}} \rhohatnlscha\ket{\bm{R}'} \psi(\bm{R}')_{\text{nl},i} = p_i  \psi(\bm{R})_{\text{nl},i}
\end{aligned}
\end{equation}
Then we use the nonlinear transformation
\begin{equation}
    \myint\hspace{-0.08cm} d\bm{\unlscha}' \detJnlscha(\bm{\unlscha}')\frac{\gaussnlscha(\bm{\unlscha},\bm{\unlscha}')}{\sqrt{\detJnlscha(\bm{\unlscha}) \detJnlscha(\bm{\unlscha}')}}
    \psi(\bm{\unlscha}')_{\text{nl},i}
    = p_i  \psi(\bm{\unlscha})_{\text{nl},i}
\end{equation}
Rescaling the eigenfunctions with the Jacobian, $\overline{\psi}(\bm{\unlscha})_{\text{nl},i} = \psi(\bm{\unlscha})_{\text{nl},i}/ \sqrt{\detJnlscha(\bm{\unlscha})} $, we end up with
\begin{equation}
\label{eq app: equation for eigenvalues rho nlscha}
     \myint \hspace{-0.08cm} d\bm{\unlscha}' \gaussnlscha(\bm{\unlscha},\bm{\unlscha}')
     \overline{\psi}(\bm{\unlscha}')_{\text{nl},i} = p_i \overline{\psi}(\bm{\unlscha})_{\text{nl},i}
\end{equation}
Because $\gaussnlscha(\bm{\unlscha},\bm{\unlscha}')$ has the form of Eq.\ \eqref{eq: u rho u' nlscha}, then Eq.\ \eqref{eq app: equation for eigenvalues rho nlscha} takes on a similar form to the eigenvalue equation of a harmonic density matrix in Cartesian coordinates. The only difference is that it is expressed in terms of the $\bm{\unlscha}$ variables, whose range of definition is $(-\infty,+\infty)^{3N}$. 
As a consequence, the eigenvalues of $\rhohatnlscha$ remain unaltered by the nonlinear change of variables, and they depend solely on the nonlinear SCHA phonons (Eq.\ \eqref{eq: def nlscha phonons}). The entropy, as given in Eq.\ \eqref{eq app: sum pi log pi}, takes on a harmonic form (see Eq.\ \eqref{eq: S entropy nlscha expression}).

\section{Nonlinear SCHA}
\label{APP: NEW Nonlinear SCHA}
In this appendix, we derive the nonlinear SCHA free energy.

\subsection{Free energy}
\label{APP SUBSEC: NEW Free energy}
The nonlinear SCHA kinetic energy is computed using the expression of $\Kgeom(\bm{\unlscha},\bm{\unlscha}')$ given by Eqs \eqref{eq: K u u'} 
\begin{equation}
\label{eq app: kinetic 1}
\begin{aligned}
    &\Tru\left[\Khatgeom \rhouhatnlscha\right] 
    = -\frac{\hbar^2}{2} \myint d\bm{\unlscha} 
    \sum_{ab=1}^{3N}\left(
    \pdv{\gtildenlscha{a}{b}}{\widetilde{\unlscha}_b} \pdv{}{\widetilde{\unlscha}_a} \right.\\
    &\left.
    + \gtildenlscha{a}{b}\pdv{}{\widetilde{\unlscha}_a}{\widetilde{\unlscha}_b}\right)
    \gaussnlscha(\bm{\unlscha},\bm{\unlscha}') \biggl|_{\bm{\unlscha}'=\bm{\unlscha}} 
    + \myint d\bm{\unlscha} \Vgeomfunc(\bm{\unlscha})  \gaussnlscha(\bm{\unlscha})\\
    & =  -\frac{\hbar^2}{2} \myint d\bm{\unlscha} 
    \sum_{ab=1}^{3N}\left[
    \pdv{\gtildenlscha{a}{b}}{\widetilde{\unlscha}_b} 
    \left(-\frac{1}{2}\sum_{i=1}^{3N}\Ytildenlscha{a}{i}\widetilde{\unlscha}_i\right) \gaussnlscha(\bm{\unlscha})\right.\\
    &\left.
    + \gtildenlscha{a}{b}\left(-\frac{1}{2}\Ttildenlscha{a}{b} + 
    \frac{1}{4}\sum_{ij=1}^{3N} \Ytildenlscha{a}{i}\Ytildenlscha{b}{j} \widetilde{\unlscha}_i \widetilde{\unlscha}_j\right) \gaussnlscha(\bm{\unlscha})\right] \\
     & + \myint d\bm{\unlscha} \Vgeom(\bm{\unlscha})  \gaussnlscha(\bm{\unlscha})
\end{aligned}
\end{equation}
where we used
\begin{subequations}
\label{eq app: derivative of log rho u u'}
\begin{align}
    \pdv{\log( \gaussnlscha(\bm{\unlscha},\bm{\unlscha}'))}{\widetilde{\unlscha}_a} \biggl|_{\bm{\unlscha}=\bm{\unlscha}'} 
    & = - \frac{1}{2} \sum_{i=1}^{3N} \Ytildenlscha{a}{i} \widetilde{\unlscha}_i  \\
    \pdv{\log( \gaussnlscha(\bm{\unlscha},\bm{\unlscha}'))}{\widetilde{\unlscha}_a}{\widetilde{\unlscha}_b} \biggl|_{\bm{\unlscha}=\bm{\unlscha}'} 
    & = 
    -\frac{\Ttildenlscha{a}{b}}{2} \notag \\
    & +\frac{1}{4} \sum_{ij=1}^{3N} \Ytildenlscha{a}{i} \Ytildenlscha{b}{j} \widetilde{\unlscha}_i \widetilde{\unlscha}_j 
\end{align}
\end{subequations}
and define the mass-rescaled tensors as
\begin{equation}
\label{eq app: mass rescaled tensors}
    \widetilde{\circ}_{ab} = \frac{\circ _{ab}}{\sqrt{m_a m_b}}
\end{equation}
We integrate by parts the first term in the last line of Eq.\ \eqref{eq app: kinetic 1} and we get
\begin{equation}
\label{eq app: kinetic 2}
\begin{aligned}
    &\Tru\left[\Khatgeom \rhouhatnlscha\right] = -\frac{\hbar^2}{2} \myint d\bm{\unlscha} 
    \sum_{ab=1}^{3N}\gtildenlscha{a}{b} 
    \frac{\left(\Ytildenlscha{a}{b} -\Ttildenlscha{a}{b}  \right)}{2}
    \gaussnlscha(\bm{\unlscha})\\
    & +\frac{\hbar^2}{8} \myint d\bm{\unlscha} 
    \sum_{abij=1}^{3N} \gtildenlscha{a}{b} \Ytildenlscha{a}{i}\Ytildenlscha{b}{j} \widetilde{\unlscha}_i \widetilde{\unlscha}_j \gaussnlscha(\bm{\unlscha}) \\
     & + \myint d\bm{\unlscha} \Vgeom(\bm{\unlscha})  \gaussnlscha(\bm{\unlscha})
\end{aligned}
\end{equation}
In Eq.\ \eqref{eq app: kinetic 2} we replace $\Ytildebmnlscha - \Ttildebmnlscha = - 2\Atildebmnlscha$ and the definition of $\Vgeomfunc(\bm{\unlscha})$ (Eq.\ \eqref{eq: V geom u kinetic}) to get
\begin{equation}
\label{eq app: kinetic 3}
\begin{aligned}
   &\Tru\left[\Khatgeom \rhouhatnlscha\right] = 
    \frac{\hbar^2}{2} \myint d\bm{\unlscha} 
    \sum_{ab=1}^{3N}\gtildenlscha{a}{b} 
    \Atildenlscha{a}{b}
    \gaussnlscha(\bm{\unlscha})\\
    & +\frac{\hbar^2}{8} \myint d\bm{\unlscha} 
    \sum_{abij=1}^{3N} \gtildenlscha{a}{b} \Ytildenlscha{a}{i}\Ytildenlscha{b}{j} \widetilde{\unlscha}_i \widetilde{\unlscha}_j \gaussnlscha(\bm{\unlscha}) \\
    & + \frac{\hbar^2}{2}\myint d\bm{\unlscha}\sum_{ab=1}^{3N}  \left[
    \pdv{\left(\gtildenlscha{a}{b}\dtildelogJdq{b}\right) }{\widetilde{\unlscha}_a}
    +\dtildelogJdq{a} \gtildenlscha{a}{b}  \dtildelogJdq{b}\right] \gaussnlscha(\bm{\unlscha}) \\
    &= \frac{\hbar^2}{2} \myint d\bm{\unlscha} 
    \sum_{ab=1}^{3N}\gtildenlscha{a}{b} \left(\frac{\Ytildenlscha{a}{b}}{4} + \Atildenlscha{a}{b} \right)
    \gaussnlscha(\bm{\unlscha})\\
     & +\frac{\hbar^2}{8} \myint d\bm{\unlscha} 
    \sum_{ab=1}^{3N} \gtildenlscha{a}{b} \left( \sum_{ij=1}^{3N}\Ytildenlscha{a}{i}\Ytildenlscha{b}{j} \widetilde{\unlscha}_i \widetilde{\unlscha}_j - \Ytildenlscha{a}{b}\right) \gaussnlscha(\bm{\unlscha}) \\
    & + \frac{\hbar^2}{2}\myint d\bm{\unlscha}\sum_{abi=1}^{3N} 
    \gtildenlscha{a}{b}\dtildelogJdq{b} \Ytildenlscha{a}{i} \widetilde{\unlscha}_i \gaussnlscha(\bm{\unlscha}) \\
    &  + \frac{\hbar^2}{2} \myint d\bm{\unlscha}\sum_{ab=1}^{3N} 
    \dtildelogJdq{a} \gtildenlscha{a}{b}  \dtildelogJdq{b}\gaussnlscha(\bm{\unlscha})\\
\end{aligned}
\end{equation}
where, in the last line, we integrate by parts and we add and subtract $\hbar^2\sum_{ab=1}^{3N} \gtildenlscha{a}{b} \Ytildenlscha{a}{b} /8$ to reproduce the kinetic energy of a harmonic oscillator in arbitrary coordinates (see Eq.\ \eqref{eq: kinetic energy of nonlinear phonons}).
In the end, the nonlinear SCHA kinetic energy, Eq.\ \eqref{eq app: kinetic 3}, can be written in the form of Eq.\ \eqref{eq: nlscha kinetic energy}
\begin{equation}
\label{eq app: kinetic 4}
\begin{aligned}
    &\Tru\left[\Khatgeom \rhouhatnlscha\right] = \frac{\hbar^2}{8} \myint d\bm{\unlscha} 
    \sum_{ab=1}^{3N} \gtildenlscha{a}{b} 
    \pdv{}{\widetilde{\unlscha}_a}{\widetilde{\unlscha}_b} \gaussnlscha(\bm{\unlscha}) \\
    & - \frac{\hbar^2}{2}\myint d\bm{\unlscha}\sum_{ab=1}^{3N} 
    \gtildenlscha{a}{b}\dtildelogJdq{b} \pdv{}{\widetilde{\unlscha}_a} \gaussnlscha(\bm{\unlscha}) \\
    & + \frac{\hbar^2}{2} \myint d\bm{\unlscha} 
    \sum_{ab=1}^{3N}\gtildenlscha{a}{b} \left(\frac{\Ytildenlscha{a}{b}}{4} + \Atildenlscha{a}{b} +   \dtildelogJdq{a} \dtildelogJdq{b}\right)
    \gaussnlscha(\bm{\unlscha})
\end{aligned}
\end{equation}

To have it in a more compact form, we introduce the kinetic kernel $\Knlscha$ as 
\begin{equation}
\label{eq app: knlscha}
    \Knlscha = \sum_{ab=1}^{3N} \Ktwonlscha{a}{b} \Ltwonlscha{a}{b}  + \sum_{a=1}^{3N} \Konenlscha{a}\Lonenlscha{a} +\Kzeronlscha
\end{equation}
where the coefficients $\Ltwobmnlscha$, $\Lonebmnlscha$ are defined by
\begin{subequations}
\label{eq app: def L coeff}
\begin{align}
    \Ltwonlscha{a}{b} & = 
    -  \pdv{ \log\left(\gaussnlscha(\bm{\unlscha})\right)}{\widetilde{\unlscha}_a}{\widetilde{\unlscha}_b} 
    \label{eq app: L two def}\\
    & = 
    \Ytildenlscha{a}{b}- \sum_{ij=1}^{3N}
    \Ytildenlscha{a}{i} \Ytildenlscha{b}{j}
    \widetilde{\unlscha}_i \widetilde{\unlscha}_j
    \notag \\
    \Lonenlscha{a}   &= 
    - \pdv{ \log\left(\gaussnlscha(\bm{\unlscha})\right)}{\widetilde{\unlscha}_a} = 
   \sum_{i=1}^{3N} \Ytildenlscha{a}{i} 
    \widetilde{\unlscha}_i  \label{eq app: L one app}
\end{align}
\end{subequations}
and $\Ktwobmnlscha$, $\Konebmnlscha$, $\Kzeronlscha$ by
\begin{subequations}
\label{eq app: K2 K1 K0 def}
\begin{align}
    & \Kzeronlscha  = \frac{\hbar^2}{2} \sum_{ab=1}^{3N}\gtildenlscha{a}{b}
    \left(\dtildelogJdq{a} \dtildelogJdq{b}
    +  \frac{\Ytildenlscha{a}{b}}{4}
    + \Atildenlscha{a}{b}\right) \\
    & \Konenlscha{a} = \frac{\hbar^2}{2}
    \sum_{b=1}^{3N} \gtildenlscha{a}{b} \dtildelogJdq{b}\\
    &\Ktwonlscha{a}{b}
     = -\frac{\hbar^2}{8} \gtildenlscha{a}{b} \\
\end{align}
\end{subequations}
The nonlinear SCHA kinetic energy is
\begin{equation} 
\label{eq app: kinetic final}
    \hspace{-0.2cm}
    \Tru\left[\Khatgeom \rhouhatnlscha\right] =\left\langle
    \Knlscha \right\rangle_\text{nl}
\end{equation}
introducing the notation
\begin{equation}
    \left\langle \circ \right\rangle_\text{nl} = \myint d\bm{\unlscha} \circ 
     \gaussnlscha(\bm{\unlscha})
\end{equation}

If the mass tensor coincides with the physical mass $\masstns{ab}=\delta_{ab} m_a$ and the transformation is linear, i.e.\ the metric tensor $\gbmnlscha$ becomes a constant and $\dlogJdqbm = \bm{0}$ ($\Konebmnlscha = \bm{0}$), we find the standard SCHA kinetic energy
\begin{equation}
\begin{aligned}
    \averagegaussnl{\Knlscha} = & \sum_{ab=1}^{3N} \Ktwonlscha{a}{b} \averagegaussnl{\Ltwonlscha{a}{b}}  +\Kzeronlscha \\
    = &\sum_{\mu=1}^{3N} \gtildenlscha{\mu}{\mu} 
    \frac{\hbar \onlscha{\mu}}{4}\left(1 + 2 \nnlscha{\mu}\right) 
\end{aligned}
\end{equation}
The entropy $\Snl$ (Eq.\ \eqref{eq: S entropy nlscha expression}) does not change but the sampling of the potential (Eq.\ \eqref{eq: nlscha potential energy}) does and reproduces the standard SCHA sampling of the phase space.

\section{Gradient of nonlinear SCHA}
\label{APP: NEW Gradient of nonlinear SCHA}
In this appendix, we compute the nonlinear SCHA gradients. In appendix \ref{APP SUBSEC: Tensor derivatives} we report the derivatives of $\Ybmnlscha$ and $\Abmnlscha$ with respect to the auxiliary force constant $\FCbmnlscha$. Then in appendix \ref{APP SUBSEC: NEW Gradient calculations} we report all the calculations for the nonlinear SCHA gradients.

\subsection{Tensor derivatives}
\label{APP SUBSEC: Tensor derivatives}
A generic tensor in nonlinear SCHA (see Eqs \eqref{eq: def Y A nlscha}) has the following form
\begin{equation}
\label{eq app: nlscha F def}
    \bm{F}_\text{nl} = \sqrtmasstnsTbm\cdot\overline{\bm{F}}_\text{nl} \cdot\sqrtmasstnsbm
\end{equation}
where $\overline{\bm{F}}_\text{nl}=\overline{\bm{F}}_\text{nl}(\Dbmnlscha)$  is a function of frequencies and eigenvectors $\onlscha{\mu}$ and $\polbmnlscha{\mu}$ (Eq.\ \eqref{eq: def nlscha phonons})
\begin{equation}
\label{eq app: nlscha F wout mass}
    \overline{F}_{\text{nl},ab} = \sum_{\mu=1}^{3N} f(\onlscha{\mu}) \polnlscha{\mu}{a} \polnlscha{\mu}{b}
\end{equation}
In general the derivative of $\bm{F}_\text{nl}$ can be expressed in terms of $\overset{-1}{\bm{F}}_\text{nl}$
\begin{equation}
    \pdv{\bm{F}_\text{nl}}{\lambda}
    = - \bm{F}_\text{nl} \cdot \pdv{\overset{-1}{\bm{F}}_\text{nl}}{\lambda} \cdot \bm{F}_\text{nl} 
\end{equation}

First, we compute the derivatives of $\overset{-1}{\bm{F}}_\text{nl}$ with respect to $\FCbmnlscha$
\begin{equation}
\label{eq app: d F d FC}
    \Lambda_{\text{nl,}abcd} = 
    \frac{1}{2}\pdv{\overset{-1}{F}_{\text{nl,}ab}}{\FCnlscha{c}{d}} = \sum_{ij=1}^{3N}\invsqrtmasstns{ai} 
    \frac{1}{2}
    \pdv{\overset{-1}{\overline{F}}_{\text{nl},ij}}{\FCnlscha{c}{d}} 
    \invsqrtmasstnsT{jb}  
\end{equation}
As $\overset{-1}{\overline{\bm{F}}}_{\text{nl}}$ depends on the dynamical matrix $\Dbmnlscha$ we write its derivative in Eq.\ \eqref{eq app: d F d FC} as
\begin{equation}
     \frac{1}{2}\pdv{\overset{-1}{\overline{F}}_{\text{nl},ab}}{\FCnlscha{c}{d}}
    = \sum_{ij=1}^{3N} \invsqrtmasstnsT{ci} 
    \frac{1}{2}
     \pdv{\overset{-1}{\overline{F}}_{\text{nl},ab}}{\Dnlscha{i}{j}}  \invsqrtmasstns{jd}
\end{equation}
where the derivative with respect $\Dbmnlscha$ is computed using the results of Ref.\ \cite{Bianco}
\begin{subequations}
\label{eq app: Lambda Bianco}
\begin{align}
    &  \overline{\Lambda}_{\text{nl,}abcd} = \pdv{\overset{-1}{\overline{F}}_{\text{nl},ab}}{\Dnlscha{c}{d}} =
    \sum_{\mu\nu=1}^{3N} \polnlscha{\mu}{a} \polnlscha{\nu}{b} \polnlscha{\mu}{c} \polnlscha{\nu}{d} \overline{L}_{\mu\nu} \\
    & \overline{L}_{\mu\nu} = \left\{ \begin{array}{lr}
    & \pdv{f(\onlscha{\mu})}{\onlscha{\mu}^2}  \quad 
    \onlscha{\mu} = \onlscha{\nu} \\
    \\
    &  \frac{f(\onlscha{\mu}) - f(\onlscha{\nu})}{\onlscha{\mu}^2 - \onlscha{\nu}^2} \quad 
    \onlscha{\mu} \neq \onlscha{\nu}
    \end{array}\right.
\end{align}
\end{subequations}

Secondly, we compute the derivative of $\bm{F}_\text{nl}$ with respect to $\sqrtmasstnsbm$. Note that $\bm{F}_\text{nl}$ (Eq.\ \eqref{eq app: nlscha F def}) depends on $\sqrtmasstnsbm$ both explicitly and implicitly via $\Dbmnlscha$ (Eq.\ \eqref{eq: def nlscha phonons}), so the derivative reads as
\begin{equation}
\label{eq app: d F d sqrt M}
\begin{aligned}
    \pdv{F_{\text{nl},ab}}{\sqrtmasstns{cd}} = & \sum_{i=1}^{3N} \overline{F}_{\text{nl},ci} \left(\sqrtmasstns{ib}\delta_{ad} + \sqrtmasstns{ia} \delta_{bd}\right) \\
    & + \sum_{ijkl=1}^{3N}   \invsqrtmasstnsT{ai}\pdv{\overline{F}_{\text{nl},ij}}{\Dnlscha{k}{l}}  \invsqrtmasstns{jb}  \pdv{\Dnlscha{k}{l}}{\sqrtmasstns{cd}} 
\end{aligned}
\end{equation}
where in the second line we use $\overline{\bm{\Lambda}}_\text{nl}$ (Eqs \eqref{eq app: Lambda Bianco})
\begin{equation}
    \pdv{\overline{F}_{\text{nl},ab}}{\Dnlscha{c}{d}}
    = - 2\sum_{ij=1}^{3N}  \overline{F}_{\text{nl},ai}
    \overline{\Lambda}_{\text{nl,}ijcd}
    \overline{F}_{\text{nl},jb}
\end{equation}
The derivative of $\Dbmnlscha$ in Eq.\ \eqref{eq app: d F d sqrt M} can be computed using the chain rule
\begin{subequations}
\label{eq app: d D d sqrt M}
\begin{align}
    & \pdv{\Dnlscha{a}{b}}{\sqrtmasstns{cd}} = -\sum_{mn=1}^{3N}
    \invsqrtmasstnsT{cm} \pdv{\Dnlscha{a}{b}}{\invsqrtmasstns{mn}}  \invsqrtmasstnsT{nd} \\
    & \pdv{\Dnlscha{a}{b}}{\invsqrtmasstns{cd}}
    = \sum_{i=1}^{3N} 
    \FCnlscha{c}{i}\left(\invsqrtmasstns{ib}\delta_{ad}
    + \invsqrtmasstns{ia} \delta_{bd}\right)
    \label{eq app: d D d sqrt M inv}
\end{align}
\end{subequations}

\deleted{\subsection{The \texorpdfstring{$\overline{\bm{\Lambda}}_\text{nl}$}{} matrix}
\label{APP SUBSEC: The Lambda matrix}
Here, we compute the derivative of matrix $\overline{\bm{F}}(\Dbmnlscha)$ with respect to $\Dbmnlscha$
\begin{equation}
\label{eq app: Lambda overline Bianco}
	\overline{\Lambda}_{\text{nl},abcd} = \frac{1}{2}\pdv{\overline{F}_{\text{nl},ab}}{\Dnlscha{c}{d}} =
	\frac{1}{2}\pdv{}{\Dnlscha{c}{d}} \sum_{\mu=1}^{3N} f_\mu \polnlscha{\mu}{a} \polnlscha{\mu}{b}
\end{equation}
where $f_\mu = f_\mu(\onlscha{\mu})$
\begin{equation}
\begin{aligned}
	\pdv{\overline{F}_{\text{nl},ab}}{\Dnlscha{c}{d}} = \sum_{\mu=1}^{3N} \left( \pdv{f_\mu}{\Dnlscha{c}{d}} \polnlscha{\mu}{a} \polnlscha{\mu}{b}
	+f_\mu
	\pdv{(\polnlscha{\mu}{a} \polnlscha{\mu}{b})}{\Dnlscha{c}{d}} \right)
\end{aligned}
\end{equation}

The derivative of the polarization vector $\polbmnlscha{\mu}$ with respect to a parameter $g$ in the nondegenrate case is
\begin{equation}
	\label{eq app: d e d g nondegenerate}
	\pdv{\polbmnlscha{\mu}}{g} = \sum_{\nu\neq\mu}^{3N} \frac{\polbmnlscha{\nu} }{\onlscha{\mu}^2 - \onlscha{\nu}^2}
	\left(\polbmnlscha{\nu} \cdot \pdv{\Dbmnlscha}{g} \cdot \polbmnlscha{\mu}\right)
\end{equation}
so the derivative with respect to $\Dbmnlscha$ is
\begin{equation}
	\pdv{\polnlscha{\mu}{a}}{\Dnlscha{c}{d}} = \sum_{\nu\neq\mu}^{3N} \polnlscha{\nu}{a}
	\frac{ \left(\polnlscha{\mu}{c} \polnlscha{\nu}{d}
		+\polnlscha{\nu}{c} \polnlscha{\mu}{d} \right)}{2(\onlscha{\mu}^2 - \onlscha{\nu}^2)}
\end{equation}
Then, the derivative of $\polnlscha{\mu}{a} \polnlscha{\mu}{b}$ is derived with the chain rule
\begin{equation}
	\label{eq app: d e mu e nu d D}
	\begin{aligned}
		\pdv{(\polnlscha{\mu}{a} \polnlscha{\mu}{b})}{\Dnlscha{c}{d}} = &
		\sum_{\nu\neq\mu}^{3N} \polnlscha{\nu}{a} \polnlscha{\mu}{b}
		\frac{ \left(\polnlscha{\mu}{c} \polnlscha{\nu}{d}
			+\polnlscha{\nu}{c} \polnlscha{\mu}{d} \right)}{2(\onlscha{\mu}^2 - \onlscha{\nu}^2)} \\
		& +  \sum_{\nu\neq\mu}^{3N}\polnlscha{\mu}{a} \polnlscha{\nu}{b}
		\frac{ \left(\polnlscha{\mu}{c} \polnlscha{\nu}{d}
			+  \polnlscha{\nu}{c}\polnlscha{\mu}{d}\right)}{2(\onlscha{\mu}^2 - \onlscha{\nu}^2)} \\
		=  & \sum_{\nu\neq\mu}^{3N} 
		\frac{ \left(\polnlscha{\nu}{a} \polnlscha{\mu}{b}+ \polnlscha{\mu}{a} \polnlscha{\nu}{b}\right)\left(\polnlscha{\mu}{c} \polnlscha{\nu}{d}
			+  \polnlscha{\nu}{c}\polnlscha{\mu}{d}\right)}{2(\onlscha{\mu}^2 - \onlscha{\nu}^2)}
	\end{aligned}
\end{equation}
The derivative of $f_\mu$ is
\begin{equation}
\label{eq app: d f mu d D}
	\pdv{f_\mu}{\Dnlscha{c}{d}} = 
	\polnlscha{\mu}{c} \polnlscha{\mu}{d}
	\pdv{f_\mu}{\onlscha{\mu}^2} 
\end{equation}

Putting Eqs \eqref{eq app: d e mu e nu d D} \eqref{eq app: d f mu d D} together, we get
\begin{equation}
	\begin{aligned}
		\pdv{\overline{F}_{\text{nl},ab}}{\Dnlscha{c}{d}} = &\sum_{\mu=1}^{3N} \left[\polnlscha{\mu}{a} \polnlscha{\mu}{b} \polnlscha{\mu}{c} \polnlscha{\mu}{d}
		\pdv{f_\mu}{\onlscha{\mu}^2}  \right.\\
		&\left. + \sum_{\nu\neq\mu}^{3N} f_\mu 
		\frac{ \left(\polnlscha{\nu}{a} \polnlscha{\mu}{b}+ \polnlscha{\mu}{a} \polnlscha{\nu}{b}\right)\left(\polnlscha{\mu}{c} \polnlscha{\nu}{d}
			+ \polnlscha{\mu}{d} \polnlscha{\nu}{c}\right)}{2(\onlscha{\mu}^2 - \onlscha{\nu}^2)} \right] \\
		= & \sum_{\mu=1}^{3N} \left[\polnlscha{\mu}{a} \polnlscha{\mu}{b} \polnlscha{\mu}{c} \polnlscha{\mu}{d}
		\pdv{f_\mu}{\onlscha{\mu}^2}  \right.\\
		&\left. + \sum_{\nu\neq\mu}^{3N}
		\polnlscha{\mu}{a} \polnlscha{\nu}{b}
		\frac{\left(\polnlscha{\mu}{c} \polnlscha{\nu}{d}
		+ \polnlscha{\nu}{c} \polnlscha{\mu}{d} \right)}{2}
		\frac{f_\mu - f_\nu}{\onlscha{\mu}^2 - \onlscha{\nu}^2} \right]
	\end{aligned}
\end{equation}
So, the $\overline{\bm{\Lambda}}_\text{nl}$ matrix is
\begin{equation}
	\label{eq app: Lambda overline final Bianco}
	\overline{\Lambda}_{\text{nl},abcd} = \frac{1}{2}
	\sum_{\mu\nu=1}^{3N} \polnlscha{\mu}{a} \polnlscha{\nu}{b}
	\left(\polnlscha{\mu}{c} \polnlscha{\nu}{d}
	+ \polnlscha{\nu}{c} \polnlscha{\mu}{d} \right)
	\left\{ \begin{array}{lr}
	& \pdv{f_\mu}{\onlscha{\mu}^2}  \quad 
	\onlscha{\mu} = \onlscha{\nu} \\
	&  \frac{f_\mu - f_\nu}{2(\onlscha{\mu}^2 - \onlscha{\nu}^2)} \quad 
	\onlscha{\mu} \neq \onlscha{\nu}
	\end{array}\right.
\end{equation}}

\subsection{Gradient calculations}
\label{APP SUBSEC: NEW Gradient calculations}
In this appendix, we compute all the nonlinear SCHA free energy gradients. First, we define the BO forces as
\begin{equation}
    \bm{f}^{\text{(BO)}} = - \pdv{V^{\text{(BO)}}}{\bm{R}}
\end{equation}

The derivative with respect to the free parameters $\freeparambmnlscha$ of the transformation is
\begin{equation}
    \pdv{\Fnl}{\freeparamnlscha{a}} = 
    \averagegaussnl{\pdv{\Knlscha}{\freeparamnlscha{a}}} + \averagegaussnl{\pdv{V^{\text{(BO)}}}{\freeparamnlscha{a}}}
\end{equation}
The derivative of the kinetic energy kernel (Eq.\ \eqref{eq app: knlscha}) is
\begin{equation}
\begin{aligned}
    & \pdv{\Knlscha(\bm{\unlscha})}{\freeparamnlscha{a}}
    =  \sum_{ij=1}^{3N} \pdv{\Ktwonlscha{i}{j}}{\freeparamnlscha{a}}
     \Ltwonlscha{j}{i}    \\
    & + \sum_{i=1}^{3N} \pdv{\Konenlscha{i}}{\freeparamnlscha{a}}
     \Lonenlscha{i}
    + \pdv{\Kzeronlscha}{\freeparamnlscha{a}}
\end{aligned}
\end{equation}
The derivatives of $\Kzeronlscha$, $\Konebmnlscha$ and $\Ktwobmnlscha$  are
\begin{subequations}
\label{eq app: derivatives of K wrt xi}
\begin{align}
    &\pdv{\Kzeronlscha}{\freeparambmnlscha}   = 
    \hbar^2 \sum_{ab=1}^{3N}
    \gtildenlscha{a}{b} \dtildelogJdq{a} \pdv{\dtildelogJdq{b}}{\freeparambmnlscha}  \\
    & +\frac{\hbar^2}{2} \sum_{ab=1}^{3N}\pdv{\gtildenlscha{a}{b}}{\freeparambmnlscha}
    \left(\dtildelogJdq{a} \dtildelogJdq{b}
    +  \frac{\Ytildenlscha{a}{b}}{4} + \Atildenlscha{a}{b}\right)  \notag \\
    & \pdv{\Konenlscha{a}}{\freeparambmnlscha}  = \sum_{b=1}^{3N}\frac{\hbar^2}{2}
    \left(
    \pdv{\gtildenlscha{a}{b}}{\freeparambmnlscha} \dtildelogJdq{b}
    +\gtildenlscha{a}{b}\pdv{\dtildelogJdq{b}}{\freeparambmnlscha} \right)\\
    &\pdv{\Ktwonlscha{a}{b}}{\freeparambmnlscha}
     = -\frac{\hbar^2}{8} \pdv{\gtildenlscha{a}{b}}{\freeparambmnlscha} 
\end{align}
\end{subequations}

The derivative of the BOES is straightforward and depends on the BO forces
\begin{equation}
    \averagegaussnl{\pdv{V^{\text{(BO)}}}{\freeparamnlscha{a}}} = 
    - \sum_{i=1}^{3N} \averagegaussnl{f^{\text{(BO)}}_i \pdv{\xinlscha_{i}(\bm{\unlscha})}{\freeparamnlscha{a}}}
\end{equation}

To compute the derivative with respect to $\FCbmnlscha$ we employ the formula introduced by Ref.\ \cite{Bianco}
\begin{equation}
\label{eq app: derivative wrt FC nlscha}
\begin{aligned}
     \pdv{\averagegaussnl{O}}{\FCnlscha{a}{b}} & =  
    \averagegaussnl{\pdv{O}{\FCnlscha{a}{b}}}  \\
    & +
     \sum_{ijk=1}^{3N} \frac{1}{2} \pdv{\Yinvnlscha{i}{j}}{\FCnlscha{a}{b}}
    \Ynlscha{i}{k} \averagegaussnl{\unlscha_k \pdv{O}{\unlscha_j}} 
\end{aligned}
\end{equation}
where the derivative of $\Ybmnlscha$ can be computed using the formalism of  appendix \ref{APP SUBSEC: Tensor derivatives}. The first term in Eq.\ \eqref{eq app: derivative wrt FC nlscha} considers the explicit dependence on $\FCbmnlscha$ while the second one takes into account the change of $\gaussnlscha(\bm{\unlscha})$ with respect to $\FCbmnlscha$. 

Using Eq.\ \eqref{eq app: derivative wrt FC nlscha}, the derivative of $\Fnl$ with respect to the auxiliary force constant is
\begin{equation}
\begin{aligned}
    & \pdv{\Fnl}{\FCnlscha{a}{b}} = \averagegaussnl{\pdv{\Knlscha}{\FCnlscha{a}{b}}}  - T \pdv{\Snl}{\FCnlscha{a}{b}}\\
    & +  \sum_{ijk=1}^{3N}  \frac{1}{2} \pdv{\Yinvnlscha{i}{j}}{\FCnlscha{a}{b}}
    \Ynlscha{i}{k} \averagegaussnl{\unlscha_k \left(\pdv{\Knlscha}{\unlscha_j} +
    \pdv{V^{\text{(BO)}}}{\unlscha_j}\right)} 
\end{aligned}
\end{equation}

The derivative of the kinetic energy kernel $\Knlscha$ with respect to $\FCbmnlscha$
\begin{equation}
    \hspace{-0.1cm} \pdv{\Knlscha}{\FCnlscha{a}{b}}
    = \sum_{ij=1}^{3N} \Ktwonlscha{i}{j} \pdv{ \Ltwonlscha{j}{i} }{\FCbmnlscha}
    + \sum_{i=1}^{3N} \Konenlscha{i}\pdv{\Lonenlscha{i}}{\FCbmnlscha}
    + \pdv{\Kzeronlscha}{\FCbmnlscha}
\end{equation}
where
\begin{subequations}
\begin{align}
    & \pdv{\Kzeronlscha}{\FCbmnlscha} =
    \frac{\hbar^2}{2} \sum_{ab=1}^{3N}\gtildenlscha{a}{b}
    \left(\frac{1}{4} \pdv{\Ytildenlscha{a}{b}}{\FCbmnlscha}
    + \pdv{\Atildenlscha{a}{b}}{\FCbmnlscha}\right)\\
    & \pdv{\Lonenlscha{i}}{\FCbmnlscha}  = 
    \sum_{m=1}^{3N} \pdv{\Ytildenlscha{i}{m}}{\FCbmnlscha} \widetilde{\unlscha}_m \\
     & \pdv{\Ltwonlscha{i}{j}}{\FCbmnlscha}   = 
    \pdv{\Ytildenlscha{i}{j}}{\FCbmnlscha}  
     \\
    & -  \sum_{mn=1}^{3N}
    \left(\pdv{\Ytildenlscha{j}{m}}{\FCbmnlscha} \Ytildenlscha{i}{n} 
    + \pdv{\Ytildenlscha{i}{m}}{\FCbmnlscha} \Ytildenlscha{j}{n}\right)
    \widetilde{\unlscha}_m \widetilde{\unlscha}_n \notag
\end{align}
\end{subequations}
The derivatives of $\Ybmnlscha$ and $\Abmnlscha$ are discussed in appendix \ref{APP SUBSEC: Tensor derivatives}.

The derivative of the kinetic energy kernel $\Knlscha$ with respect to $\widetilde{\bm{\unlscha}}$
\begin{equation}
\begin{aligned}
    & \pdv{\Knlscha(\bm{\unlscha})}{\widetilde{\unlscha}_a}
    =  \sum_{ij=1}^{3N}  \pdv{}{\widetilde{\unlscha}_a} 
    \left(\Ktwonlscha{i}{j} \Ltwonlscha{j}{i}\right)
    \\
    & + \sum_{i=1}^{3N} \pdv{}{\widetilde{\unlscha}_a}
    \left(\Konenlscha{i}\Lonenlscha{i} \right) + \pdv{\Kzeronlscha}{\widetilde{\unlscha}_a}
\end{aligned}
\end{equation}
The derivatives of $\Kzeronlscha$, $\Konebmnlscha$, and $\Ktwobmnlscha$  are 
\begin{subequations}
\label{eq app: derivatives of K wrt u}
\begin{align}
    &\pdv{\Kzeronlscha}{\widetilde{\unlscha}_i}   = 
    \hbar^2 \sum_{ab=1}^{3N}
    \gtildenlscha{a}{b} \dtildelogJdq{a} \pdv{\dtildelogJdq{b}}{\widetilde{\unlscha}_i}  \\
    & +\frac{\hbar^2}{2} \sum_{ab=1}^{3N}\pdv{\gtildenlscha{a}{b}}{\widetilde{\unlscha}_i}
    \left(\dtildelogJdq{a} \dtildelogJdq{b}
    +  \frac{\Ytildenlscha{a}{b}}{4} + \Atildenlscha{a}{b}\right)  \notag\\
    & \pdv{\Konenlscha{a}}{\widetilde{\unlscha}_i}  = \sum_{b=1}^{3N}\frac{\hbar^2}{2}\left(
    \pdv{\gtildenlscha{a}{b}}{\widetilde{\unlscha}_i} \dtildelogJdq{b}
    +\gtildenlscha{a}{b}\pdv{\dtildelogJdq{b}}{\widetilde{\unlscha}_i} \right)\\
    &\pdv{\Ktwonlscha{a}{b}}{\widetilde{\unlscha}_i}
     = -\frac{\hbar^2}{8} \pdv{\gtildenlscha{a}{b}}{\widetilde{\unlscha}_i}
\end{align}
\end{subequations}
where all the terms can be computed using Eqs \eqref{eq app: first derivative of d} \eqref{eq app: diff inv metric tensor wrt u}. The derivatives of $\Lonebmnlscha$  and $\Ltwobmnlscha$ are
\begin{subequations}
\begin{align}
      &\pdv{\Lonenlscha{i}}{\widetilde{\unlscha}_j}  \hspace{-0.1cm}
     =  \Ytildenlscha{i}{j} \\
     &\pdv{\Ltwonlscha{i}{j}}{\widetilde{\unlscha}_k} \hspace{-0.1cm}
     =  - \hspace{-0.2cm} \sum_{m=1}^{3N}\left( 
    \Ytildenlscha{j}{k} \Ytildenlscha{i}{m} 
    + \Ytildenlscha{i}{k} \Ytildenlscha{j}{m} \right) \widetilde{\unlscha}_m
\end{align}
\end{subequations}

The derivative of the BOES is
\begin{equation}
    \pdv{V^{\text{(BO)}}}{\widetilde{\unlscha}_a} = -\sum_{b=1}^{3N}
    \frac{f^{\text{(BO)}}_b}{\sqrt{m_b}} \Jtildenlscha{b}{a}
\end{equation}

The derivative of the entropy $\Snl$ with respect to $\FCbmnlscha$ is computed using the chain rule
\begin{equation}
    \pdv{\Snl}{\FCnlscha{a}{b}} 
    = \sum_{ij=1}^{3N} \invsqrtmasstnsT{ai} \pdv{\Snl}{\Dnlscha{i}{j}} \invsqrtmasstns{jb}
\end{equation}
and the derivative with respect to $\Dbmnlscha$
\begin{equation}
\begin{aligned}
     \pdv{\Snl}{\Dnlscha{a}{b}} 
     & = \sum_{\mu=1}^{3N} 
     \pdv{\onlscha{\mu}{}^2}{\Dnlscha{a}{b}} 
     \frac{1}{2\onlscha{\mu}} \pdv{\Snl}{\onlscha{\mu}} \\
     & = \sum_{\mu=1}^{3N} 
     \frac{\polnlscha{\mu}{a}\polnlscha{\mu}{b}}{2\onlscha{\mu}}
     \pdv{\nnlscha{\mu}}{\onlscha{\mu}}\pdv{\Snl}{\nnlscha{\mu}}
\end{aligned}
\end{equation}
where the derivatives of $\nnlscha{\mu}$ and of $\Snl$ are
\begin{subequations}
\begin{align}
    \pdv{\nnlscha{\mu}}{\onlscha{\mu}} & = - \beta \hbar\nnlscha{\mu} 
    (1 + \nnlscha{\mu}) \\
    \pdv{\Snl}{\nnlscha{\mu}} & =  k_\text{B} \beta \hbar \onlscha{\mu}
\end{align}
\end{subequations}

The derivative with respect to $\sqrtmasstnsbm$ is computed using a similar formula to Eq.\ \eqref{eq app: derivative wrt FC nlscha}
\begin{equation}
\label{eq app: derivative wrt sqrt mass}
\begin{aligned}
     \pdv{\averagegaussnl{O}}{\sqrtmasstns{ab}} & =  
    \averagegaussnl{\pdv{O}{\sqrtmasstns{ab}}}  \\
    & +
     \sum_{ijk=1}^{3N} \frac{1}{2} \pdv{\Yinvnlscha{i}{j}}{\sqrtmasstns{ab}}
    \Ynlscha{i}{k} \averagegaussnl{\unlscha_k \pdv{O}{\unlscha_j}} 
\end{aligned}
\end{equation}
Using Eq.\ \eqref{eq app: derivative wrt FC nlscha}, the derivative of $\Fnl$ with respect to the auxiliary force constant is
\begin{equation}
\begin{aligned}
    & \pdv{\Fnl}{\sqrtmasstns{ab}} = \averagegaussnl{\pdv{\Knlscha}{\sqrtmasstns{ab}}}  
    - T \pdv{\Snl}{\sqrtmasstns{ab}}\\
    & +  \sum_{ijk=1}^{3N} \frac{1}{2} \pdv{\Yinvnlscha{i}{j}}{\sqrtmasstns{ab}}
    \Ynlscha{i}{k} \averagegaussnl{\unlscha_k \left(\pdv{\Knlscha}{\unlscha_j} +
    \pdv{V^{\text{(BO)}}}{\unlscha_j}\right)} 
\end{aligned}
\end{equation}
The derivative of the kinetic energy kernel $\Knlscha$ with respect to $\sqrtmasstnsbm$
\begin{equation}
    \hspace{-0.1cm} \pdv{\Knlscha}{\sqrtmasstnsbm}
    = \sum_{ij=1}^{3N} \Ktwonlscha{i}{j} 
    \pdv{ \Ltwonlscha{j}{i} }{\sqrtmasstnsbm}
    + \sum_{i=1}^{3N} \Konenlscha{i}
    \pdv{\Lonenlscha{i}}{\sqrtmasstnsbm}
    + \pdv{\Kzeronlscha}{\sqrtmasstnsbm}
\end{equation}
where
\begin{subequations}
\begin{align}
    & \pdv{\Kzeronlscha}{\sqrtmasstnsbm} =
    \frac{\hbar^2}{2} \sum_{ab=1}^{3N}
    \gtildenlscha{a}{b}
    \left(\frac{1}{4} 
    \pdv{\Ytildenlscha{a}{b}}{\sqrtmasstnsbm}
    + 
    \pdv{\Atildenlscha{a}{b}}{\sqrtmasstnsbm}
    \right)\\
    & \pdv{\Lonenlscha{i}}{\sqrtmasstnsbm}  = 
    \sum_{m=1}^{3N} \pdv{\Ytildenlscha{i}{m}}{\sqrtmasstnsbm} \widetilde{\unlscha}_m \\
     & \pdv{\Ltwonlscha{i}{j}}{\sqrtmasstnsbm} = \pdv{\Ytildenlscha{i}{j}}{\sqrtmasstnsbm}  \\
    & -  \sum_{mn=1}^{3N}
    \left(\pdv{\Ytildenlscha{j}{m}}{\sqrtmasstnsbm} \Ytildenlscha{i}{n} 
    + \pdv{\Ytildenlscha{i}{m}}{\sqrtmasstnsbm} \Ytildenlscha{j}{n}\right)
    \widetilde{\unlscha}_m \widetilde{\unlscha}_n \notag
\end{align}
\end{subequations}
The derivatives with respect to $\sqrtmasstnsbm$ of $\Ybmnlscha$ and $\Abmnlscha$ are discussed in appendix \ref{APP SUBSEC: Tensor derivatives}.

The derivative of the entropy is
\begin{equation}
    \pdv{\Snl}{\sqrtmasstns{ab}}=
    \sum_{ij=1}^{3N}
    \pdv{\Snl}{\Dnlscha{i}{j}} 
    \pdv{\Dnlscha{i}{j}}{\sqrt{\mathcal{M}}_{ab}}
\end{equation}
where the derivative of $\Dbmnlscha$ is given in Eqs \eqref{eq app: d D d sqrt M}.

\section{Computational details}
\label{APP: Computational details}

\subsection{Exact diagonalization}

The exact diagonalization (ED) of the 1D Schrodinger equation is performed as a diagonalization of $N \times N$ matrix where $N=2000$ is the size of the uniform Cartesian grid between $\pm 5$ Bohr. We employed the \texttt{numpy} function \texttt{numpy.linalg.eigh} \cite{numpy}.
To get the free energy at temperature $T$ we included the excited states up to $10 k_\text{B} T$ checking that are normalized and orthogonal. 

\begin{figure}[!htb]
    \centering
    \begin{minipage}[c]{1.0\linewidth}
    \includegraphics[width=1.0\textwidth]{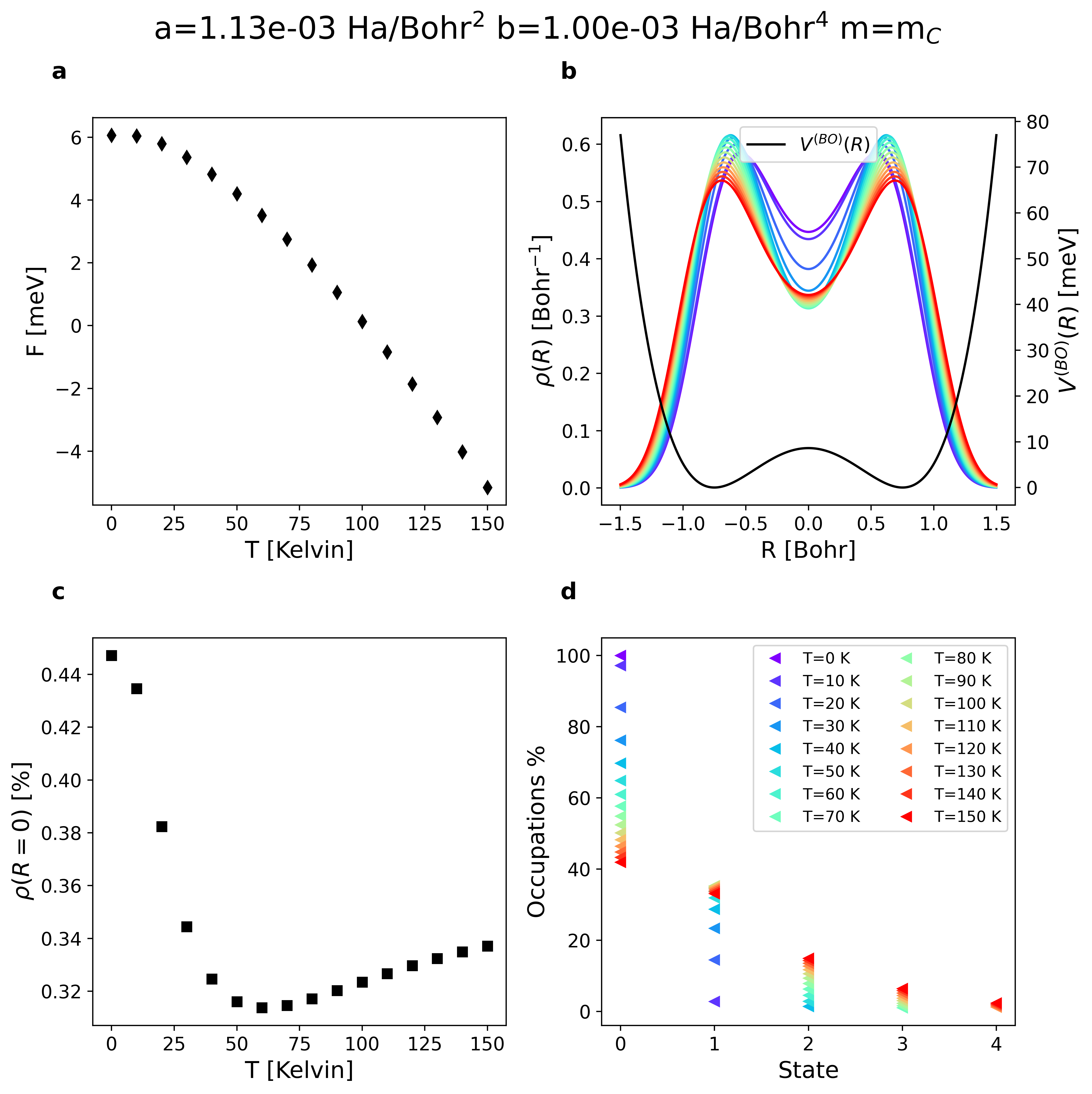}
    \end{minipage}
    \caption{Exact results for $a=1.13 \cdot 10^{-3}$ Ha/Bohr$^2$ $b=1.00 \cdot 10^{-3}$ Ha/Bohr$^4$ $m=m_\text{\ch{C}}$. Fig.\ a shows the free energy as a function of temperature $T$. Fig.\ b reports $V(R)$ with $\rho(R)$ as a function of $T$. In Fig.\ c we plot the probability at $R=0$ Bohr as a function of $T$. Fig.\ d plots the eigenstates' occupations for each $T$.}
    \label{fig:ED 1}
\end{figure}

\begin{figure}[!htb]
    \centering
    \begin{minipage}[c]{1.0\linewidth}
    \includegraphics[width=1.0\textwidth]{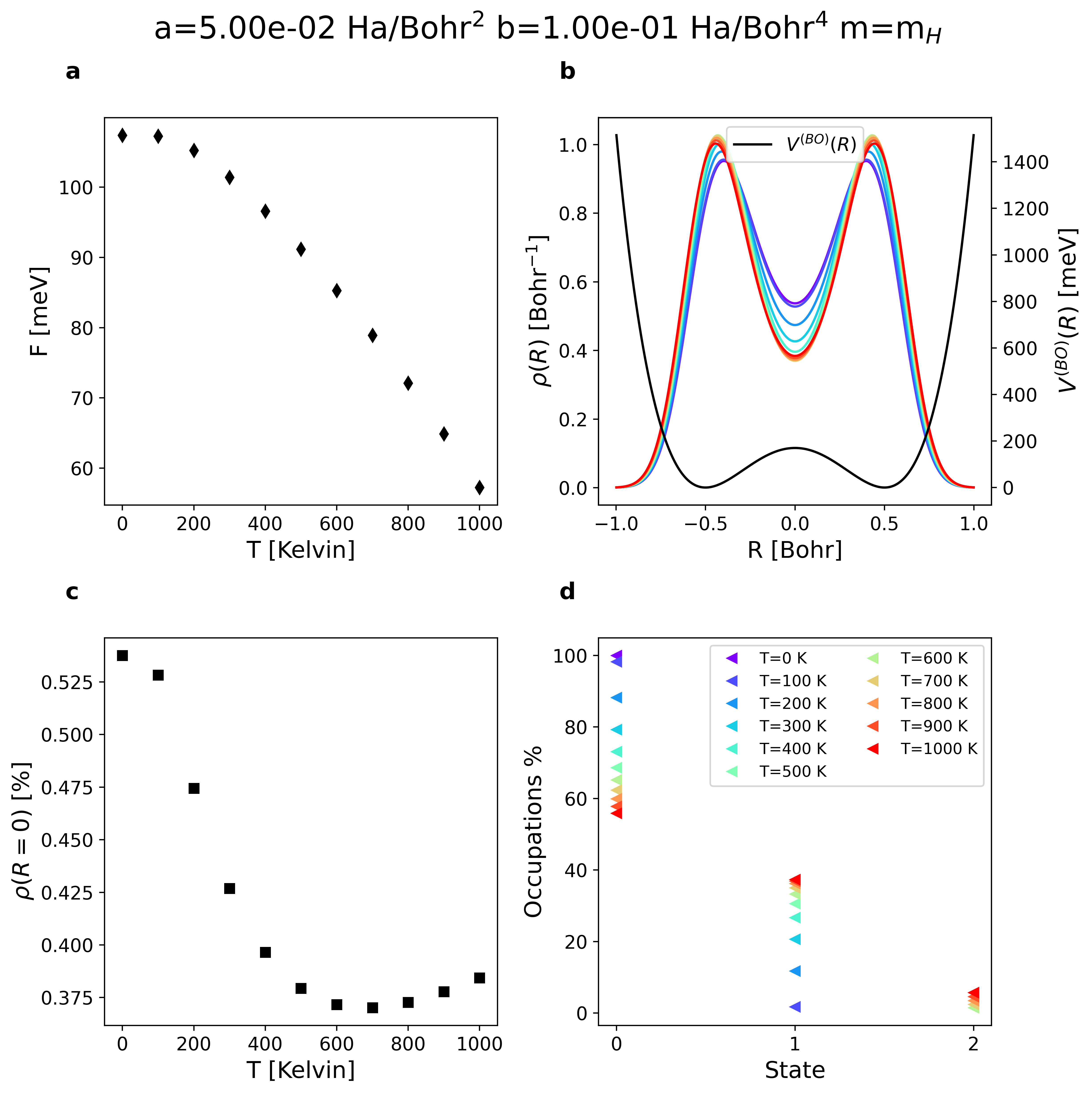}
    \end{minipage}
    \caption{Exact results for $a=0.05 $ Ha/Bohr$^2$ $b=0.1$ Ha/Bohr$^4$ $m=m_\text{\ch{H}}$. Fig.\ a shows the free energy as a function of temperature $T$. Fig.\ b reports $V(R)$ with $\rho(R)$ as a function of $T$. In Fig.\ c we plot the probability at $R=0$ Bohr as a function of $T$. Fig.\ d plots the eigenstates' occupations for each $T$.}
    \label{fig:ED 2}
\end{figure}

In Figs \ref{fig:ED 1} \ref{fig:ED 2} we report the exact results for $a=1.13 \cdot 10^{-3}$ Ha/Bohr$^2$ $b=1.00 \cdot 10^{-3}$ Ha/Bohr$^4$ $m=m_\text{\ch{C}}$ and for $a=0.05 $ Ha/Bohr$^2$ $b=0.1$ Ha/Bohr$^4$ $m=m_\text{\ch{H}}$. 
Note that $\rho(R=0\text{ Bohr})$ (Fig.\ \ref{fig:ED 1} c and Fig.\ \ref{fig:ED 2} c) is not a monotonous function of $T$. Initially, it decreases as we populate the first excited state with a node in $R=0$ Bohr. Then, upon heating, $\rho(R=0\text{ Bohr})$ increases as the temperature populates the excited states with non-zero probability in $R=0$ Bohr.

\newpage

\subsection{SCHA and NLSCHA simulations}
The SCHA and NLSCHA minimizations were performed on a 1D uniform grid in $\unlscha$-space of size $3000$ between $\pm 5$ Bohr using the conjugate gradient (CG) implemented in \texttt{scipy.minimize.optimize} \cite{scipy}. None of the stochastic techniques were used. The SCHA/NLSCHA minimizations were done both by increasing and decreasing the temperature and for each new temperature, the initial solution was taken from the previous one. We observed hysteresis in the NLSCHA free parameters and chose to present the solution with the lowest free energies.

\begin{figure}[!htb]
    \centering
    \begin{minipage}[c]{1.0\linewidth}
    \includegraphics[width=1.0\textwidth]{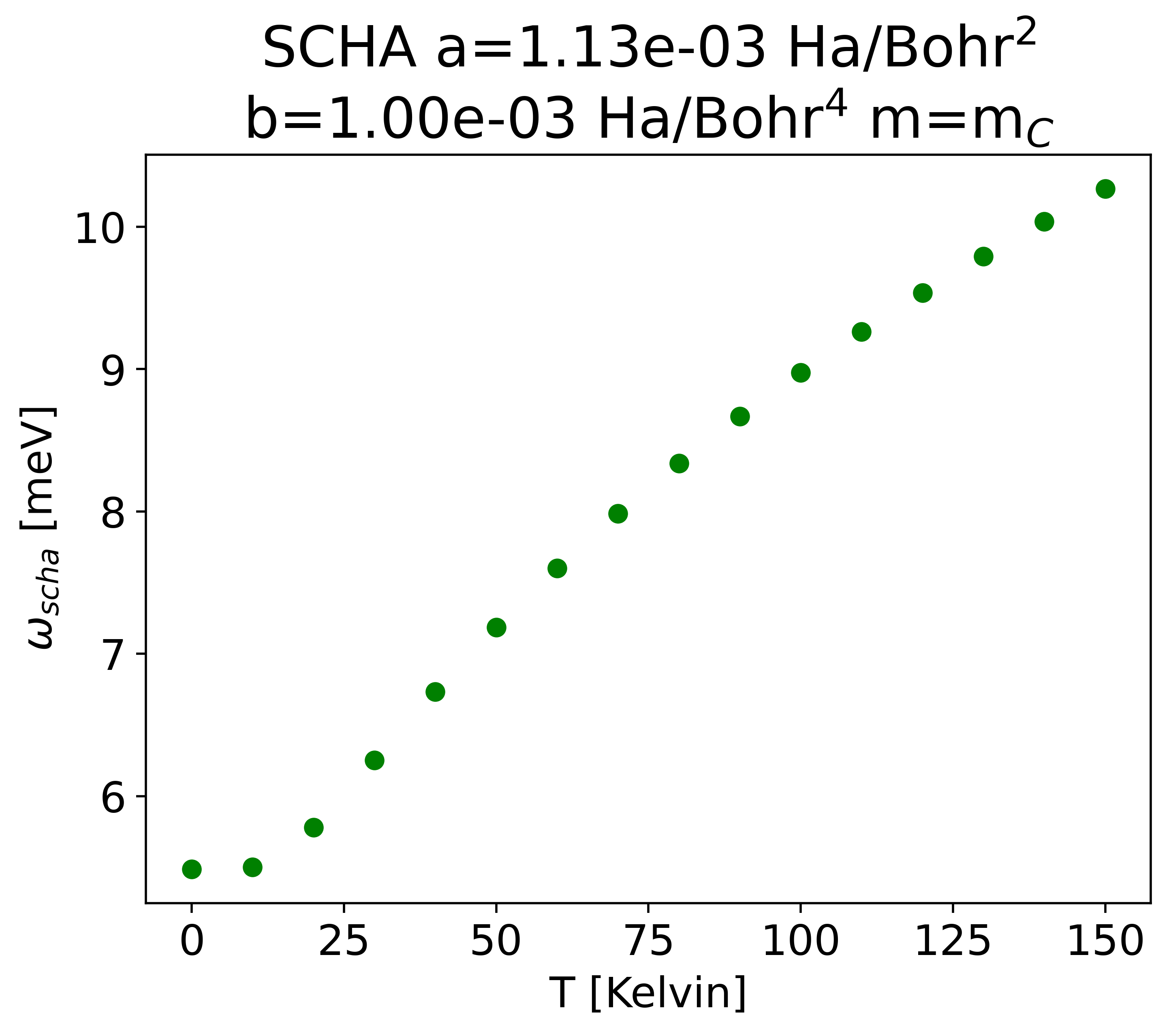}
    \end{minipage}
    \caption{SCHA auxiliary frequency for $a=1.13 \cdot 10^{-3}$ Ha/Bohr$^2$ $b=1.00 \cdot 10^{-3}$ Ha/Bohr$^4$ $m=m_\text{\ch{C}}$. }
    \label{fig: SCHA 1}
\end{figure}

\begin{figure}[!htb]
    \centering
    \begin{minipage}[c]{1.0\linewidth}
    \includegraphics[width=1.0\textwidth]{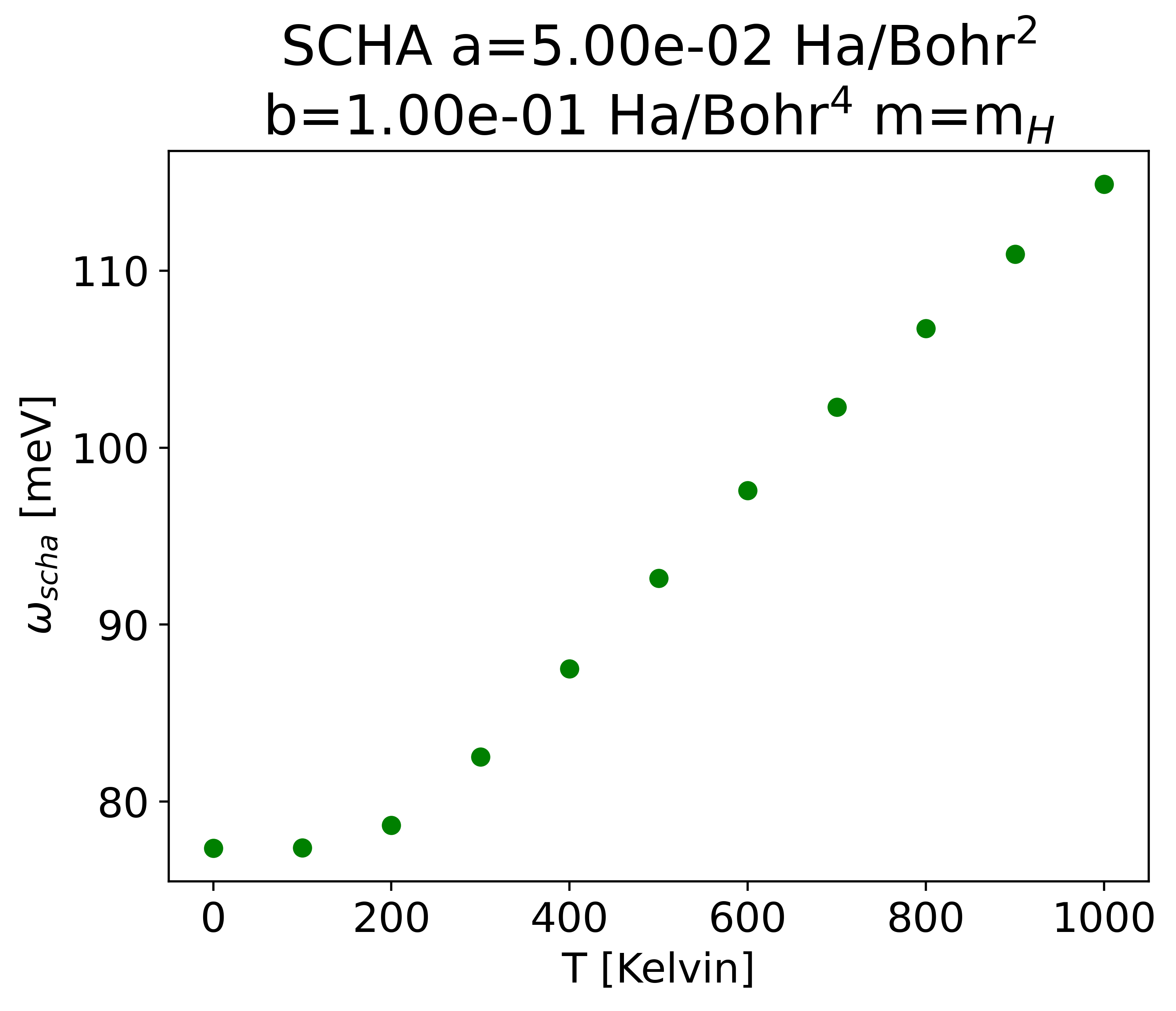}
    \end{minipage}
    \caption{SCHA auxiliary frequency for $a=0.05 $ Ha/Bohr$^2$ $b=0.1$ Ha/Bohr$^4$ $m=m_\text{\ch{H}}$. }
    \label{fig: SCHA 2}
\end{figure}
\begin{figure}[!htb]
    \centering
    \begin{minipage}[c]{1.0\linewidth}
    \includegraphics[width=1.0\textwidth]{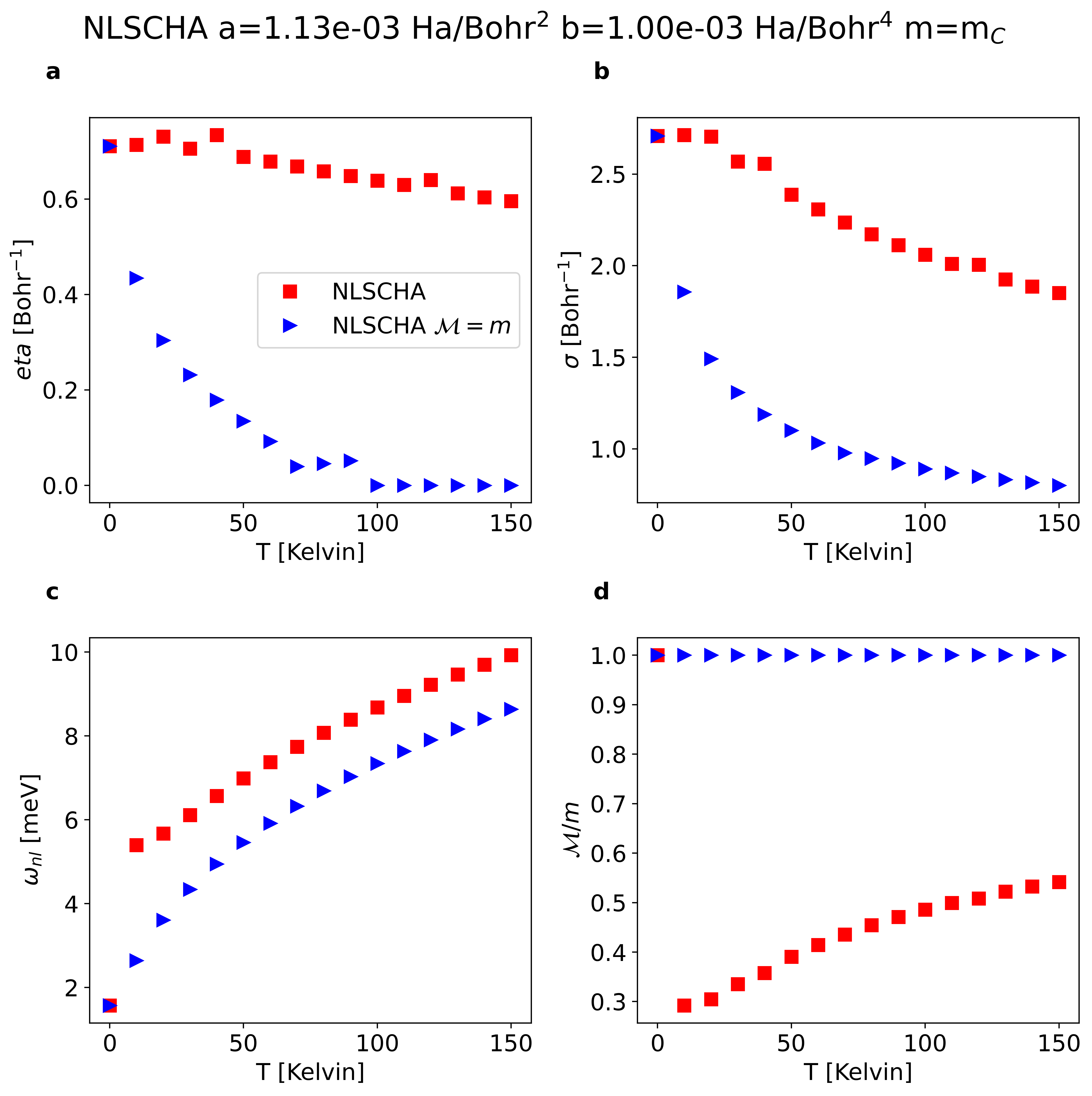}
    \end{minipage}
    \caption{NLSCHA parameters for $a=1.13 \cdot 10^{-3}$ Ha/Bohr$^2$ $b=1.00 \cdot 10^{-3}$ Ha/Bohr$^4$ $m=m_\text{\ch{C}}$. The nonlinear transformation used is Eq.\ \eqref{eq: toy model transformation}. Figs a-d show the free parameters, $\eta$, $\sigma$, $\omega_\text{nl}$ and $\masstns{}/m$.}
    \label{fig: NLSCHA 1}
\end{figure}
\begin{figure}[!htb]
    \centering
    \begin{minipage}[c]{1.0\linewidth}
    \includegraphics[width=1.0\textwidth]{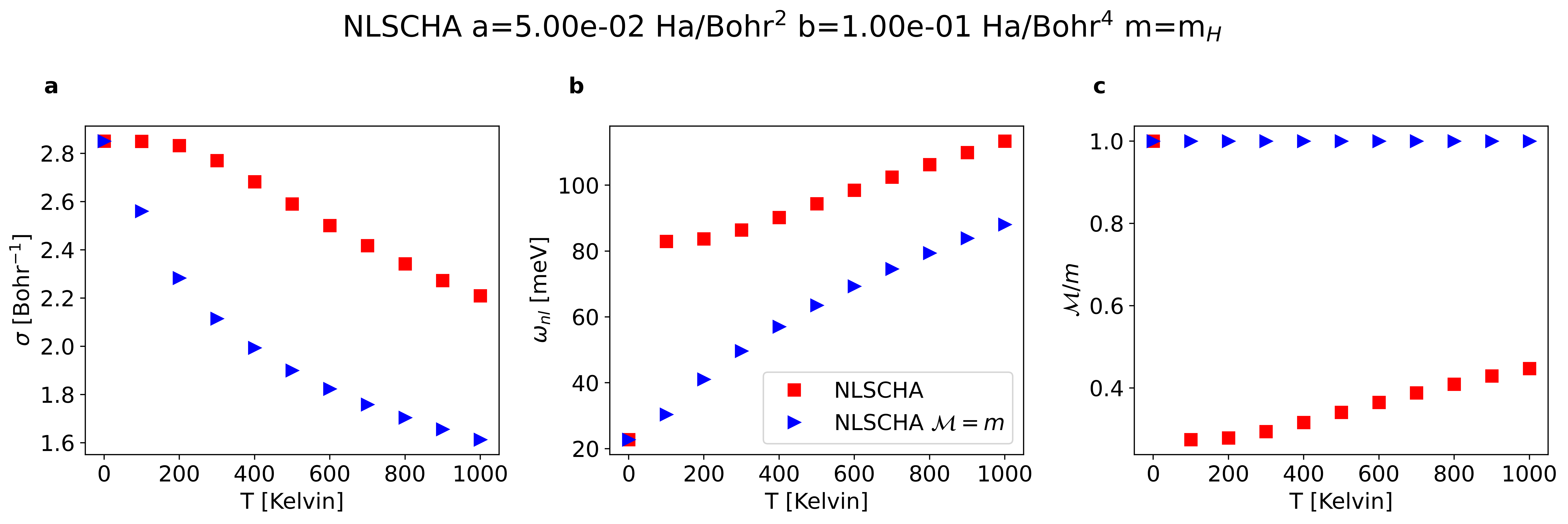}
    \end{minipage}
    \caption{NLSCHA parameters for $a=0.05 $ Ha/Bohr$^2$ $b=0.1$ Ha/Bohr$^4$ $m=m_\text{\ch{H}}$. The nonlinear transformation used is Eq.\ \eqref{eq: toy model transformation} with $\eta=0$ Bohr$^{-1}$. Figs a-c show the free parameters, $\sigma$, $\omega_\text{nl}$ and $\masstns{}$.}
    \label{fig: NLSCHA 2}
\end{figure}

In Figs \ref{fig: SCHA 1} \ref{fig: SCHA 2} we report the SCHA auxiliary frequency. In Figs \ref{fig: NLSCHA 1} \ref{fig: NLSCHA 2} we report the NLSCHA free parameters for $a=1.13 \cdot 10^{-3}$ Ha/Bohr$^2$, $b=1.00 \cdot 10^{-3}$ Ha/Bohr$^4$, $m=m_\text{\ch{C}}$ and for $a=0.05 $ Ha/Bohr$^2$, $b=0.1$ Ha/Bohr$^4$, $m=m_\text{\ch{H}}$.

\newpage
\section{Is $\Snl$ the correct entropy?}
\label{APP: dF dT}
In this appendix, we show that the analytical expression of the entropy $\Snl$ (Eq.\ \eqref{eq: entropy in u space}) coincides with the temperature derivative of the free energy $\Fnl$ computed at the minimum. First, we compute the total derivative of $\Fnl$
\begin{equation}
\label{eq app: dF dT}
\begin{aligned}
    \frac{d \Fnl}{d T} = & \pdv{\Fnl}{\FCbmnlscha}\cdot\pdv{\FCbmnlscha}{T} +
    \pdv{\Fnl}{\masstnsbm}\cdot\pdv{\masstnsbm}{T}+
    \pdv{\Fnl}{\freeparambmnlscha}\cdot\pdv{\freeparambmnlscha}{T}  \\
    & + \pdv{\Fnl}{T}
\end{aligned}
\end{equation}
The first line of Eq.\ \eqref{eq app: dF dT} considers the implicit dependence of the NLSCHA free parameters with $T$. The second line of Eq.\ \eqref{eq app: dF dT} contains the explicit temperature dependence. 
We evaluate Eq.\ \eqref{eq app: dF dT} using the NLSCHA parameters that self-consistently minimize $\Fnl$. Thanks to Eqs \eqref{eq: nlscha equilibrium conditions} we are left with the explicit $T$ dependence of $\Fnl$
\begin{equation}
\label{eq app: dF dT min}
     \frac{d \Fnl}{d T}\biggl|_{(0)} =   \pdv{\Fnl}{T} \biggl|_{(0)} =
     \left(\sum_{\mu=1}^{3N}\pdv{\Fnl}{\nnlscha{\mu}} \pdv{\nnlscha{\mu}}{T}
    - \Snl \right) \biggl|_{(0)} 
\end{equation}
where we note that all the terms in $\Fnl$ depends on $T$ through $\nnlscha{\mu}$ (Eq.\ \eqref{eq: BE occupation}) with the exception of the $T$ multiplying the entropy $\Snl$. There are two ways to have
\begin{equation}
\label{eq app: dF dT results}
    \frac{d \Fnl}{d T} \biggl|_{(0)} = - \Snl{}|_{(0)} 
\end{equation}
The first solution is to use the occupations as free parameters
\begin{equation}
\label{eq app: dF dn = 0}
    \pdv{\Fnl}{\nnlscha{\mu}} \biggl|_{(0)} = 0
\end{equation}
but, as discussed in section \ref{SEC: Entropy discussion}, we would relax the hypothesis that a harmonic Hamiltonian generates the NLSCHA trial density matrix.
The second solution is to have
\begin{equation}
\label{eq app: dn dT = 0}
    \pdv{\nnlscha{\mu}}{T} = 0
\end{equation}
Eq.\ \eqref{eq app: dn dT = 0} implies defining effective thermal energies on each mode $\beta^{-1}_\mu >0 $ (Eq.\ \eqref{eq: BE occupation tmode}). Here, we prove that the $\{\beta_{\mu}\}$ are equivalent to introducing an effective mass tensor $\masstnsbm$ keeping the standard Bose-Einstein occupations (Eq.\ \eqref{eq: BE occupation}). Indeed, the mode-dependent temperatures enter only as multiplicative factors of $\onlscha{\mu}$ so we can replace $\beta_\mu$ with $\beta$ by changing the auxiliary frequencies
\begin{equation}
\label{eq app: beta mu into omega mu}
\begin{aligned}
    & \beta_\mu \hbar \onlscha{\mu} =
    \beta \hbar\frac{\beta_\mu}{\beta}  \onlscha{\mu}\\
    &=
     \beta \hbar \sqrt{\sum_{ab=1}^{3N} \left(\frac{\beta_\mu}{\beta} \polnlscha{\mu}{a}\right) \Dnlscha{a}{b} \left(\frac{\beta_\mu}{\beta} \polnlscha{\mu}{b}\right)},
\end{aligned}
\end{equation}
So $\beta_\mu/\beta$ can be reabsorbed as they change the polarization vectors. The most general way to take into account the rescaling of $\Dbmnlscha$ (Eq.\ \eqref{eq app: beta mu into omega mu}) is to optimize both the force constant $\FCbmnlscha$ and a non-diagonal mass tensor $\masstnsbm$.

\begin{figure}[!htb]
    \centering
    \begin{minipage}[c]{1.0\linewidth}
    \includegraphics[width=1.0\textwidth]{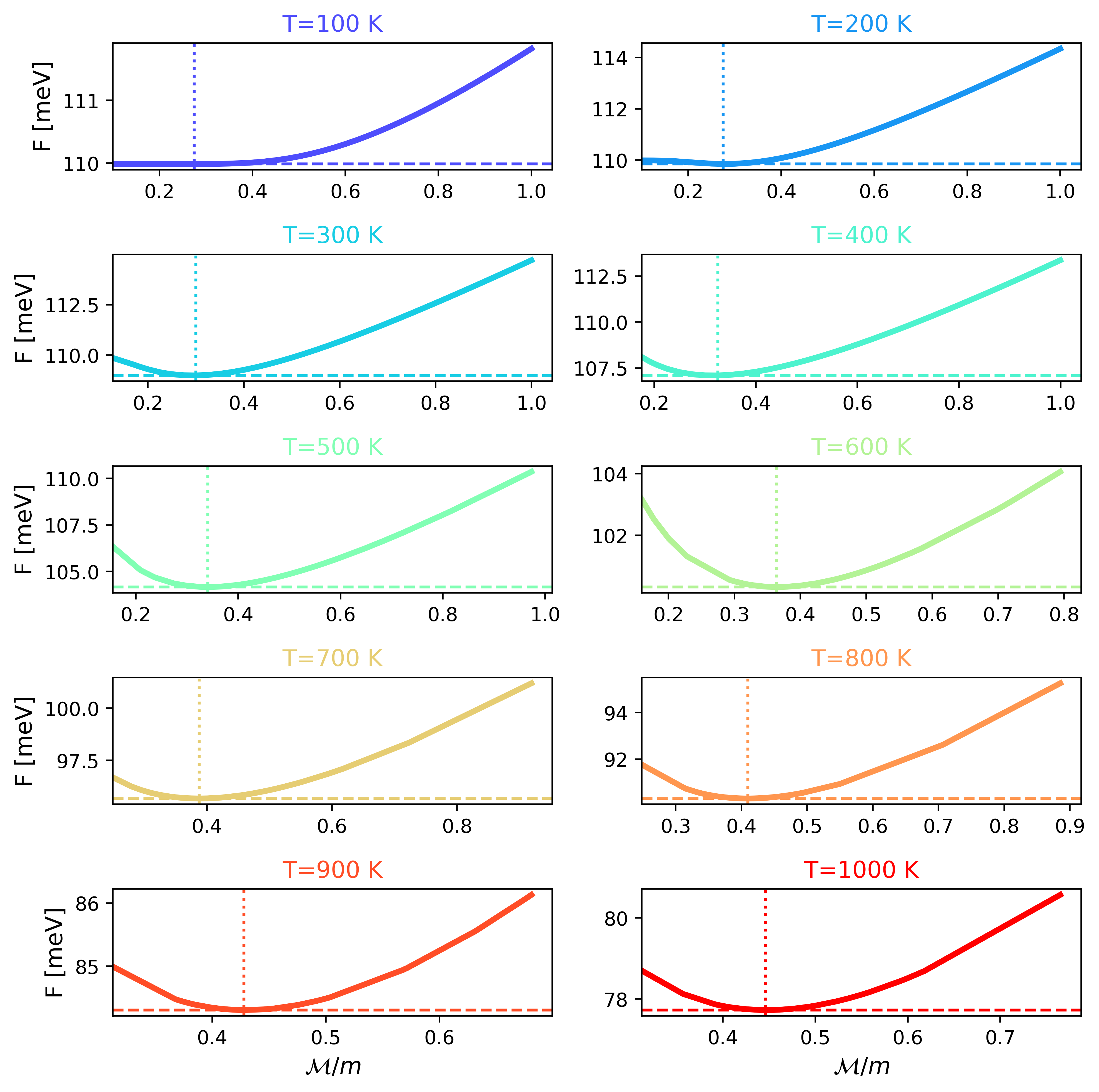}
    \end{minipage}
    \caption{The NLSCHA free energy $\Fnl$ ($a=0.05 $ Ha/Bohr$^2$ $b=0.1$ Ha/Bohr$^4$ $m=m_\text{\ch{H}}$) as a function of the ratio $\mathcal{M}/m$. The nonlinear transformation is given in Eq.\ \eqref{eq: toy model transformation} with $\eta=0$ Bohr$^{-1}$. The horizontal/vertical dotted lines indicate the minimum $\Fnl$ and the corresponding value of $\masstns{}/m$.}
    \label{fig:F(T) vs M}
\end{figure}

In Fig.\ \ref{fig:F(T) vs M} we report $\Fnl$ as a function of the ratio $\masstns{}/m$ for $a=0.05 $ Ha/Bohr$^2$ $b=0.1$ Ha/Bohr$^4$ $m=m_\text{\ch{H}}$. The nonlinear transformation used is Eq.\ \eqref{eq: toy model transformation} with $\eta=0$ Bohr$^{-1}$. As the temperature increases, the value of $\masstns{}/m$ corresponding to the minimum of $\Fnl$ deviates from $1$ and decreases. So at finite temperatures, $\pdv{\Fnl}{\masstns{}}|_{\masstns{}=m}\neq 0$ and
$\frac{d \Fnl}{d T}|_{(0),\masstns{}=m}\neq -T \Snl|_{(0),\masstns{}=m}$ (see Fig.\ \ref{fig:F(T) vs M} c-d).

\newpage
\bibliography{apssamp}

\end{document}